\documentclass[11pt,a4paper]{amsart}

\usepackage[a4paper,textwidth=160mm,textheight=240mm]{geometry}
\usepackage[T1]{fontenc}
\usepackage[utf8]{inputenc}
\usepackage{lmodern}
\usepackage{amssymb,etoolbox,graphicx,microtype}
\usepackage{needspace}
\pretocmd{\part}{\Needspace{12\baselineskip}}{}{}
\usepackage[numbers,sort&compress]{natbib}

\usepackage{xurl}
\usepackage[hidelinks]{hyperref}
\theoremstyle{plain}
\newtheorem{theorem}{Theorem}[section]
\newtheorem{lemma}[theorem]{Lemma}
\newtheorem{proposition}[theorem]{Proposition}
\newtheorem{corollary}[theorem]{Corollary}
\theoremstyle{definition}
\newtheorem{definition}[theorem]{Definition}
\newenvironment{acks}{\section*{Acknowledgements}}{}
\date{}
\usepackage{chngcntr}
\usepackage{multicol}
\usepackage{tikz-cd}
\usepackage[shortlabels]{enumitem}
\makeatletter
\@addtoreset{equation}{section}
\renewcommand\paragraph{%
  \@startsection{paragraph}{4}{\z@}%
    {-.7\baselineskip}{.3\baselineskip}%
    {\normalfont\bfseries}}
\makeatother
\numberwithin{equation}{section}
\counterwithin{figure}{section}

\providecommand{\uppsi}{\psi}
\newcommand{\diag}[1]{\csname diag-#1\endcsname}
\expandafter\def\csname diag-trianglewCPO\endcsname{%
\begin{tikzcd}[ampersand replacement=\&,column sep=large,row sep=large]
\catL(f)(Y) \arrow[r,"f'"] \arrow[d,swap,"\catL(f\leq g)_Y"] \& X\\
\catL(g)(Y) \arrow[ur,swap,"g'"] \&
\end{tikzcd}}
\expandafter\def\csname diag-initial-algebra-universal-property\endcsname{%
\begin{tikzpicture}[baseline=(current bounding box.center)]
\node (A) at (0,0) {$\Set^\op$};
\node (B) at (4,0) {$\Set^\op$};
\node (C) at (2,1.7) {$\catCat$};
\draw[->] (A) -- node[above left] {$\Vect^{(-)}$} (C);
\draw[->] (B) -- node[above right] {$\Vect^{(-)}$} (C);
\draw[->] (A) -- node[below] {$(-\sqcup\mathrm1)$} (B);
\draw[double,->] (1.55,.8) -- node[above] {$\uppsi$} (2.45,.8);
\end{tikzpicture}}
\expandafter\def\csname diag-triangle-monad-morphism\endcsname{%
\csname diag-initial-algebra-universal-property\endcsname}
\expandafter\def\csname diag-freyd-category-morphismm\endcsname{%
\begin{tikzcd}[ampersand replacement=\&,column sep=huge,row sep=large]
\catC \arrow[r,"\hat F"] \& \catC'\\
\catV \arrow[u,"j"] \arrow[r,swap,"F"] \&\catV' \arrow[u,swap,"j'"]
\end{tikzcd}}
\expandafter\def\csname diag-freyd-category-morphism-CHAD\endcsname{%
\begin{tikzcd}[ampersand replacement=\&,column sep=huge,row sep=large]
\PVMAN \arrow[r,"\DSempartial{}"] \&\PFam{\Vect^\op}\\
\VMAN \arrow[u,"\pman"] \arrow[r,swap,"\DSemtotal{}"] \&\Fam{\Vect^\op} \arrow[u,swap,"\pfam"]
\end{tikzcd}}
\expandafter\def\csname diag-correctness-total-language-partial\endcsname{%
\begin{tikzcd}[ampersand replacement=\&,column sep=huge,row sep=large]
\Syn \arrow[r,"\DSynrevt"] \arrow[d,swap,"\sem{-}"] \&\TPSYN \arrow[d,"\semt{-}"]\\
\FFMAN \arrow[r,swap,"\DSemCHAD{}"] \&\FFFVECT
\end{tikzcd}}
\expandafter\def\csname diag-correctness-total-language\endcsname{%
\begin{tikzcd}[ampersand replacement=\&,column sep=huge,row sep=large]
\Syntotal \arrow[r,"\DSynrevt{}"] \arrow[d,swap,"\sem{-}"] \&\Sigma_{\CSynt}\LSynt^{op} \arrow[d,"\semt{-}"]\\
\VMAN \arrow[r,swap,"\DSemtotal{}"] \&\Fam{\Vect^\op}
\end{tikzcd}}
\expandafter\def\csname diag-kleisli-morphism-pair\endcsname{%
\begin{tikzcd}[ampersand replacement=\&,column sep=huge,row sep=large]
\catC \arrow[r,"\lift H"] \&\catC'\\
\catV \arrow[u,"J"] \arrow[r,swap,"H"] \&\catV' \arrow[u,swap,"J'"]
\end{tikzcd}}
\expandafter\def\csname diag-basic-wcpo-chain\endcsname{%
\begin{tikzcd}[ampersand replacement=\&,column sep=large]
\leastelement\arrow[r]\&q(\leastelement)\arrow[r]\&\cdots\arrow[r]\&q^n(\leastelement)\arrow[r]\&\cdots
\end{tikzcd}}

\usepackage{xcolor}
\usepackage{amsmath,amsthm}
\usepackage{mathtools}
\usepackage{suffix}
\usepackage{amsfonts}
\newcommand{\inferrule}[2]{\frac{\begin{gathered}#1\end{gathered}}{\begin{gathered}#2\end{gathered}}}
\usepackage{enumitem}
\usepackage{twoopt}
\usepackage{amsthm}
\makeatletter
\RequirePackage{amsmath,amsfonts,suffix}

\RequirePackage{suffix}

\newcommand\generic@explainabove[3]{%
   \mathrel{%
		\overset{%
			\overset{\raisebox{0.5ex}{%
       \makebox[0pt][c]{%
				\begin{minipage}{\textwidth}%
				 \center%
				 \scriptsize%
         {#2}%
				\end{minipage}%
       }}}%
      {%
			 \raisebox{1ex}{${#3}$}
			 }}%
     {#1}}}

\newcommand\generic@explainbelow[3]{%
	\mathrel{%
		\underset{%
			\underset{%
				\makebox[0pt][c]{\scriptsize%
					\raisebox{-1ex}{%
						\begin{minipage}{\textwidth}%
							\center%
							{#2}%
						\end{minipage}%
 				}}
			}{%
				\raisebox{-1ex}{${#3}$}%
			}
		}{%
			#1%
		}
	}
}

\newcommand\explain[2]{\generic@explainabove{#1}{#2}{\downarrow}}
\WithSuffix\newcommand\explain*[2]{\generic@explainbelow{#1}{#2}{\uparrow}}

\newcommand\explainfar[2]{\generic@explainabove{#1}{#2}{\Big\downarrow}}
\WithSuffix\newcommand\explainfar*[2]{\generic@explainbelow{#1}{#2}{\Big\uparrow}}

\WithSuffix\newcommand\explain+[2]{\mathbin{\smash{\explain{#1}{#2}}}}
\WithSuffix\newcommand\explain-[2]{\mathbin{\smash{\explain*{#1}{#2}}}}

\RequirePackage{optparams}

\DeclareSymbolFont{pxfontssymbolsC}{U}{pxsyc}{m}{n}
\DeclareMathSymbol{\coloneqq}{\mathrel}{pxfontssymbolsC}{66}
\DeclareMathSymbol{\eqqcolon}{\mathrel}{pxfontssymbolsC}{67}
\DeclareMathSymbol{\medcirc}{\mathbin}{pxfontssymbolsC}{7}

\newcommand\parent[1]{\left({#1}\right)}

\newcommand\pair[2]{\left({#1}, {#2}\right)}
\WithSuffix\newcommand\pair*[2]{(#1, #2)}
\newcommand\triple[3]{\seq{{#1}, {#2}, {#3}}}
\WithSuffix\newcommand\triple*[3]{\seq*{{#1}, {#2}, {#3}}}

\newcommand\set[1]{\left\{#1\right\}}

\newcommand\ok@seq[3]{}%
\long\def\ok@seq[#1][#2]#3{\left\langle#3\right\rangle_{#1}^{#2}}
\newcommand\seq{\optparams{\ok@seq}{[{}][{}]}}
\newcommand\ok@seqstar[3]{}%
\long\def\ok@seqstar[#1][#2]#3{(#3)_{#1}^{#2}}
\WithSuffix\newcommand\seq*{\optparams{\ok@seqstar}{[{}][{}]}}
\newcommand\ok@coseq[3]{}%
\long\def\ok@coseq[#1][#2]#3{\left[#3\right]_{#1}^{#2}}
\newcommand\coseq{\optparams{\ok@coseq}{[{}][{}]}}

\newcommand\cross\times
\newcommand\inv[1]{{#1}^{-1}}
\WithSuffix\newcommand\inv*[1]{\inv{\parent{#1}}}
\WithSuffix\newcommand\inv-[1]{#1^{-1}}

\newcommand\isomorphic\cong
\newcommand\nisomorphic\ncong

\newcommand\reals{\mathbb{R}}

{\let\TEMP\varphi
\global\let\varphi\phi
\global\let\phi\TEMP

\let\TEMP\varepsilon
\global\let\varepsilon\epsilon
\global\let\epsilon\TEMP
}

\let\subset\subseteq

\newcommand\union\cup
\newcommand\intersect\cap
\newcommand\Intersect\bigcap

\newcommand\xto\xrightarrow
\newcommand\xfrom\xleftarrow

\mathchardef\mhyphen="2D

\newcommand\id[1][{}]{\mathrm{id}_{#1}}

\newcommand\Cat[1]{\mathcal{#1}\/}
\newcommand\Category[1]{{\mathbf{ #1}}}

\newcommand\Set{\Category{Set}}

\newcommand\Vect{\Category{Vect}}

\newcommand\diagonal\Delta
\newcommand\diagonalmap\delta

\newcommand\projection\pi
\newcommand\prodmorphism\seq

\newcommand\compose\circ

\newcommand\injection\iota
\newcommand\coprodmorphism\coseq

\makeatother

\usepackage{tocvsec2}  %

\usepackage{listings}

\AtEndPreamble{%
	\theoremstyle{definition}
    \newtheorem{remark}[theorem]{Remark}}
\AtEndPreamble{%
	\theoremstyle{definition}
	}

\newcommand\obj[1]{\mathrm{ob}#1}

\newcommand\LOp{\mathsf{LOp}}

\newcommand\concINDEX{\mathfrak{FV}\mathsf{ect}}
\newcommand\lop{\mathsf{lop}}
\newcommand\gdefinedby{::=}

\newcommand{\defcolon}{\colon\!\!=}
\newcommand{\defeqq}{\stackrel {\mathrm{def}}=}
\newcommand{\defeq}{\stackrel {\mathrm{def}}=}

\newcommand{\initiall}{\mathsf{0}}

\newcommand{\terminall}{\mathsf{1}}

\newcommand\domain[1]{\texttt{domain}\left(#1\right) }

\newcommand{\wcpo}{pointed\, $\omega$-cpo }
\newcommand{\wcpos}{pointed\, $\omega$-cpos }

\newcommand{\wCpo}{\mathbf{\boldsymbol\omega Cpo}_{\leastelement } }

\newcommand\inCC[1]{{\left( #1 \right)_{\leastelement}}}

\newcommand{\ww}{\omega}

\newcommand\lift[1]{\overline{#1}}

\newcommand\colim{\mathrm{colim}}

\newcommand\lhi[1]{\mathcal{H}_{\left( #1\right)_{\bot} }}

\newcommand\letin[3]{\mathbf{let}\,#1=\,#2\,\mathbf{in}\,#3}

\newcommand{\smashproduct}{\star }

\newcommand\LDomain[1]{\mathbf{LDom}(#1)}
\newcommand\LCDomain[1]{\mathbf{CDom}(#1)}

\newcommand\coproje[1]{\iota _{#1} }
\newcommand\iiterativeFn{Iterative Freyd category}
\newcommand\iterativeFn{iterative Freyd category}
\newcommand\iterativeF{iterative Freyd category }
\newcommand\iterativeFsn{iterative Freyd categories}
\newcommand\iterativeFs{iterative Freyd categories }
\newcommand\cpairL{[ }
\newcommand\cpairR{] }

\newcommand\pairfreydR{  \left|\!) \right. }
\newcommand\pairfreydL{ \left.\right(\!\!|  }

\newcommand\cISO{\varsigma}

\newcommand\weirdcoherence{\kappa }

\newcommand\KiAlph[1]{
	\ifcase #1\or \kappa\or \kappa'\or \kappa''\else \@ctrerr \fi%
}

\newcommand\TyAlph[1]{
	\ifcase #1\or \tau\or \sigma\or \rho\else \@ctrerr \fi%
}
\newcommand\ty[1]{{\TyAlph{#1}}}

\newcommand\CTyAlph[1]{
	\underline{\ifcase #1\or \tau\or \sigma\or \rho\else \@ctrerr \fi}%
}
\newcommand\cty[1]{{\CTyAlph{#1}}}

\newcommand\TySchAlph[1]{
	\ifcase #1\or \phi\or \psi \or \upsilon\else \@ctrerr \fi%
}

\newcommand\var[1]{{\VarAlph{#1}}}
\newcommand\VarAlph[1]{%
	\ifcase #1\or x\or y\or z\else \@ctrerr \fi%
}

\newcommand\TVarAlph[1]{%
	\ifcase #1\or \alpha\or \beta\or \gamma\else \@ctrerr \fi%
}

\newcommand\LTVarAlph[1]{%
	\computationtype{\ifcase #1\or \alpha\or \beta\or \gamma\else \@ctrerr \fi%
	}
}

\newcommand\val[1]{%
	\ifcase #1\or v\or w\or u\else \@ctrerr \fi%
}

\newcommand\trm[1]{{\TermAlph{#1}}}
\newcommand\TermAlph[1]{%
	\ifcase #1\or t\or s\or r\else \@ctrerr \fi%
}

\newcommand\ID{\mathrm{id}} %

\newcommand\catV{\Cat{V}}

\newcommand\catC{\Cat{C}}
\newcommand\catD{\Cat{D}}
\newcommand\catL{\Cat{L}}

\newcommand\catCat{\mathsf{Cat}}
\newcommand\CAT{\catCat}

\newcommand\catCatb{\mathsf{Cat}_{\ww\leastelement}}

\newcommand\catSet{\mathsf{Set}}
\newcommand\catPSet{\mathsf{PSet}}

\newcommand\op{\mathrm{op} }

\newcommand\AAlg{\textrm{-}\mathrm{Alg}}

\newcommand\fixpoint{\mu }  %

\newcommand\foldd[1]{ \mathtt{fold}\left( #1 \right)  }

\newcommand\foldingS[1]{ \mathbf{fd}_{\mathfrak{S}yn}\left( #1 \right)  }
\newcommand\folding[1]{ \mathbf{fd}\left( #1 \right)  }
\newcommand\foldingg[2]{ \mathbf{fd}^{\left( #2 \right)}\left( #1 \right)  }

\newcommand\inAl[1]{\fixpoint{\left( #1\right) } }
\newcommand\minAl[1]{\mathsf{in}_{\left( #1\right) } }

\newcommand\iterationg{{\mathtt{it}_{\Sigma} \, }}
\newcommand\otimesg{\otimes _{\Sigma } }

\newcommand\foldersname{iterative folders}

\newcommand\iterationn{{\mathtt{it} \, }}  %

\newcommand\leastelement{\bot}                              %

\newcommand\diagk[1]{\mathrm{diag}_{#1} }
\newcommand\pproj[1]{\overline{\pi}_{#1}}
\newcommand\proj[1]{\pi_{#1}}            %

\newcommand\pairL{\left( }
\newcommand\pairR{\right) }

\newcommand{\RR}{\mathbb{R}}
\newcommand{\NN}{\mathbb{N}}

\newcommand{\recipp}{\mathsf{recpr}}
\newcommand{\normRR}[1]{\mathsf{norm}_{#1}}
\newcommand{\signR}{\mathsf{sign}}
\newcommand{\orientationF}[1]{\mathsf{norma}_{#1}}

\newcommand{\deciderr}[3]{\prescript{}{#3}{\mathsf{decid}^{#1}_{#2}} }

\newcommand\tangentreals{\underline{\mathbf{reals}} }
\newcommand\creals{\tangentreals }

\renewcommand\reals{\mathbf{real}}            %

\newcommand\cnst[1]{\underline{#1}}           %

\newcommand\Op{\mathrm{Op}}                   %

\newcommand\sigmoid{\varsigma}

\newcommand\syncat[1]{\mspace{-25mu}\synname{#1}}
\newcommand\synname[1]{\qquad\text{#1}}

\newenvironment{syntax}[1][]{%
	\(
	\begin{array}[t]{#1l@{\quad\!\!}*3{l@{}}@{\,}l}
	}{
	\end{array}
	\)%
}

\newcommand\gor{\mathrel{\lvert}}
\newcommand\vor{\mathrel{\big\lvert}}
\newcommand\lvar{\mathsf{v}}

\newcommand\Init{\mathbf{0}}

\newcommand\tTuple[1]{\langle #1\rangle}

\newcommand\pMatch[5][\,]{\mathbf{case}\,#2\,\mathbf{of}#1\tPair{#3}{#4}\To#5}
\newcommand\pletin[4]{\pMatch{#3}{#1}{#2}{#4}}
\newcommand\tMatch[4][\,]{\mathbf{case}\,#2\,\mathbf{of}#1\tTuple{#3}\To#4}
\newcommand\lrvMatch[3][\,]{\mathbf{case}\,#2\,\mathbf{of}#1\big\{#3\big\}}
\newcommand\vMatch[3][\,]{\mathbf{case}\,#2\,\mathbf{of}#1\{#3\}}

\newcommand\tIt[3][]{\mathbf{iterate}_{#1}(#2\To #3)}
\newcommand\To{\to}
\newcommand\tRoll[1][]{\mathbf{roll}_{#1}\,}

\newcommand\tFold[4][]{\mathbf{fold}_{#1}\,#2\, \mathbf{with}\ #3\to #4}

\newcommand\tInl{\tIni[1]}
\newcommand\tInr{\tIni[2]}
\newcommand\tIni[1][]{\mathbf{in}_{#1}\,}
\newcommand\tFst{\tPrj[1]}
\newcommand\tSnd{\tPrj[2]}
\newcommand\tPrj[1][]{\mathbf{pr}_{#1}\,}

\newcommand\tPair[2]{\langle #1, #2\rangle}

\newcommand\tZero{\underline{0}}

\WithSuffix\newcommand\t*{\,{\mathop{\times}}\,}
\WithSuffix\newcommand\t+{\,{\mathop{\sqcup}}\,}

\newcommand\Unit{\mathbf{1}}
\newcommand\lUnit{\computationtype{\Unit}}

\newcommand\lfun[1]{\underline{\lambda} #1.}
\newcommand\lapp[2]{#1\bullet #2}
\newcommand\zero{ { 0}^{\mathbf v}}

\newcommand\subst[2]{#1{}[#2]}
\newcommand\sfor[2]{^{#2}\!/\!_{#1}}

\newcommand\tinf{\vdash}
\newcommand\ctx{\Gamma}
\newcommand\freeGinf[3][]{#1\tinf #2 : #3}
\newcommand\Ginf[3][]{\ctx #1\tinf #2 : #3}

\newcommand\freeeq[1]{\stackrel{\# #1}{=}}
\newcommand\beeq{\stackrel{\beta\eta}{=}}

\newcommand\computationtype[1]{\underline{#1}}

\newcommand\CSyn{\mathbf{CSyn}}
\newcommand\LSyn{\mathbf{LSyn}}

\newcommand\TPSYN{\mathfrak{P}\mathbf{TSyn}}
\newcommand\TLSynit{\iterationn _{\Sigma\mathbf{LSyn}}}
\newcommand\LSynTF{\mathfrak{L}\mathfrak{S}\mathbf{yn}}

\newcommand\CSynV{\mathbf{CSyn}_V}
\newcommand\CSynC{\mathbf{CSyn}_C}

\newcommand\LSynV{\mathbf{LSyn}_V}

\newcommand\CSynit{\mathbf{CSyn}_\iterationn}

\newcommand\LSynC{\mathbf{LSyn}_C}

\newcommand\CSynj{\mathbf{CSyn}_j}

\newcommand\LSynt{\mathbf{LSyn}_{tot}}
\newcommand\CSynt{\mathbf{CSyn}_{tot}}

\newcommand\DSynrevt{\underleftarrow{\mathbb{D}} }
\newcommand\DSynrev{\underleftarrow{\mathbb{D}} }
\newcommand\Dsynrevarg[2][]{\DSynrev^{#1}(#2)}

\newcommand\DSynmacrot[2][]{\DSynrev^{#1}(#2)}

\newcommand\ltype{\mathrm{ltype}}

\newcommand\DSem[1]{\partial ^\ast \left(  #1 \right)  }
\newcommand\DSempart[1]{\partial^\ww \left(  #1 \right)  }

\newcommand\DSemipart[2]{\partial _{#2}^\ww  \left(  #1 \right) }

\newcommand\DSemi[2]{\partial _{#2} \left(  #1 \right) }

\newcommand\DSemtotal[1]{\mathfrak{D} #1 }
\newcommand\DSemCHAD[1]{\prescript{}{cha}{\mathfrak{D} #1} }

\newcommand\DSempartial[1]{\mathfrak{D}_{\ww}#1 }

\newcommand\pman{\mathfrak{p} }
\newcommand\pfam{\mathfrak{j} }

\newcommand\FFSET{\Cat{FS}et }
\newcommand\forgettingagain{\mathfrak{U}}
\newcommand\FFMAN{\Cat{FM}an }
\newcommand\FFFVECT{\Cat{F}\Fam{\Vect ^\op }}
\newcommand\VMAN{\Cat{VM}an }
\newcommand\PVMAN{\mathfrak{P}\Cat{M}an }
\newcommand\OMAN{\mathfrak{O}\Cat{M}an }
\newcommand\MAN{\Cat{M}an }

\newcommand\Euclid{\mathfrak{E} }

\newcommand\Syntotal{\mathbf{Syn}_{tot}}
\newcommand\Syn{\mathbf{Syn}}
\newcommand\SynV{\Syn_V}
\newcommand\SynC{\Syn_C}
\newcommand\SynJ{\Syn_J}
\newcommand\CSynJ{\CSyn_J}

\newcommand\Synit{\Syn_\mathsf{it} }  %

\newcommand\SynTT{\Syn_{\otimes }}
\newcommand\CSynTT{\CSyn_{\otimes }}

\newcommand\objects{\mathrm{ob}}

\newcommand\semt[1]{[\hspace{-2.5pt}[#1]\hspace{-2.5pt}]  }
\newcommand\semtl[1]{\overline{[\hspace{-2.5pt}[#1]\hspace{-2.5pt}]}  }
\newcommand\semtlu[1]{\underline{\overline{[\hspace{-2.5pt}\underline{[#1]}\hspace{-2.5pt}]}}  }

\newcommand\sem[1]{[\hspace{-2.5pt}[#1]\hspace{-2.5pt}] }
\newcommand\sempp[1]{\tilde{[\hspace{-2.5pt}[\hspace{-2.5pt}[#1]\hspace{-2.5pt}]\hspace{-2.5pt}]} }

\newcommand\Tagen[1]{\mathcal{T}^\ast_{#1}}
\newcommand\TagentS[2]{\mathcal{T}^\ast_{#2}{#1}}
\newcommand\Tagentt[2]{\mathcal{T}_{#2}{#1}}

\newcommand\Fam[1]{ \mathbf{Fam}\left( #1\right)   }
\newcommand\PFam[1]{ \mathfrak{P}\mathbf{Fam}\left( #1\right)   }

\makeatletter
\let\c@equation\c@figure    %
\makeatother

\newcommand\ndots{\cdot\cdot\cdot\cdot\cdot\cdot}

\newcommand\nndots{\ndots\ndots\ndots}

\newcommand\vGamma{\overline{\Gamma}}

\newcommand\tCoProj[1]{\mathbf{coproj}_{#1}\,}

\newcommand\idx[2]{\mathbf{idx}(#1; #2)\,}

\newcommand\pvar{p}

\begin{document}

\title[CHAD for iteration]{Unraveling the iterative CHAD}

\author[F. Lucatelli Nunes]{Fernando Lucatelli Nunes}
\address{CMUC, Department of Mathematics, University of Coimbra, 3000-143 Coimbra, Portugal}
\email{fln@uc.pt}

\author[G. Plotkin]{Gordon Plotkin}
\address{LFCS, School of Informatics, University of Edinburgh, UK}
\email{gdp@inf.ed.ac.uk}

\author[M. Vákár]{Matthijs Vákár}
\address{Department of Information and Computing Sciences, Utrecht University, Netherlands}
\email{m.i.l.vakar@uu.nl}

\settocdepth{subsection}

\begin{abstract}
Combinatory Homomorphic Automatic Differentiation (CHAD) was originally formulated as a semantics-driven source-to-source transformation for reverse-mode AD of total (terminating) functional programs. In this work, we extend CHAD to encompass programs featuring constructs such as partial (potentially non-terminating) operations, data-dependent conditionals (e.g., real-valued tests), and iteration constructs (i.e. while-loops), while maintaining CHAD's core principle of structure-preserving semantics.

Our fundamental contribution is the introduction of iteration-extensive indexed categories, which provide a principled integration of iteration into dependently typed programming languages.
This integration is achieved by requiring that iteration in the base category lifts to parameterized initial algebras in the indexed category, yielding a fibred iteration on the op-Grothendieck construction that models while-loops and other iteration constructs in the total category of this construction, which corresponds to the category of containers of our dependently typed language.

Through the idea of iteration-extensive indexed categories, we extend the CHAD transformation to looping programs as the unique structure-preserving functor in a suitable sense. Specifically, it is the unique iterative Freyd category morphism from the iterative Freyd category corresponding to the source language to the  category of containers obtained from the target language, such that each primitive operation is mapped to its (transposed) derivative.
We establish the correctness of this extended transformation via the universal property of the syntactic categorical model of the source language, showing that the differentiated programs compute correct reverse-mode derivatives of their originals.

Our work advances the study of fixpoint operators for iteration within the setting of indexed categories and uses the developed theory to extend CHAD to programs that use iteration.
\end{abstract}


\maketitle

\tableofcontents
\newpage

\setcounter{secnumdepth}{5}

\setcounter{secnumdepth}{-1}
\section*{Introduction}
In machine learning and scientific computation, we often need to know how a program's result changes with its input. Reverse-mode automatic differentiation (AD) is particularly useful when a scalar objective depends on many parameters \cite{margossian2019review}. We study how to extend Combinatory Homomorphic Automatic Differentiation (CHAD) to programs with iteration while retaining its principle of structure-preserving program transformation. This leads us to a general question about iteration and dependent types, whose solution supplies the CHAD extension.

\paragraph{From derivatives to CHAD}
For a differentiable function $f:\RR^n\to\RR^m$, the derivative
$f'(x):\RR^n\to\RR^m$ is the linear map describing how input
variations affect the output to first order:
\[
f(x+h)\approx f(x)+f'(x)h.
\]
An input variation $h$, regarded as a vector based at $x$, is a
\emph{tangent vector} at $x$. Its image $f'(x)h$ is the corresponding
tangent vector at $f(x)$. These vectors belong to the \emph{tangent
	spaces} at the respective points, which in this setting are copies
of $\RR^n$ and $\RR^m$.

A \emph{cotangent vector} at a point is a linear functional on its
tangent space: it assigns a scalar to each tangent variation.
For example, suppose that a differentiable objective
$\ell:\RR^m\to\RR$ depends on the output $y=f(x)$. Its derivative
$\ell'(y)$ is a cotangent vector at $y$: for an input variation $k$ to $\ell$,
the scalar $\ell'(y)k$ gives the first-order change in the objective.

The derivative also determines how to turn an output cotangent into
an input cotangent. Given an output cotangent $\alpha$, we obtain
an input cotangent by assigning to each input variation $h$ the
scalar $\alpha(f'(x)h)$. This defines the \emph{coderivative}, or
dual derivative:
\[
\DSemi{f}{x}(\alpha)=\alpha\circ f'(x).
\]
Thus, the derivative sends tangent vectors from the input to the
output, whereas the coderivative sends cotangent vectors from the
output back to the input. In the objective-function example, the
chain rule gives
\[
\DSemi{f}{x}\bigl(\ell'(f(x))\bigr)
=\ell'(f(x))\circ f'(x)
=(\ell\circ f)'(x).
\]

The transpose describes this backward operation in coordinates.
Represent an output cotangent $\alpha$ by the column vector
$v\in\RR^m$ of its coefficients, so that
$\alpha(k)=v^{\mathsf T}k$, where $(-)^{\mathsf T}$ denotes matrix
transpose. Writing $f'(x)$ also for the derivative matrix, we have
\[
\bigl(\DSemi{f}{x}(\alpha)\bigr)(h)
=v^{\mathsf T}\bigl(f'(x)h\bigr)
=\bigl(f'(x)^{\mathsf T}v\bigr)^{\mathsf T}h.
\]
Hence, the input cotangent is represented by
$f'(x)^{\mathsf T}v$: the matrix of the coderivative is the
transpose of the derivative matrix.

Forward-mode automatic differentiation propagates tangent
variations forwards through a computation; reverse mode propagates
cotangent sensitivities backwards.

CHAD implements reverse AD as a \emph{semantics-driven, structure-preserving source-to-source transformation} \cite{vakar2021chad,lucatellivakar2021chad,DBLP:journals/toplas/VakarS22,2023arXiv230705738S}, following the perspective of Elliott \cite{DBLP:journals/pacmpl/Elliott18} and the pursuit of purely functional reverse AD at compile time \cite{pearlmutter2008reverse,krawiec2022provably}. It translates source code into target code returning a pair: the original result, called the \emph{primal value}, and a linear backward function representing the coderivative at the input. In Euclidean coordinates, the intended behaviour is
\[
 x\longmapsto\left(f(x),\,v\longmapsto f'(x)^{\mathsf T}v\right).
\]
The backward function need not construct the derivative matrix. For a scalar output, applying it to $1$ gives the gradient under the usual Euclidean identification of cotangents with vectors.

Structure preservation means that each source construct has a corresponding operation for combining transformed programs. Sequential composition, for instance, combines primal computations forwards and backward maps in reverse order. Once the transformations of the primitive operations are chosen, these requirements determine the transformation of composite programs. Categorically, CHAD is the uniquely determined structure-preserving functor extending those choices.

\paragraph{Iteration and dependent types}
Consider an iterative solver that repeatedly improves an approximation until a stopping criterion is met. If its result contributes to an objective, we want the sensitivity of the approximation actually computed to the solver's inputs and parameters. Similar questions arise for Taylor approximations to special functions and optimization algorithms.

A loop body acts on a \emph{state}, the information passed between successive steps. If it returns, it produces either a final result or a new state with which to continue. A \emph{variant type}, also called a sum type, expresses this choice by tagging a value with its alternative. Writing $A$ for states and $B$ for results, we represent the body and the complete loop by
\[
 f:A\to B\sqcup A,
 \qquad
 \mathtt{iterate}\ f:A\to B.
\]
Here both arrows denote possibly partial computations. The loop returns when the body produces a value in $B$; it is undefined if a body evaluation is undefined or no step returns a final result. This encompasses \texttt{while}-loops with an input-dependent number of steps.

Variant types require dependent types in the target language if we retain the usual CHAD interpretation of cotangents. A value in $\RR\sqcup\RR^2$ contains either a real number or a pair of real numbers, and its cotangent space is correspondingly one- or two-dimensional. Keeping these usual spaces means that the cotangent type follows the branch selected by the primal value. This is a \emph{dependent type}, a type that varies with a value. Such dependence already arises in expressive total CHAD, which handles sums, functions and (co)inductive types using a linear dependent target and correctness proofs by logical relations \cite{lucatellivakar2021chad}.

For a total example without iteration, let $g:\RR\sqcup\RR\to\RR\sqcup\RR^2$ be
\[
 g(\operatorname{inl}x)=\operatorname{inl}(x^2),\qquad
 g(\operatorname{inr}x)=\operatorname{inr}(x,x^2).
\]
Its backward maps at $\operatorname{inl}x$ and $\operatorname{inr}x$ are, respectively,
\[
 \lambda\longmapsto 2x\lambda,\qquad
 (u,v)\longmapsto u+2xv.
\]
The backward map therefore accepts a scalar or a pair according to the branch returned.

Our source language remains simply typed. Preserving its iteration requires a corresponding operation in the dependent target, where the backward maps must fit the cotangent types at successive states and at the final result. The problem consequently becomes more general: how can we describe and compute with data whose types depend on the result of a possibly non-terminating loop? We retain the usual cotangent interpretation here; Section~\ref{sec:related-work} discusses its relation to subsequent simply typed reverse AD.

Suppose that each possible result $b$ has an associated type $Y(b)$. At an input where the loop returns $b$, the required type is $Y(b)$. In our concrete models, this family can be described by considering executions that return after one step, then those requiring two steps, and so on. Every terminating input is eventually assigned the type at its result. Inputs with no result receive the empty set for set-valued families, or the zero space for vector-space-valued families. Thus, building the family from the body agrees with selecting the type at the result of the whole loop.

The backward computation follows the same execution in reverse. For a three-step loop, a cotangent at the result passes through the backward map of the third step, then the second, and finally the first. Our theory provides an operation that assembles these maps for any finite execution, respecting their dependent types. This operation is a \emph{fold}. Our theory formulates it for general dependent families, supplying a principle for dependent computation as well as the backward operation needed by CHAD.

\paragraph{The categorical principle}
An indexed category $\catL:\catC^{op}\to\catCat$ records type dependence: $\catC$ models ordinary types and programs, while $\catL(C)$ models dependent types and terms in a context $C$. A program $f:D\to C$ induces a substitution functor $\catL(f):\catL(C)\to\catL(D)$ \cite{jacobs1999categorical}. In our family models, its value on a family $P$ at $x$ is $P(f(x))$ where $f(x)$ is defined, and the initial object otherwise.

The op-Grothendieck construction $\Sigma_\catC\catL^{op}$ combines a primal type $C$ and a dependent family $L\in\catL(C)$ into a container $(C,L)$ \cite{abbott2003categories,foster2007lenses}. For CHAD, $L$ is the cotangent family, and a container map sends primal values forwards and cotangents backwards. This is the lens perspective on the returned pair. It also expresses the duality with forward mode: passing from the Grothendieck to the op-Grothendieck construction reverses the linear maps in the fibres while retaining the direction of the primal computation.

To give this category of containers iteration, we study inductive types arising from substitution. Inductive types are commonly interpreted by initial algebras $\mu F$ of endofunctors $F:\catC\to\catC$. More generally, $F:\catC'\times\catC\to\catC$ may have \emph{parameterized initial algebras}, with carrier functor $\mu F:\catC'\to\catC$ \cite{lehmannSmyth1981,freyd1991algebraically,fiore2004axiomatic}. The parameter allows the inductive type to vary, as a list type varies with its type of entries. Familiar examples use products, coproducts and suitable function spaces \cite{zbMATH01850600,initialalgebrasPatricia}; here we apply this established theory to substitution functors.

For $f:A\to A$, the initial algebra of $\catL(f)$ is trivial when substitution preserves initial objects. A loop body gives a more interesting case. If $\catL$ is extensive, a family over $B\sqcup A$ is specified by a family over each branch, and substitution along $f:A\to B\sqcup A$ induces the composite
\[
 H_f:\catL(B)\times\catL(A)
 \xrightarrow{\ \cong\ }\catL(B\sqcup A)
 \xrightarrow{\ \catL(f)\ }\catL(A).
\]
The first arrow is inverse to the comparison induced by the coproduct injections. The parameter in $\catL(B)$ supplies the family at final results; the argument in $\catL(A)$ supplies the family for continuation. The body selects the appropriate branch after one step.

Our central principle, \emph{iteration extensivity}, relates the resulting parameterized initial algebra to substitution along the whole loop:
\[
 \mu H_f\cong\catL(\mathtt{iterate}\ f):\catL(B)\to\catL(A).
\]
The left-hand side builds the family inductively from the body; the right-hand side selects it at the loop's result. We show that this holds for families valued in any category with an initial object, extending substitution to partial functions by assigning that object where the computation is undefined.

The associated fold supplies iteration in the category of containers. Specializing to cotangents gives iterative CHAD: the primal component iterates and the backward component uses an \emph{iterative folder}. Assuming correct primitive derivatives, the transformed program returns exactly when the source program returns, yielding its original result and transpose-derivative function. Correctness follows from the universal property of the source language: interpreting the transformed program and differentiating the original meaning agree on primitives and both preserve the language structure. The general theory thus provides the construction and justification of the CHAD extension.

\paragraph{Contributions}
More precisely, we make the following contributions:
\begin{itemize}
\item we revisit extensive indexed categories \cite{lucatellivakar2021chad,LV24b}, motivated by extensive categories \cite{zbMATH00166168}, and clarify their relation to $\Sigma$-bimodels for sums, which additionally require initial and terminal objects in every fibre (Lemma~\ref{lemma49});

\item we show that initial algebras (resp.\ terminal coalgebras) of substitution along endomorphisms are trivial in the presence of indexed initial (resp.\ terminal) objects, and study the parameterized case induced by $f:A\to B\sqcup A$;

\item we identify several levels of structure on indexed categories that induce iteration in their (op-)Grothendieck constructions, making dependent iteration compatible with the category of containers;

\item we introduce iteration-extensive indexed categories, relating substitution along a loop to parameterized initial algebras of substitution along its body (Subsect.~\ref{subsection:iteration-extensive-categories});

\item we show that every iteration-extensive indexed category induces iteration with a universal property on its op-Grothendieck construction (Theorem~\ref{theo:iteration-extensive-indexed-category});

\item for any category $\mathcal K$ with an initial object, we show that $\mathcal K^{(-)}_\leastelement:\catPSet^{op}\to\catCat$ is iteration-extensive, characterize the resulting iteration in the Kleisli category of the maybe-monad on $\Fam{\mathcal K^{op}}$, and give further examples;

\item we introduce iterative indexed categories, equipped with the \foldersname{}, as a less structured setting suited to target syntax (Subsect.~\ref{syn:container-folder-iteration}), and establish a bijection between these operations and fibred raw iterations in the category of containers (Lemma~\ref{lem:bijection-iterative-indexed-categories-Grothendieck});

\item we introduce finite-$\wCpo$-coproduct-extensive indexed categories as a more structured setting for concrete iteration (Subsect.~\ref{subsect:concrete-iterations-wcpo});

\item we develop Freyd versions of each notion, whose op-Grothendieck constructions are iterative Freyd categories interpreting call-by-value languages with tuples, cotuples, and iteration;

\item using $\Vect^{(-)}_\leastelement:\catPSet^{op}\to\catCat$ and the semantics of reverse derivatives in $\PFam{\Vect^{op}}$, we derive a reverse CHAD code transformation and correctness proof for programs with iteration.
\end{itemize}

The definitions serve three purposes. Iteration-extensivity supplies a semantic universal property; iterative folders isolate the operations needed by the target syntax; and the $\wCpo$-enriched models identify the induced iteration with ordinary least-fixed-point iteration, obtained from finite unfoldings. Much of the work lies in identifying the required structures and their compatibility, after which several proofs become short. The Freyd versions also account for parameters carried unchanged through a loop, retaining the computation's possible divergence in the dependent semantics.

\paragraph{Conventions and outline}
We write $A\sqcup B$ for a coproduct, $\iota_A$ and $\iota_B$ for its injections, and $[f,g]$ for the induced cotuple. In sets and manifolds, coproducts are disjoint unions. The notation $\catCat$ denotes categories, functors and, when needed, natural transformations; we fix sufficiently large universes. We write $\Fam{\mathcal K}$ for the category of families of objects of $\mathcal K$, recalled in Section~\ref{sec:semantics-CHAD}.

Unless stated otherwise, indexed categories $\catL:\catC^{op}\to\catCat$ are strictly indexed, and the same convention applies to their structured variants. In particular, binary-coproduct-extensive indexed categories preserve binary products in $\catC^{op}$, as in \cite[Lemma~32]{lucatellivakar2021chad}. We use strict indexing for exposition; in the non-strict setting, the substitution laws hold up to coherent natural isomorphism. The corresponding forward-mode constructions use the appropriate dual hypotheses, notably terminal coalgebras in place of initial algebras. Their relation to predicate transformers is a further direction of interest.

The paper has three parts. Part~\ref{part:background} revisits total CHAD and its correctness argument. Part~\ref{part:iteration} develops iteration and dependent types, beginning with the partial concrete models in Section~\ref{sec:denotational-partial-features} and proceeding to the general indexed theory in Section~\ref{sec:initial-algebras}. Part~\ref{part:iterative-chad} applies this theory: Sections~\ref{sec:source-language} and~\ref{sec:target-language} present the source and target languages, and Section~\ref{sec:correctness-iterative-CHAD} derives the transformation and proves correctness. Section~\ref{sec:related-work} discusses related work and further questions.

\resettocdepth

\setcounter{secnumdepth}{5}
\part{CHAD for expressive total languages revisited}\label{part:background}

We begin by revisiting CHAD for total languages, where the central idea can be seen without the additional problem of non-termination. A transformed program computes its original result together with a linear map that propagates cotangents backwards. When programs are composed, their primal computations are composed in the usual order, while their backward maps are composed in the opposite order. Products collect the component cotangents, whereas a variant carries the cotangent space of its selected branch.

This part recalls how this picture is expressed by the categorical semantics of the source and target languages. Section~\ref{sec:semantics-CHAD} describes the concrete models and the CHAD-derivative functor. Section~\ref{section:denotational-correctness-of-CHAD-total} then recalls the program transformation and its correctness argument. The essential point is that the transformation and its mathematical specification preserve the same structure and agree on primitive operations. The universal property of the source language extends this agreement to all programs. This is the principle we retain when introducing iteration.

\section{Total CHAD: concrete semantics}\label{sec:semantics-CHAD}
Reverse mode Combinatory Homomorphic Automatic Differentiation (CHAD) is an automatic differentiation technique realized as a program transformation. As such, it is executed at compile-time as a source-code transformation between a \textit{source language} and a \textit{target language}. We direct the reader to \cite{lucatellivakar2021chad} for a comprehensive understanding of CHAD within the setting of total languages and to \cite{2023arXiv230705738S} for its efficient implementation.

Herein, we commence by revisiting the fundamental principle of the (concrete) denotational semantics of CHAD for total languages. Specifically, we draw attention to the detailed treatment of the semantics of the target language and differentiable functions within the context of CHAD for total functions, as explicated in \cite[Section~6 and 10.4]{lucatellivakar2021chad}.

\subsection{Grothendieck constructions and the free coproduct completion}
Given the reliance of our target language's concrete denotational semantics on (op-)Grothendieck constructions, we direct the reader to \cite[Section~6]{lucatellivakar2021chad} and \cite{LV24b} for a detailed discussion within our context.

The role of an indexed category is to describe objects that vary with a parameter. For each object $M$ of a base category $\catC$, it provides a category $\mathcal{A}(M)$ of such objects over $M$. A morphism $f:M\to N$ gives a substitution functor $\mathcal{A}(f):\mathcal{A}(N)\to\mathcal{A}(M)$. The Grothendieck construction puts the parameter and the object over it into a single pair. In the op-Grothendieck construction, a morphism moves the parameter forwards and the object over it backwards. This is precisely the direction of the primal computation and the cotangent computation in reverse mode AD.

Let $\mathcal{A} :  \catC ^\op \to \catCat $ be a (strictly) indexed category. We denote by $\Sigma_{\catC }\mathcal{A} $ the Grothendieck construction of $\mathcal{A}$. More importantly for our setting, we denote by
$\displaystyle\Sigma_{\catC }\mathcal{A} ^\op  $
the op-Grothendieck construction, that is to say, the Grothendieck construction of the
(strictly) indexed category $\mathcal{A} ^\op $ defined by
$\mathcal{A}^\op(f) =  \left( \mathcal{A}(f) \right) ^\op $, which can be seen as the composition
$\op \circ \mathcal{A}: \catC ^\op \to \catCat $. More explicitly, the objects of $\displaystyle\Sigma_{\catC }\mathcal{A} ^\op $ are pairs $\left( M\in \textrm{obj}\left( \catC  \right), X\in \mathcal{A}\left( M \right)    \right)$, and a morphism $\left( M, X \right) \to \left( N, Y \right) $ consists of a pair
\begin{equation}
 \left( f: M\to N , f':  \mathcal{A}(f)(Y) \to X \right)\in \catC\left( M, N \right)\times \mathcal{A}\left( M \right)(\mathcal{A}(f)(Y), X) .
\end{equation}

Recall that, if $(f, f'): (M, X)\to (N,Y) $ and $(g, g'): (N, Y)\to (K,Z) $ are morphisms of $\displaystyle\Sigma_{\catC }\mathcal{A} ^\op  $, the composition in $\displaystyle\Sigma_{\catC }\mathcal{A} ^\op  $ is given by
\begin{equation}
(g,g')\circ (f, f') = \left( g\circ f, f'\circ \mathcal{A}\left( f\right) (g') \right).
\end{equation}

Furthermore, as many of the Grothendieck constructions we encounter can be understood as free coproduct completions, we recall that the free coproduct completion $\Fam{\catC ^\op }$ of the opposite of the category $\catC $
is the op-Grothendieck construction of the indexed category
 \begin{equation}
 	\catC ^{(-)} : \catSet ^\op \to \catCat; \qquad A\mapsto \catC ^A, \qquad \left(f:A\to B\right) \mapsto
 	\left( \catC ^f:\catC ^B\to \catC ^A\right) ,
 \end{equation}
where, for each set $A$ treated as a discrete category, we denote by $\catC ^A$ the category of functors from $A$ to  the category $\catC $, and we denote by $\catC ^f$ the functor $\catC ^B\to \catC ^A$ defined by $X\mapsto X\circ f $.
For the interested reader, we suggest consulting \cite[Section~9.2]{lucatellivakar2021chad}, \cite[Section~1]{2024arXiv240310447L}, \cite{zbMATH07186728}, \cite[Section~8.5]{2023arXiv230908084P}, and \cite[Section~1.2]{2023arXiv230504042P} for fundamental definitions and further exploration.

Explicitly, objects of $\Fam{\catC ^\op } $ take the form of pairs $\left( A, X \right) $, where $A$ is a set and $X$ denotes a functor $X: A\to \catC  $, treating $A$ as a discrete category; \textit{i.e.}, $A$ is a set and $X$ is an $A$-indexed family of objects in $\catC $. A morphism
$\mathbf{f}: \left( A, X \right) \to \left( B, Y \right) $ in $\Fam{\catC^\op }$ is a pair
\begin{equation}
	\mathbf{f} = \left( f, f' \right)\in \catSet \left(A, B\right)\times \mathsf{NAT}\left(  Y\circ f , X\right)
\end{equation}
where $f: A\to B$ is a function and $f' $ is a family
\begin{equation}
	f'= \left( f'_a:  Y\left(  f (a)\right)\to  X (a)   \right) _{a\in A}
\end{equation}
of morphisms in $\catC $. Finally, recall that, if $\mathbf{f}: \left( A, X \right) \to \left( B, Y \right) $
and $\mathbf{g}: \left( B, Y \right) \to \left( C, Z \right) $ are morphisms in $\Fam{\catC ^\op } $,
\begin{equation}
	\mathbf{g}\circ\mathbf{f}=(g,g')\circ (f, f') = \left( g\circ f, (gf)' \right),
\end{equation}
where $(gf)'_a = f_{a}'\circ g_{f(a)}' : Z(g(f(a)))\to X(a) $.

\subsection{Differentiable functions}
We consider the category $\VMAN$ of differentiable manifolds of variable dimension, called herein \textit{manifolds} for short, and differentiable functions between them. Variable dimension means that different connected components may have different finite dimensions; we allow formal disjoint unions of these components. The category $\VMAN$ can be realized as the free coproduct completion $\Fam{\MAN }$ of the usual category $\MAN$  of connected differentiable manifolds of finite dimension.

For the total fragment considered in this section, in order to provide a concrete semantics for CHAD, we only need Euclidean spaces and (free) coproducts (copairing) of Euclidean spaces. Hence, the reader is free to think of $\VMAN $ as the free coproduct completion  $\Fam{\Euclid} $ of the category $\Euclid$ consisting of the Euclidean spaces $\RR^n$ and differentiable functions between them. For the partial semantics below, we use $\VMAN$ so that arbitrary open subsets are also available.

Recall that, if $f : M\to N$ is a morphism in $\Cat{VM}an$, we have a notion of derivative of $f$, \textit{e.g.} \cite{zbMATH06034615} for the classical case. More precisely, the derivative of
$f $ at $w\in M $ is given by a linear transformation
\begin{equation}
	f'(w) : \Tagentt{w}{M} \to  \Tagentt{f(w)}{N}
\end{equation}
where $\Tagentt{w}{M} \cong \RR^{n}$ and $\Tagentt{f(w)}{N} \cong \RR^{m}$ are the tangent spaces of $M$ at $w$ and of $N$ at $f(w)$, respectively.
The \textit{coderivative} (or the reverse-mode derivative) of $f$, denoted simply as $\DSemi{f}{w}$ herein, is the transpose (or dual) of the linear transformation  $f'(w)$; namely:
\begin{equation}
	\DSemi{f}{w} : \TagentS{f(w)}{N} \to \TagentS{w}{M}
\end{equation}
is defined between the linear duals\footnote{Recall that the linear dual $V^*$ of an $\mathbb{R}$-vector space is defined as the vector space $V\multimap \mathbb{R}$ of linear transformations from $V$ to $\mathbb{R}$.} of the respective tangent spaces, called \textit{cotangent spaces} $\TagentS{f(w)}{N}$ and $\TagentS{w}{M}$, by precomposing $f'(w)$.
Observe that given our identifications  $\Tagentt{w}{M} \cong \RR^{n}$ and $\Tagentt{f(w)}{N} \cong \RR^{m}$, $f'(w)$ has a representation as an $m\times n$-matrix and $\DSemi{f}{w}$ is then represented by its matrix transpose.
We denote by $\DSem{f}$ the family of coderivatives, defined as
\begin{equation}
	\DSem{f}\defcolon \left( \DSemi{f}{w} : \TagentS{f(w)}{N} \to \TagentS{w}{M} \right) _{w\in M} .
\end{equation}
With this notation, the basic tenet of CHAD is to pair up the primal $f$ with $\DSem{f}$; namely we denote
\begin{equation}
	\DSemtotal{f} = \left( f_0, \DSem{f} \right),
\end{equation}
where $f_0$ is the underlying function of $f$.
The pair $\left( f_0, \DSem{f} \right) $, which we denote by   $\left( f, \DSem{f} \right) $ by abuse of language, is called herein the (semantic) \textit{CHAD-derivative}: it is realized as a morphism
in the category $\Fam{\Vect ^\op } $ which plays the role of the concrete denotational semantics of the target language of CHAD, as explained below.
For readers familiar with differential geometry, this is essentially the usual cotangent bundle functor.

\subsection{CHAD-derivatives as morphisms}
We start by recalling some basic aspects of the concrete model for the target language. Our focus lies on the category $\Fam{\Vect ^\op }$, which represents the free coproduct completion of the opposite of the category of vector spaces (and linear transformations).

Recall that $\Fam{\Vect ^\op }$ emerges as the \textit{op-Grothendieck construction} of the indexed category
\begin{equation}
	\Vect ^{(-)} : \catSet ^\op \to \catCat; \qquad A\mapsto \Vect ^A, \qquad \left(f:A\to B\right) \mapsto
	\left( \Vect ^f:\Vect ^B\to \Vect ^A\right).
\end{equation}
Explicitly, objects of $\Fam{\Vect ^\op }$ are pairs $\left( A, X \right) $ where $A$ is a set and
$X$ is  a functor $X: A\to \Vect $. A morphism
$\mathbf{f}: \left( A, X \right) \to \left( B, Y \right) $ in $\Fam{\Vect ^\op }$ is a pair $\mathbf{f} = \left( f, f' \right)$
where $f: A\to B$ is a function and $f' $ is a family of linear transformations
\begin{equation}
	f'= \left( f'_a: Y\left(  f (a)\right)  \to X (a)  \right) _{a\in A}.
\end{equation}

We observe that, for each object $M$ of $\Cat{VM}an$, we can define an object $\left( M, \Tagen{M} \right)$ where
$M$ is the underlying set of $M$ and $\Tagen{M}$ is the functor ($M$-indexed family)
\begin{equation}
	\Tagen{M} : M\to \Vect, \qquad  w\mapsto  \TagentS{w}{M}
\end{equation}
of the cotangents on $M$. Moreover, by the above, it is clear that, whenever $f : M\to N$ is a morphism in $\Cat{VM}an$, the CHAD-derivative $\DSemtotal{f} = \left( f, \DSem{f} \right) $
is a morphism
\begin{equation}
	\DSemtotal{f} = \left( f, \DSem{f} \right) : \left( M, \Tagen{M} \right)\to \left( N, \Tagen{N}  \right)
\end{equation}
between the objects $\left( M, \Tagen{M} \right)$ and $\left( N, \Tagen{N}  \right)$  in $\Fam{\Vect ^\op } $, where, by abuse of notation, the first coordinate of $ \left( f, \DSem{f} \right)$ is the underlying function of $f$.

\subsection{CHAD-derivative as a structure-preserving functor}
The basic principle of CHAD is to exploit structure-preserving functors and, more than that, the unique ones arising from the universal properties of the source languages (or, more precisely, from the freely generated categorical structures over their signatures of primitive types and operations). The reason that we can follow this approach is that the (semantic) CHAD-derivative happens to be a functor that preserves the categorical structure required by the source language.
Once we have demonstrated that the CHAD-derivative preserves the relevant categorical structure, we can  implement it as a code transformation on the syntax and derive a simple correctness proof from the fact that denotational semantics (and logical relations) also give structure preserving functors.

More precisely, within the context of first-order functional  languages with variant and product types, Theorem \ref{theo:products-coproducts-preserving} is pivotal to proceed with our denotational correctness proof.

Recall that if $\left( A, X \right) $ and $\left( B, Y \right) $ are objects of $\Fam{\Vect ^\op } $, then $\left( A\sqcup B, \cpairL X,Y\cpairR \right) $ is the coproduct and $\left( A\times B, X\underline{\times} Y \right) $, where $X\underline{\times} Y (w,w') = X(w)\times Y(w') $,  is the product (see, for instance, \cite[Section~10.2]{lucatellivakar2021chad} and \cite{LV24b}). These constructions are instances of the corresponding constructions in $\Fam{\catC^\op}$: coproducts concatenate families, while products use coproducts in $\catC$ at each pair of indices. In $\Vect$, finite coproducts are also products. Thus a pair of primal values carries a pair of cotangents, whereas a variant carries the cotangent space of its chosen branch.

We use the canonical identifications of the cotangent space of a product with the product of the cotangent spaces, and of a coproduct with its corresponding component. With these identifications understood, the preservation statements below are written strictly. We can, then, verify that the CHAD derivative preserves products and coproducts:

\begin{theorem}\label{theo:products-coproducts-preserving}
The association
\begin{equation}
	\DSemtotal{ }:  \VMAN \to \Fam{\Vect ^\op}, \qquad  	 M \mapsto \left( M, \Tagen{M}   \right) , \qquad f \mapsto 	\left( f, \DSem{f} \right)
\end{equation}
defines a (strictly) bicartesian functor.
\end{theorem}
\begin{proof}
 The \textit{chain rule} for derivatives implies that $\DSemtotal{ }$ is indeed a functor. The preservation of products and coproducts follow from basic properties of derivatives, namely that derivative of a tupled function is the tupling of the derivatives, and that derivatives only depend on local information.
\end{proof}

\begin{definition}[CHAD-derivative]\label{def:CHAD-derivative-functor-total}
Herein, the functor $\DSemtotal{ }$ established in Theorem \ref{theo:products-coproducts-preserving} is called the \textit{(total) CHAD-derivative functor}.
\end{definition}

\section{Denotational correctness of CHAD for total languages}\label{section:denotational-correctness-of-CHAD-total}
We can now briefly recall the correctness result of CHAD for total languages with  variant and product types.
CHAD is realized as structure-preserving program transformation between the source language, a simply-typed (i.e., non-dependently typed) language, and the op-Grothendieck construction of the target language, which is a dependently typed language.

Type dependency already occurs in this total setting. A value of a variant type carries the cotangent space of the branch it belongs to: for instance, a value in $\reals^n\sqcup\reals^m$ carries either an $n$-dimensional or an $m$-dimensional cotangent. The target language expresses this dependence on the primal value. We start by recalling our source language and target language, and their respective categorical semantics.

\subsection{Source language}\label{sec:source-language-total}
We consider a standard total first-order functional programming language with variant and product types. This is constructed over ground types $\reals ^n $ for arrays of real numbers with static length $n$ (for $n\in\NN$), and sets of primitive operations $\Op _{\mathbf{m}}^{\mathbf{n}} $  for each $(k_1, k_2)\in\NN\times\NN$ and each  $\left( \mathbf{n}, \mathbf{m} \right)\in\NN ^{k_1}\times \NN ^{k_2} $. For convenience, we denote by $$\displaystyle\Op \defcolon \bigcup _{\left( k_1, k_2\right)\in\NN \times \NN }\bigcup _{\left( \mathbf{n}, \mathbf{m}\right)  \in \NN ^{k_1} \times \NN ^{k_2}} \Op  ^{\mathbf{n}}_{\mathbf{m}} $$ the set of all primitive operations.
If $\mathbf{n} = \left( n_1 , \ldots , n_{k_1}\right)  $ and $\mathbf{m} = \left( m_1 , \ldots , m_{k_2}\right)  $,
 a primitive operation $\op $ in $\Op ^{\mathbf{n}}_{\mathbf{m}} $  intends to implement a differentiable function
\begin{equation}
	\sem{\op} : \RR^{\mathbf{n} }\to \RR_{\mathbf{m}}
\end{equation}
where we denote $\RR^{\mathbf{n} }\defcolon \RR ^{n_1} \times \cdots \times \RR ^{n_{k_1}} $ and $\RR_{\mathbf{m}}\defcolon \RR ^{m_1} \sqcup\cdots \sqcup \RR ^{m_{k_2}} $. Here $\sqcup$ is disjoint union, and we omit parentheses around singleton tuples of dimensions.
The reader can keep the following examples of primitive operations in mind:
\begin{itemize}
	\item constants $\cnst{c}\in \Op^{()}_{(n)}$ for each $c\in \RR^n$, for which  $\sem{\cnst{c}} = \cnst{c}$;
	\item elementwise addition and product $(+),(*)\!\in\!\Op^{\left( n,n\right) }_n$
	and matrix product $(\cdot )\!\in\!\Op^{\left( n\cdot m, m\cdot r\right)}_{n\cdot r}$;
	\item operations for summing all the elements in an array: $\mathsf{sum}\in\Op^{n}_1$;
	\item some non-linear functions like the sigmoid function $\sigmoid\in \Op_{1}^1$.
\end{itemize}

Primitive operations with coproducts in the codomain will be more important below, after adding partial features to our source language, in Section \ref{sec:source-language}.
The reader interested in the detailed grammar, typing rules, and $\beta\eta$-equational theory for this standard functional programming language can consult Section~\ref{sec:source-language}.

\subsection{Categorical semantics of the source total language} \label{sec:semantics-source-language-total}
A bicartesian category $\catC $ is \textit{distributive} if, for each triple of objects
$A, B, C $ in $\catC $, the canonical morphism
\begin{equation}
\cpairL A\times\coproje{B}, A\times\coproje{C} \cpairR  : \left( A\times B\right)\sqcup \left( A\times C \right) \to A\times \left(B\sqcup C\right)
\end{equation}
is invertible. Alternatively, we can demand that there is an isomorphism $\left( A\times B\right)\sqcup \left( A\times C \right) \cong A\times \left(B\sqcup C\right) $ natural in $A, B, C$ (see, for instance, \cite{zbMATH06039246}).

While not strictly required in our context, for the sake of simplicity and convenience, we consistently adopt the assumption that \textit{distributive categories} have chosen (finite) products and coproducts. Conveniently,  we require that \textit{structure-preserving functors} between these distributive categories strictly preserve finite products and coproducts, that is to say, \textit{(strictly) bicartesian functors}.

The primitive types and operations of the source language
can be framed within a structured graph/computad, \textit{e.g.} \cite{zbMATH03523837}. We can take the freely generated \textit{distributive category} $\Syntotal $ over this structured computad (primitive types and operations) as the abstract categorical semantics, represented by the terms of the source language modulo its equational theory.\footnote{See \cite{zbMATH03523837,zbMATH00417345,zbMATH04105188} for works on freely generated categorical structures and their framing in two-dimensional monad theory; we intend to provide a detailed exposition of these foundations in our precise setting in future work.}
More explicitly, the universal property of $\Syntotal $ can be described as follows:
\begin{theorem}\label{theo:universal-property-source-language}
The distributive category $\Syntotal $ corresponding to our (total) source language has the following universal property. Given any distributive category $\catC $, for each pair
$	\left( K , \mathtt{h}   \right) $
where $K = \left(  K_ n \right) _ {n\in\NN} $ is a family of objects in $\catC $ and
$\mathtt{h} = \left( h_\op \right) _{\op\in\Op} $ is a consistent family of morphisms
in $\catC$, with $h_\op:K_{n_1}\times\cdots\times K_{n_{k_1}}\to K_{m_1}\sqcup\cdots\sqcup K_{m_{k_2}}$ for $\op\in\Op^{\mathbf{n}}_{\mathbf{m}}$, there is a unique structure-preserving functor (strictly bicartesian functor)
	\begin{equation}
		H : \Syntotal \to \catC
	\end{equation}
such that $H(\reals ^n ) = K _n $ and $H(\op ) = h_\op $ for any $n\in\NN$ and any primitive operation $\op\in\Op $.
\end{theorem}
\begin{proof}
Interpret each type constructor by the corresponding chosen product or coproduct, and each term constructor by its categorical operation. The product, coproduct, and substitution equations hold in every distributive category, so this interpretation descends to the equational theory of the total fragment. It preserves the chosen structure by construction. Conversely, any such functor has this action on constructors, and its action on the primitive types and operations is prescribed; induction on types and terms gives uniqueness.
\end{proof}

In order to state and prove correctness of CHAD, we need to give a concrete semantics to our
source language. For the purposes of this paper, we define the concrete semantics of the source language on the category $\VMAN$ of (differentiable) manifolds (with variable dimensions).
 More precisely, products and disjoint unions of manifolds distribute by the usual canonical diffeomorphism $M\times(N\sqcup K)\cong(M\times N)\sqcup(M\times K)$, and $M\times\emptyset=\emptyset$. Thus $\VMAN$ is a distributive category, and  by the universal property of $\Syntotal$, we can define the concrete semantics as a structure-preserving functor
\begin{equation}\label{eq:sematics-FUNCTOR}
	\sem{-}: \Syntotal\to\VMAN .
\end{equation}

\begin{theorem}[Semantics Functor]
There is only one (strictly) bicartesian functor \eqref{eq:sematics-FUNCTOR} such that, for each
$n\in \NN $, $\sem{\reals ^n} = \RR ^n $ and, for each $\left( \mathbf{n}, \mathbf{m}\right) \in \NN ^{k_1}\times \NN ^{k_2} $,
 $\sem{\op} $ is the differentiable function (morphism of $\VMAN$)
 $ \RR^{n_1}\times \cdots \times \RR ^{n_{k_1}}\to\RR^{m_1}\sqcup \cdots \sqcup \RR ^{m_{k_2}} $
 that $\op $ intends to implement.
\end{theorem}
\begin{proof}
	Since $\VMAN$ is a distributive category, the result follows from the universal property of $\Syntotal $ established in Theorem \ref{theo:universal-property-source-language}.
\end{proof}

\subsection{Target total language and its categorical semantics}\label{sec:target-language-total}
The target language of CHAD is a variant of the dependently typed enriched effect calculus \citep[Chapter 5]{vakar2017search}. Its cartesian types, linear types, and terms are generated by the
grammar of Fig. \ref{fig:sl-terms-types-kinds} and \ref{fig:tl-terms-types-kinds}, making the target language
a proper extension of the source language.
We describe the full language in more detail in Section~\ref{sec:target-language}.

We work with linear operations $\lop\in\LOp^{\mathbf{r};\mathbf{n}}_{\mathbf{m}}$,
which are intended to represent functions that are linear (in the sense of
respecting $\zero$ and $+$) in the
arguments corresponding to $\mathbf{n}=(n_1,\ldots,n_{k_1})$, with parameters of dimensions $\mathbf{r}$. The tuple $\mathbf{m}=(m_1,\ldots,m_{k_2})$ specifies the output dimensions. Thus $\mathbf{r}$ records the cartesian inputs, $\mathbf{n}$ the linear inputs, and $\mathbf{m}$ the linear outputs.

We
write
$$
\LDomain{\lop}\defeq \tangentreals^{n_1}\t*\ldots\t* \tangentreals^{n_{k_1}}
\qquad\text{and}\qquad
\LCDomain{\lop}\defeq \tangentreals^{m_1}\t*\ldots\t* \tangentreals^{m_{k_2}}
$$
for $\lop\in\LOp^{\mathbf{r};\mathbf{n}}_{\mathbf{m}}$.
The details of the typing rules and equational theory are described in Section~\ref{sec:target-language}, and it is
a fragment of the target language presented in \cite{lucatellivakar2021chad}.

To serve as a practical target language for the automatic derivatives of all programs from the source language, we make the following assumption:
for each $\left( k_1, k_2\right) \in\NN\times\NN $, each $\left( \mathbf{n}, \mathbf{m}\right) \in \NN ^{k_1}\times \NN ^{k_2} $, and each $\op\in \Op_{\mathbf{m}}^{\mathbf{n}}$, we have a chosen program
\begin{equation}
\begin{array}{l}
x_1:\reals^{n_1},\ldots,x_{k_1}:\reals^{n_{k_1}};\\\lvar:
\vMatch{\op(x_1,\ldots,x_{k_1})}{\tIni[1] \_\To \tangentreals^{m_{1}}\mid\cdots\mid \tIni[k_2] \_\to
\tangentreals^{m_{k_2}}}\\
\quad\vdash D\op(x_1,\ldots,x_{k_1})(\lvar):\tangentreals^{n_1}\times\cdots\times \tangentreals^{n_{k_1}}
\end{array}
\end{equation}
that intends to implement a function that gives the (transpose) derivative of $\op$.
In particular, in case $k_2=1$, we have that
\begin{equation*}
	\begin{array}{l}
	x_1:\reals^{n_1},\ldots,x_{k_1}:\reals^{n_{k_1}};\lvar: \tangentreals^{m_1}\vdash D\op(x_1,\ldots,x_{k_1})(\lvar):\tangentreals^{n_1}\times\cdots\times \tangentreals^{n_{k_1}}.
	\end{array}
	\end{equation*}

The initial model of the total target theory has an underlying strict indexed category $\LSynt:\CSynt^\op\to\catCat$, which is a $\Sigma$-bimodel for \textit{tuple and variant types} in the terminology of \cite{lucatellivakar2021chad}. Initiality is with respect to the full typed structure, including dependent pairs and linear-function types (see \citep[Chapter~5]{vakar2017search} and \cite[Section~7]{lucatellivakar2021chad}).

This provides us with Lemma \ref{lem:basic-property-bicart-closed-total} on the op-Grothendieck construction $\Sigma_{\CSynt }\LSynt^{op}$ of the syntactic categorical semantics  $\LSynt :  \CSynt ^\op \to \catCat $ of the target language. This is particularly useful, since the basic tenet of CHAD is to follow the recipe given by the concrete semantics, and give the AD macro as a structure preserving functor from the source language category and the op-Grothendieck construction of the target language, following the concrete semantics given by Theorem \ref{theo:products-coproducts-preserving}.

\begin{lemma}\label{lem:basic-property-bicart-closed-total}
$\Sigma_{\CSynt }\LSynt^{\op}$ is a distributive category.
\end{lemma}
\begin{proof}
	We refer the reader to \cite[Section~6]{lucatellivakar2021chad} for more details.
\end{proof}

Recalling that $\Fam{\Vect^\op}$ is the op-Grothendieck construction of $\Vect^{(-)}$, we can finally define the concrete categorical semantics of our target language, in terms of
a structure-preserving indexed functor $\LSynt \to \Vect ^{(-)}$ . If the reader is not familiar with indexed functors, we stress that the induced strictly bicartesian functor \eqref{eq:canonically-induced-target-language-reverse-UP} plays the most important role in our development.\footnote{We refer the reader, for instance, to \cite[Section~6.9]{lucatellivakar2021chad} for indexed functors. }

Before establishing the concrete semantics of our target language in Lemma \ref{lem:concrete-semantics-target-language-total}, we introduce the following notation for each  $\mathbf{n} = \left( n_1 , \ldots , n_k\right) \in \NN ^{k} $.
We denote by
\begin{equation}
	\overline{\RR ^\mathbf{n} }: \RR ^{n_1}\times \cdots \times \RR ^{n_k} \to \Vect, \qquad   \overline{\RR _\mathbf{n} }: \RR ^{n_1}\sqcup \cdots \sqcup \RR ^{n_k}\to \Vect
\end{equation}
respectively, the constant functor/family equal to $\RR^{n_1}\times \cdots \times\RR^{n_k}  $
and the family/functor defined by $\overline{\RR _\mathbf{n} }(x) = \RR^{n_t} $ if $x\in \RR^{n_t}$.

\begin{lemma}[Concrete semantics of the target language]\label{lem:concrete-semantics-target-language-total}
	The strict indexed category $\Vect ^{(-)} : \catSet ^\op\to\catCat$ is a $\Sigma$-bimodel for tuple and variant types.
	Together with families of sets for Cartesian dependent types, dependent sums, and sets of linear maps for linear-function types, this gives an interpretation of the full total target theory. Its underlying indexed functor $\LSynt \to \Vect ^{(-)}$ induces a structure-preserving (in particular, strictly bicartesian) functor
	\begin{equation} \label{eq:canonically-induced-target-language-reverse-UP}
		\semt{-}:  \Sigma_{\CSynt }\LSynt^{op} \to\Fam{\Vect ^\op }
	\end{equation}
	such that
	\begin{enumerate}[a)]
		\item \label{UP-TARGET1}  for each $n$-dimensional array $\reals^n\in\Syntotal$,
		$\semt{\reals^n}\defeqq \RR ^n\in\objects\left( \catSet\right) $;
		\item \label{UP-TARGET2} for each $n$-dimensional array $\reals^n\in\Syntotal$,
		$$\semtl{\reals^n} \defeqq \overline{\RR ^n} \in  \Vect^{\RR ^n} .  $$
		As consequence, for each $\mathbf{n} = \left( n_1 , \ldots , n_k\right) \in \NN ^{k} $,
		we have that $$\semtl{\reals^\mathbf{n} }\defeqq \semtl{\reals^{n_1}\times \cdots \times\reals^{n_k}}  = \overline{\RR ^\mathbf{n} } \in  \Vect^{\RR ^\mathbf{n} }; $$
				\item \label{UP-TARGET3} for each primitive $\op\in \Op^{\mathbf{n}}_\mathbf{m}$, where
		$\mathbf{n} = \left( n_1, \ldots , n_{k_1} \right) $ and $\mathbf{m} = \left( m_1, \ldots , m_{k_2}\right) $:
		\begin{enumerate}[i)]
			\item  $\semt{\op} :\RR^{n_1}\times\cdots\times \RR^{n_{k_1}}\To \RR^{m_1}\sqcup\cdots\sqcup \RR^{m_{k_2}}$ is the map in $\Set$ corresponding to the operation that $\op$ intends to implement;
			\item $\semt{D\op}\in\Vect ^{\left( \semt{\reals^{n_1}}  \times\cdots\times \semt{\reals^{n_{k_1}}}  \right)} \left( \semtlu{\reals_{\mathbf{m}} } \circ  \semt{\op}, \semtl{\reals^{\mathbf{n}}}  \right) $, where
			$\semtlu{\reals_{\mathbf{m}} } = \overline{\RR _\mathbf{m}}$; the morphism $\semt{D\op}$ is the family of linear transformations that $D\op$ intends to implement.
		\end{enumerate}
		\end{enumerate}
	\end{lemma}
	\begin{proof}
		We refer the reader to \cite[Section~6 and Section~7]{lucatellivakar2021chad} for more details.
	\end{proof}

\subsection{Macro and correctness}\label{sub:correctness-macro-total}

CHAD is a program transformation between the source language discussed in Section~\ref{sec:source-language-total}
and the target language described in Section~\ref{sec:target-language-total}.
We firstly establish how to correctly implement the CHAD-derivatives of the primitive (types and) operations,
respecting the concrete semantics defined in Section~\ref{sec:semantics-source-language-total} and Lemma \ref{lem:concrete-semantics-target-language-total}. Then, we define CHAD as the only \textit{structure-preserving transformation} between the source language and the total category (op-Grothendieck construction) obtained from the target language that extends the implementation on the primitive operations. This allows for a straightforward correctness proof (presented in Theorem \ref{theo:fundamental-correctness-total-language}).

As discussed earlier, our present approach starts with primitive operations that are differentiable over the entire domain. For clarity and conciseness, we present the macro in this simplified setting. However, the approach can be generalized to accommodate primitive operations with singularities, or even more complex settings, without affecting the correctness proof. This broader generalization is left to future work.

We refer to Subsect.~\ref{sec:MACRO} for the precise definition of our macro $\Dsynrevarg{-}$, restricting its attention to the total language fragment of the source language and target languages (for a detailed revision of CHAD for total languages, we refer the reader to \cite[Section~8]{lucatellivakar2021chad}). At the categorical semantics level, the (total fragment) of the macro defined in Subsect.~\ref{sec:MACRO} corresponds to the following functor.
\begin{theorem}
	There is only  one (strictly) bicartesian functor
\begin{equation}
\DSynrevt :  \Syntotal \to \Sigma_{\CSynt }\LSynt^{op}
\end{equation}
	such that, for each $n\in\NN $,
	\begin{equation}
		\DSynrevt \left( \reals ^n \right) = \left( \reals ^n , \tangentreals^n\right),
	\end{equation}
	 and, for each $\op\in\Op$,   $\DSynrevt \left(\op \right) $ implements the CHAD-derivative
	of the primitive operation $\op\in\Op$ of the source language, that is to say,
	\begin{equation}
\DSynrevt \left(\op \right) = \left( \op, D\op\right) ,
\end{equation}
Here $\tangentreals^n$ denotes the constant linear type over the cartesian type $\reals^n$. If the chosen primitive derivative programs are correct, then
\[
\semt{\DSynrevt(\op)}=\DSemtotal{\sem{\op}},
\]
where $\DSemtotal{}$ is the \textit{CHAD-derivative functor} defined in Definition~\ref{def:CHAD-derivative-functor-total}. The universal property specifies the images of the syntactic generators; the displayed semantic equality expresses correctness of those chosen images.
The functor $\DSynrevt $ corresponds to the total fragment of the macro $\DSynmacrot$ defined in Subsect.~\ref{sec:MACRO}.
\end{theorem}

\begin{proof}
Since $\Sigma_{\CSynt }\LSynt^{op} $ is a distributive category, the result follows from the universal property of $\Syntotal$ (Theorem \ref{theo:universal-property-source-language}).
\end{proof}

Finally, we can state and prove correctness of CHAD for the total language setting above.
We assume that the chosen programs $D\op$ correctly implement the coderivatives of the primitive operations, as in the equality above. The correctness diagram then compares two routes from a source program to a primal and a cotangent computation. One first differentiates the program and then interprets it; the other first interprets the program and then differentiates the resulting function. Both routes preserve products and variants and agree on the primitive operations. The universal property says that this agreement extends to every program.

\begin{theorem}[Correctness of CHAD for total first-order programs] \label{theo:fundamental-correctness-total-language}
	The diagram below commutes.
\begin{equation}\label{deag1:correctness-total-CHAD}
	\diag{correctness-total-language}
\end{equation}
\end{theorem}
\begin{proof}
By the universal property of the distributive category $\Syntotal$ established in Theorem \ref{theo:universal-property-source-language}, since $\Fam{\Vect ^\op }$ is a distributive category, we have that there is only one (strictly) bicartesian functor
\begin{equation}
\mathfrak{C} : \Syntotal \to \Fam{\Vect ^\op },
\end{equation}
such that
\begin{enumerate}[c1)]
	\item for each $n\in \NN $, $\mathfrak{C} \left( \reals ^n\right) = \left( \RR ^n, \overline{\RR ^n} \right)  $;\label{condition1:correctness-total-functor}
	\item $ \mathfrak{C}\left( \op \right) = \DSemtotal{\sem{\op}}   $ for each $\op \in \Op $.\label{condition2:correctness-total-functor}
\end{enumerate}
Since both $\DSemtotal{\sem{-}}: \Syntotal \to \Fam{\Vect ^\op }  $ and $ \semt{\DSynrevt \left( - \right)}:  \Syntotal \to \Fam{\Vect ^\op } $ are strictly bicartesian functors such that \ref{condition1:correctness-total-functor} and \ref{condition2:correctness-total-functor} hold,
we get that $\DSemtotal{\sem{-}} = \mathfrak{C} = \semt{\DSynrevt \left( - \right)}$. That is to say, Diagram
\eqref{deag1:correctness-total-CHAD} commutes.
\end{proof}

\Needspace{4\baselineskip}
\begin{corollary}
	For any well-typed program
$	x: \ty{1} \vdash t: \ty{2}, $
we have that $\semt{\DSynmacrot{t}} = \DSemtotal{\sem{t}} $.
\end{corollary}
\begin{proof}
	It follows from the fact that $\semt{\DSynrevt \left( t \right)} = \DSemtotal{\sem{t}} $ for any morphism
	$t$ of $\Syntotal$.
\end{proof}

\part{Iteration and dependent types}\label{part:iteration}

A loop repeatedly performs a computation until it returns a result, using a variant to distinguish a final result from a new state with which to continue. For a terminating loop, the backward map composes the cotangent maps of the steps performed in reverse order. The concrete partial models introduced next explain how a loop and its derivative behave, including what happens when a computation does not return.

We then study the indexed structure that allows this behaviour to be expressed in a dependently typed language. The question is how reindexing along a possibly partial loop relates to reindexing along its body. Section~\ref{sec:denotational-partial-features} establishes the concrete semantics, while Section~\ref{sec:initial-algebras} develops the more general answer in terms of parameterized initial algebras. Their folds express how to handle a return and how to pass through one further step. We show that this construction agrees with iteration in our concrete models, and then identify the folder operations needed by the target language. Thus, the general account of dependent iteration supplies the means to extend CHAD.

\section{Iterative CHAD: basic concrete semantics}\label{sec:denotational-partial-features}
The goal of the present work is to extend CHAD to accommodate languages with partial features. Particularly, we focus on iteration constructs, which are relevant in many scientific computing and machine learning applications where we want to compute derivatives.
Extensions to languages with general recursive features, including CHAD for PCF~\cite{zbMATH03576672} and FPC~\cite{fiore1994axiomatisation}, are left for future work.

CHAD is fundamentally principled by a concrete denotational semantics that adheres to the principle of computing CHAD-derivatives in a functorial and \textit{structure-preserving} way.
We aim to follow the same tenet while introducing partial features to our framework.
The initial step, therefore, is to define the concrete denotational semantics for our extended setting with iteration.

Establishing denotational semantics for derivatives of partially defined functions presents several possibilities, especially regarding differentiability over domains~\cite{2023arXiv230210636H} and the so-called \textit{if-problem}~\cite{beck1994if}.
For the sake of clarity and conciseness, we
extend our previous work on CHAD for total languages by considering primitive operations that are ``differentiable on their entire domain'' and by following Abadi-Plotkin solution of side-stepping the if-problem via partiality~\cite{abadi-plotkin2020}, as we outline below.

Our approach aligns with our previous work on AD in dual-numbers style~\cite{2022arXiv221007724L, 2022arXiv221008530L}, and is the natural \textit{extension} of our work on CHAD for total languages~\cite{lucatellivakar2021chad}. Furthermore, our approach is suitable for several practical AD applications, particularly in the context of our implementations~\cite{smeding2022, 2023arXiv230705738S}.
Although outside of our scope, we claim that our proofs about CHAD for iteration could be adapted to various concrete semantics, including alternative ones such as \cite{2023arXiv230210636H} along similar lines to the development for dual numbers AD in \cite[Section 10]{2022arXiv221007724L}.
Adapting CHAD to these concrete semantics will provide us with different approaches to the if-problem and
non-differentiability. A detailed exploration of these extensions is deferred to future work.

\subsection{Derivatives of partially defined functions, and the if-problem}
We extend CHAD to languages with partial features by closely following our approach for total languages. Specifically, we adapt the previously established concrete semantics to accommodate the partial constructs. For clarity and conciseness of our exposition, we focus here on applying CHAD to a (fine-grain) call-by-value language, noting that similar work for call-by-push-value and call-by-need languages is  possible, albeit with additional technical details.\footnote{We refer the reader to \cite{levy2003modelling, levy2012call, vakar2017search, 2022arXiv221008530L, 2022arXiv221007724L} for more details on the semantics of call-by-value and call-by-push-value languages.}

The primary challenge in this extension arises from addressing partiality in the presence of differentiability. We need to formally define the denotational semantics for the derivatives of functions that are only partially defined. In essence, while extending the concrete semantics presented in Section \ref{sec:semantics-CHAD}, our work is about providing a domain structure
to $\VMAN $.

Extending the concrete semantics to a call-by-value language involves providing semantics not just for values but also for computations that exhibit partiality. As discussed, we have two possible approaches to extend our concrete semantics in this setting.

The first approach is to extend the category $\VMAN$ to include more morphisms, effectively relaxing the semantics of values to encompass functions that are not differentiable over their entire domains, which would
add partial differentiability to our total language, or even consider more intricate alternative semantics along the lines of \cite{2023arXiv230210636H}.
\textit{We emphasize that CHAD is compatible with alternative semantics since the basic \textit{structure-preserving} tenet of CHAD holds in any suitable semantics for automatic differentiation}.

Nevertheless, we adopt a second approach that we outline below: it consists of maintaining our existing choice for the semantics of values -- restricting them to total, differentiable functions (morphisms of $\VMAN$) -- and handling partiality and non-differentiability at the level of computations by considering a standard domain structure of open subsets. %
By adopting this approach, we are able to give a straightforward exposition of the basic principles of CHAD for iteration, ensuring compatibility with our previous expositions on CHAD~\cite{lucatellivakar2021chad, DBLP:journals/toplas/VakarS22}.

\subsubsection{Freyd-categorical structure over $\VMAN$}
Freyd categories provide the standard categorical semantics for (fine-grain) call-by-value languages. We quickly recall the definition of \textit{Freyd categories}\footnote{What we call Freyd category herein is usually called \textit{distributive Freyd category}, \textit{e.g.} \cite{DBLP:journals/entcs/Staton14}.} below,
although we refer to \cite{levy2003modelling, DBLP:journals/entcs/Staton14} for further details.

\begin{definition}[Freyd category]\label{def:distributive-Freyd-category}
A \textit{(distributive) Freyd category}, herein, is a quadruple $\left( \catV , \catC , j, \otimes \right) $ where $\catV$ is a distributive category, $\catC$ is a category with finite coproducts, and:
\begin{itemize}
	\item $\otimes $ is an action\footnote{Recall that, by currying, an action is equivalently a strong monoidal functor from the cartesian category $\left( \catV , \times , \mathrm{1} \right) $ to the monoidal category of endofunctors $\left( \left[\catC , \catC  \right], \circ , \id \right)$. } of the cartesian category $\catV  $ on the \textit{category} $\catC $ that distributes over (finite) coproducts in the sense that $A\otimes (-)$ preserves (finite) coproducts, denoted as
	\begin{equation}
		\left( - \otimes -\right) :\catV\times \catC \to \catC ;
	\end{equation}
\item $j:\catV\to \catC $ is a strictly (finite) coproduct-preserving functor that is identity on objects such that
$j\left( A\times B \right) = A\otimes B $ for any pair $A,B\in\obj\ {(\catV)}$, and
$j(f\times g)=f\otimes j(g)$ for morphisms $f,g$ of $\catV$. We require $j$ to respect the unit and associativity isomorphisms of these actions.
\end{itemize}

A \textit{Freyd category morphism} $\left( \catV , \catC , j, \otimes \right)\to\left( \catV' , \catC' , j', \otimes ' \right)  $ between Freyd categories is a pair of functors $F:\catV\to\catV'$ and $\hat{F}:\catC\to\catC'$ such that $F$ is strictly bicartesian, Diagram~\eqref{defdiag:Freyd-categoriesmorphisms} commutes, and $\hat{F}(f\otimes g)=F(f)\otimes'\hat{F}(g)$ for morphisms $f$ of $\catV$ and $g$ of $\catC$. We also require preservation of the unit and associativity isomorphisms of the actions.\footnote{It should be noted that, under the conditions of the definition, we can conclude that $\catC$ has finite coproducts and $\hat{F}$ preserves finite coproducts, \textit{e.g.} \cite{zbMATH07041646}.} It should be noted that Freyd categories and Freyd category morphisms form a category.
\begin{equation}\label{defdiag:Freyd-categoriesmorphisms}
	\diag{freyd-category-morphismm}
\end{equation}
\end{definition}

Here $\catV$ describes values, while $\catC$ describes computations. The functor $j$ regards a value as a computation, and $A\otimes(-)$ carries a value of type $A$ alongside a computation. Thus $\catC$ need not be cartesian: copying and discarding are supplied by the value category, and the action makes them available in a computation context.

The Kleisli viewpoint applies when $j$ has a right adjoint $U$ \cite[Section~4]{levy2003modelling}. Indeed,
\[
\catC(jA,jB)\cong\catV(A,UjB)
\]
identifies $\catC$ with the Kleisli category of the induced monad $T=Uj$, and the Freyd action induces its strength.

The aim of this section is to extend the (semantic) CHAD-derivative functor presented in Theorem \ref{theo:products-coproducts-preserving} to a
Freyd category morphism, establishing how the CHAD-derivative interacts with non-termination, non-differentiability, and iteration.

The starting point is to extend the semantics of the source language to include computations that involve non-termination and non-differentiable functions. In order to do so, we consider the domain structure induced by open subsets on $\VMAN$.
\begin{definition}[Open Domain Structure]
We consider the subcategory $\OMAN$ of $\VMAN $ consisting of all objects of $\VMAN$ and open embeddings as morphisms. An open embedding identifies its source diffeomorphically with an open subset of its target; we represent it by the corresponding inclusion $V\subset M$.

The pair $\left( \VMAN , \OMAN \right) $ is a domain structure for $\VMAN $ (see, for instance, \cite[Def.~3.1.1]{fiore2004axiomatic}). It gives rise to the category of partial maps $\PVMAN$, and
we roughly denote $\PVMAN = \left( \VMAN , \OMAN \right) $.
\end{definition}

In other words, $\PVMAN$ is the category of manifolds, where the morphisms are differentiable functions that are only defined on open subsets. Explicitly, the category $\PVMAN$ has manifolds as objects. A morphism $\mathtt{f}: M\to N $ in $\PVMAN$ is a pair  $\mathtt{f} = \left( V, f\right) $ where $V\subset M $ is an open subset, and $f: V\to N $
is a morphism in $\VMAN $. In this setting,  $\domain{\mathtt{f}}\coloneqq V$ is called the domain
of $\mathtt{f}$, while $ f $ is the underlying function.

Composition of morphisms in $\PVMAN$ is given by the usual pullback/span style. Explicitly, given morphisms
$\mathtt{f} = \left( V, f\right) : M\to N $  and $\mathtt{g} = \left( W, g\right) : N\to K $, the composition
is given by
\begin{equation}
\mathtt{g}\circ \mathtt{f} = \left( V\cap f^{-1}(W), \mathtt{gf}\right)
\end{equation}
where $\mathtt{gf}: V\cap f^{-1}(W)\to K $ is defined by $\mathtt{gf}(x) = g(f(x)) $
for each $x\in V\cap f^{-1}(W)$.

Finally, let
$\pman : \VMAN \to \PVMAN $ be the obvious inclusion functor that sends $f:M\to N$ to $(M, f)$. We have the Freyd category
\begin{equation}\label{eq:freyd-category-source-language-semantics}
\FFMAN\coloneqq 	\left( \VMAN , \PVMAN , \pman , \otimes \right) ,
\end{equation}
where, for each morphism $f: M\to N $ of $\VMAN$  and each morphism $\mathtt{g} = \left( W, g\right) : R\to K $ of $\PVMAN$,
$$f\otimes\left( W, g\right)\coloneqq\left(M\times W,f\times g\right):M\times R\to N\times K.$$
$\FFMAN$ serves as a concrete model
for our (fine grain) call-by-value language.

\subsubsection{Enrichment of $\PVMAN$.}
The order used below compares computations by how much of their result is defined. An increasing chain gives successively more information, and its least upper bound collects all the information obtained at finite stages. This will allow us to interpret a loop by its finite terminating executions. We recall the standard domain-theoretic constructions needed for this purpose; see \cite[Section~3.2.4]{abramskyJung1994}.

Recall that a \wcpo (pointed and $\ww$-chain complete partially ordered set)
is a partially ordered set that has a least element, usually denoted by $\bot$, and colimits (least upper bounds) of all $\ww$-chains (countable ascending sequences). A morphism between \wcpos is a monotonic function that preserves colimits of $\omega$-chains and the bottom element. We denote the category of \wcpos and their morphisms by $\wCpo$.

\begin{remark}
It should be noted that $\wCpo$ has coproducts and products. More precisely, by denoting $\left( - \right) _0 : \wCpo \to \catCat $ the functor that takes each \wcpo to the underlying category of its order, we have the following. Thus $A_0$ denotes a category, while $\bot_A$ denotes its least element.
For each $\left(A, B\right) \in \obj{\left( \wCpo\right) }\times \obj{\left(\wCpo\right)}$, we have that
$\left( A\times B \right) _0 = \left( A \right) _0\times\left( B \right) _0 $,
and $\left( A\sqcup B \right) _0  $ is obtained by taking the coproduct of $A_0$ and $B_0$
and identifying the initial objects of $A_0$ and $B_0$.
Explicitly, $A\times B$ is the product of sets with the product order in which $(a,b)\leq_{A\times B} (a',b')$ iff $a\leq_A a'$ and $b\leq_B b'$, where $\bot_{A\times B} = (\bot_A,\bot_B)$.
$A\sqcup B$ is the wedge sum of pointed sets with the union of the orders.
\end{remark}

We consider the category $\wCpo$ equipped with a monoidal product known as the smash product, denoted by $\smashproduct$. Given \wcpos $A$ and $B$, the smash product $A\smashproduct B $ is
constructed by taking the cartesian product $A\times B$ and, then, identifying all pairs
where at least one of the components is $\bot $. Formally, we define an equivalence relation $\sim$ on $A \times B$ generated by:
\[
 \left(\bot,b\right)\sim\left(a,\bot\right)\qquad\text{for any }(a,b)\in A\times B.
\]
The elements of $A\smashproduct B $ are the equivalence classes of this relation, while the
partial order is the order on $A\times B$ modulo this equivalence relation. We can think about $A\smashproduct B $  as
$$\left( \left( A- \left\{ \bot\right\}\right) \times \left( B- \left\{ \bot\right\}\right)\right)\cup \left\{ \bot \right\} $$
with $\bot $ as the least element and the usual cartesian product order on the rest of the elements.
Finally, the identity/unit object of this smash product is
given by the \wcpo with two elements  $ \mathrm{2} \defeqq \left\{ \bot , \top \right\}$, the bottom and top elements.

Herein, an $\wCpo$-enriched category $\catC $  is a category enriched over the monoidal category $(\wCpo, \smashproduct, \mathrm{2})$.
Concretely, this means that for any objects $A, B, C$ in $\catC $, the hom-object $\catC\left( A, B\right) $
is an object of $\wCpo $, and the composition operations
$$ \circ : \catC \left(B, C \right)\smashproduct \catC \left( A, B\right) \to \catC \left( A, C\right) $$
are $\wCpo $-morphisms.\footnote{For the general theory of enriched categories, we refer to \cite{MR0280560,zbMATH04165146}; for the domain-theoretic setting used here, see \cite{abramskyJung1994,fiore2004axiomatic}.} \textit{In the setting above, we denote by $\bot$ the least element in the hom-object $\catC\left( A, B\right) $ for any $\left( A, B\right) \in \obj{(\catC)}\times \obj{(\catC)}$.}
\begin{theorem}[{\cite[Section~3.2.4]{abramskyJung1994}}]
	The monoidal category $(\wCpo, \smashproduct, \mathrm{2})$ is (monoidal) closed, where the order in the set $\wCpo\left( W, Y \right) $ of $\wCpo$-morphisms is defined pointwise. In other words, $\wCpo $ is $\wCpo $-enriched.
\end{theorem}

\begin{lemma}
	$\PVMAN$ is naturally $\wCpo$-enriched when endowed with the domain-order for the morphisms. That is to say, given $\mathtt{f}, \mathtt{g}: M\to N $, we say that $\mathtt{f}\leq \mathtt{g} $ if  $\domain{\mathtt{f}}\subset \domain{\mathtt{g}}$ and
	$f(x) = g(x) $ for any $x\in \domain{\mathtt{f}} $.
\end{lemma}
\begin{proof}
	Given an $\ww$-chain of domains (open subsets of a manifold $M$), the union of all of them is
	an open set. Therefore, the colimit of
	$$\mathtt{f}_0 = \left( U_0, f_0\right) \leq \mathtt{f}_1 = \left( U_1, f_1\right) \leq \mathtt{f}_2 = \left( U_2, f_2\right)  \leq \cdots \leq \mathtt{f}_n = \left( U_n, f_n\right) \leq \cdots $$
	in $\PVMAN\left( M, N \right) $
	is defined by
	\begin{equation}
	\mathtt{f} = \left( \bigcup _{t\in\NN} U_t, f \right)  : M\to N, \qquad f(x) \coloneqq f_t(x), \mbox{ if }x\in U_t,
	\end{equation}
which is also a morphism of $\PVMAN$: the functions agree on overlaps, and differentiability is a local property. Composition preserves these unions in each variable and is strict in each variable, giving the required enriched composition.
The least element of $\PVMAN\left( M, N \right)$ is given by the only morphism $\bot \coloneqq \left( \emptyset , \emptyset\to N\right) : M\to N $.
\end{proof}

\begin{definition}[$\ww$-Freyd category]
	A quadruple $\left( \catV , \catC , j , \otimes\right) $ is an \textit{$\ww$-Freyd category}
	if it is a Freyd category such that $\catC$ is an $\wCpo$-enriched category and its chosen finite coproducts are $\wCpo$-coproducts.

	A pair $\left( F, \hat{F}\right) $ is an \textit{$\ww$-Freyd category morphism}
	if $\left( F, \hat{F}\right) $ is a Freyd category morphism between $\ww$-Freyd categories,
	and $\hat{F}$ is an $\wCpo$-enriched functor.
\end{definition}

\begin{theorem}\label{theo:ww-Freyd-category-Manifolds}
$\FFMAN \coloneqq 	\left( \VMAN , \PVMAN , \pman , \otimes \right) $ is an $\ww$-Freyd category.
\end{theorem}
\begin{proof}
The enrichment was established above. Restriction along the coprojections identifies a hom-object out of a disjoint union with the product of the two hom-objects, including their orders and least upper bounds. Thus the chosen coproducts are enriched.
\end{proof}

\subsubsection{A word on the if-problem}
By defining the domain structure $\PVMAN = \left( \VMAN, \OMAN \right)$, we inherently establish our approach to the \textit{if-problem},~\cite{beck1994if} which concerns handling singularities arising from branches in conditional expressions.

In our semantics, we restrict ourselves to open domains. Therefore, in our language, any construct of the form \texttt{if-then-else} can only define functions whose domains are open sets and, hence, undefined in singularities.

This approach follows the solution proposed by Abadi and Plotkin~\cite{abadi-plotkin2020} and is general enough to encompass our practical goals. We emphasize that we have made this choice for clarity and simplicity -- however, as stressed above, CHAD is also compatible with other semantic frameworks
such as that of \cite{2023arXiv230210636H} or \cite[Section 10]{2022arXiv221007724L}.

\subsection{Concrete semantics for the target language}\label{subsect:concrete-semantics-FamVectop}
In order to follow our structure-preservation principle, we define our (concrete) CHAD-derivative as a structure-preserving functor in our extended setting -- that is, we define CHAD as a suitable Freyd category morphism. In order to do so, we extend the concrete categorical semantics of our target language into a suitable Freyd category that handles
partially defined differentiable functions.

\subsubsection{Basics of the concrete semantics of the target language}
The construction is the familiar one of adjoining an extra result $\bot$ for an undefined computation. Here a result also carries a vector space, so the extra result carries the zero vector space. On $\Fam{\Vect^\op}$ this is the monad $\mathcal{T}=(-)\sqcup\mathbf{1}$. The indexed description below records the same construction separately on the set of results and on their vector spaces; its Kleisli category will give the required partial computations.

Although we do not assume any knowledge on indexed monads to further understand this work, we stress that, denoting by $\mathrm{1}$ the set containing a single element $\bot$, we model computations in our target language by considering the indexed monad\footnote{The interested reader can find more about indexed functors, for instance, in \cite{lucatellivakar2021chad, vakar2017search}.}  $\mathfrak{T}$ on the indexed category $\Vect ^{(-)}: \Set^{\op}\to \catCat $
induced by the monad $\left(  -\sqcup \mathrm{1} \right): \Set \to \Set $.  More precisely,
$\mathfrak{T}$ is defined by the indexed functor
\begin{equation}
\diag{triangle-monad-morphism}
\end{equation}
with the obvious multiplication and unit. Here,
$\uppsi $ is the natural transformation
defined pointwise by
\begin{equation}
\uppsi _A : \Vect ^{A}\to \Vect ^{\left( A\sqcup  \mathrm{1} \right)}, \qquad \left( X : A\to \Vect\right)  \mapsto \left( \cpairL X, \overline{0}  \cpairR : A\sqcup\mathrm{1}\to\Vect \right)
\end{equation}
where $\overline{0} : \mathrm{1} \to \Vect$ is the single family consisting of the zero object of $\Vect$, and  $\cpairL X, \overline{0} \cpairR : A \sqcup \mathrm{1} \to \Vect$ is the family defined by $X$ and $\overline{0}$.

More importantly to our exposition, the (indexed) monad $\mathfrak{T}$ induces the monad
\begin{equation}
	\mathcal{T}\left( M, X\right) = \left(M, X \right)  \sqcup \mathbf{1} = \left( M\sqcup \mathrm{1}, \cpairL X, \overline{0}  \cpairR\right)
\end{equation}
on %
$\Fam{\Vect ^\op}$, where we denote by $\mathbf{1}$ the terminal object $\left( \mathrm{1}, \overline{0} \right)$ of $\Fam{\Vect^\op}$.
The Kleisli category of the monad $\mathcal{T}$, denoted by $\PFam{\Vect^\op}$, is the standard way of modelling partial computations
on $\Fam{\Vect ^\op} $, by adjoining an extra point to the codomain of morphisms, capturing non-termination in the usual way.

Explicitly, recall that a morphism $\mathit{f} : \left( A, X \right) \to \left( B, Y \right)$ in $\PFam{\Vect^\op}$ corresponds to a morphism $$\left( \mathtt{f}, \mathtt{f}'\right) : \left( A, X \right)\to \left( B \sqcup \mathrm{1}, \cpairL Y, \overline{0}  \cpairR \right) $$
in $\Fam{\Vect^\op}$.
In this setting, we define the domain of $\mathit{f}$ (or $\mathtt{f}$) as $\domain{\mathit{f}}\coloneqq\mathtt{f}^{-1}(B)$; that is, the set of all $a \in A$ such that $\mathtt{f}(a) \in B$. We say that $\mathit{f}$ (or $\mathtt{f}$) is \textit{defined} at $a \in A$ if $\mathtt{f} (a) \neq \bot$; we say that $\mathit{f}$ (or $\mathtt{f}$) is undefined at $a$ otherwise. It should be noted that, when $\mathit{f}$ is undefined at $a\in A $, we have that $\mathtt{f}'_a$ is the unique morphism $0\to X(a) $ from the zero vector space.

\begin{lemma} \label{lem:domain-structure-Fam}
	There is a bijection between morphisms $\PFam{\Vect ^\op}\left( \left( A, X\right) , \left( B, Y \right) \right)$ and pairs $\left( V, \hat{f} \right) $ where $V\subset A$ and $\hat{f} = \left( f, f'\right) $ is a morphism
	 $\left( V, X|_{V} \right) \to \left( B, Y \right)$ in $\Fam{\Vect ^\op} $, where we denote by $X|_{V}$ the restriction of $X$ to $V$.
\end{lemma}
\begin{proof}
It is enough to see that we have such a correspondence when we set
\begin{equation}
\mathbf{f}\mapsto \left( \domain{\mathbf{f}}, f ,f'\right)
\end{equation}
where $\left( f, f'\right) $ is the appropriate restriction of the morphism
$\left( \mathtt{f}, \mathtt{f}'\right) : \left( A, X \right)\to \left( B \sqcup \mathrm{1}, \cpairL Y, \overline{0}  \cpairR \right)$  that corresponds to $\mathit{f}$ in $\PFam{\Vect^\op}$. Explicitly,
$f\coloneqq \mathtt{f}|_{V}$ and $f'_a\coloneqq \mathtt{f}'_a $ for each $a\in V\subset A$.
\end{proof}

\textit{Henceforth, we denote by  $\mathbf{f} = \left( V, f, f'\right) : \left( A, X \right) \to \left( B, Y \right) $ the morphisms of $\PFam{\Vect ^\op}$.} With this notation,  if
 $\mathbf{f} = \left( V, f, f'\right): \left( A, X \right) \to \left( B, Y \right) $
 and
 $\mathbf{g}  = \left( W, g, g'\right): \left( B, Y \right) \to \left( C, Z \right) $
 are morphisms of $\PFam{\Vect^\op}$, the composition is given by
 \begin{equation}
 	\mathbf{g}\circ \mathbf{f} \coloneqq \left( V\cap f^{-1}(W), \left(gf\right) _{VW}, \left( gf\right) '\right)
 \end{equation}
where $\left( \left(gf\right) _{VW}, \left( gf\right) '\right): \left( V\cap f^{-1}(W), X_{V\cap f^{-1}(W)}\right) \to \left( C, Z\right) $ is defined by $\left(gf\right) _{VW}(a)\coloneqq g(f(a)) $ and  $\left( gf\right) '_a \coloneqq f'_a\circ g'_{f(a)} $ for each $a\in V\cap f^{-1}(W)$.

The above, together with Lemma \ref{lem:domain-structure-Fam}, shows that $\PFam{\Vect ^\op}$ is obtained from the domain structure induced by subset inclusions (with identities as second coordinate) 	 in $\Fam{\Vect ^\op }$. More importantly to our setting, denoting by $\pfam : \Fam{\Vect^\op} \to \PFam{\Vect^\op}$ the standard Kleisli inclusion functor, we have the Freyd category
\begin{equation}\label{eq:FFVECT}
	\FFFVECT\coloneqq	\left(\Fam{\Vect^\op}, \PFam{\Vect^\op} , \pfam , \otimes \right),
\end{equation}
where, for each morphism $(f,f'): (A, X)\to (B,Y)$ in $\Fam{\Vect ^\op}$ and each morphism
$(W, g,g'): (C, Y_0)\to (D,Z)$ in $\PFam{\Vect ^\op}$,
 $$(f,f')\otimes (W, g, g')  \coloneq \left( A\times W, (f,f')\times (g,g')\right)  .$$

\subsubsection{$\PFam{\Vect^\op}$ as an op-Grothendieck construction} \label{subsect:opGrothendieck-construction-PFAM}
The category
$\catPSet$ of sets and partially defined functions can be seen as a full subcategory of the category $\catCatb $ of the categories with
initial objects and colimits of $\ww$-chains, and structure-preserving functors as morphisms. More precisely, we have fully faithful functors
\begin{equation}\label{eq:inclusion-PSet-Catb}
	 \catPSet\xto{\inCC{-}}\wCpo \to \catCatb ,
\end{equation}
in which the second arrow is the usual inclusion, while the first one is defined as follows. It takes
each set $A$ to the pointed \wcpo $\inCC{A}$ given by $ A\sqcup \left\{ \leastelement \right\}$, the usual flat domain, where $\leastelement $ is the bottom element, and $x\leq y $ implies either $x=y$ or $x=\leastelement$; each partially defined morphism $f: A\to B $ is taken, then, to the $\wCpo $-morphism
\begin{equation}
	\inCC{f}: \inCC{A}\to \inCC{B}
\end{equation}
defined by $\inCC{f}\left( x\right) = f(x)$ in case $x$ is in the domain of definition of $f$, and $\inCC{f}\left( x\right) = \leastelement $ otherwise.

By abuse of notation, we also denote by $\inCC{-}: \catPSet\to\catCatb  $ the composition of the functors \eqref{eq:inclusion-PSet-Catb}.

The category $\PFam{\Vect^\op}$ can be seen as an op-Grothendieck construction of
the indexed category $\Vect _{\leastelement} ^{\left( -\right) } $, which
is defined by
 \begin{equation}\label{def:Vectpartial}
\Vect _{\leastelement} ^{\left( -\right) } \defeq \catCatb\left( \inCC{-}, \Vect \right) :  \catPSet ^\op \to \catCat .
\end{equation}

\subsubsection{$\wCpo$-enrichment} \label{def:wcpoFAMVECTop}   $\PFam{\Vect^\op}$
naturally inherits an $\wCpo$-enrichment
induced by the category of sets with partially defined functions. More precisely,
we have that
$$\left( \mathtt{f}, \mathtt{f}'\right) \leq \left( \mathtt{g}, \mathtt{g}'\right) $$
in $\PFam{\Vect^\op}\left( \left( A, X\right) , \left( B,Y\right)    \right) $
if the following conditions are satisfied:
\begin{itemize}
	\item  $\domain{\mathit{f}}\subset \domain{\mathit{g}} $;
	\item for each $a\in\domain{\mathit{f}} $, $\mathtt{f}(a) = \mathtt{g}(a)$ and $\mathtt{f}'_a = \mathtt{g}'_a$.
\end{itemize}
This standard ordering reflects the idea that $\left( \mathtt{g}, \mathtt{g}'\right) $ extends $\left( \mathtt{f}, \mathtt{f}'\right) $ while fully agreeing in the common domain.

\begin{theorem}\label{theo:ww-Freyd-category-Fam}
With the above, $\PFam{\Vect^\op}$ is an $\wCpo$-enriched category and, hence, $\FFFVECT$ is an
$\ww$-Freyd category.
\end{theorem}
\begin{proof}
The least morphism has empty domain. For an increasing chain, take the union of the domains and use the components of any stage at which the input is defined. These components agree at later stages, so the result is well-defined. The formula for composition shows that composition preserves these unions and the least morphisms. Restriction to each summand also identifies the hom-object out of a coproduct with the product of the hom-objects, so coproducts are enriched.
\end{proof}

The order construction does not depend on a special property of vector spaces. More generally, for a category $\catD$ with an initial object, the same argument equips the Kleisli category $\Fam{\catD^\op}_{(-)\sqcup\mathbf{1}}$ with an $\wCpo$-enrichment. Further assumptions, such as finite coproducts in $\catD$, supply the product and Freyd structure used here.

\subsection{Partial CHAD-derivative as a morphism}
Let $\mathtt{g} = \left( V, g \right) : M\to N $ be a morphism in $\PVMAN $, we define the morphism
\begin{equation}
	\DSempartial{\mathtt{g}} = \left( \mathtt{g}_0 , \DSempart{\mathtt{g} }\right)   : \left( M , \Tagen{M}\right) \to \left( N , \Tagen{N}\right),
\end{equation}
called the CHAD-derivative of $\mathtt{g}$. We start by considering the
(total) CHAD-derivative $$\DSemtotal{g} :\left( V , \Tagen{V}\right) \to \left( N , \Tagen{N}\right) $$
of $g: V\to N$. We define, then, $\DSempartial{\mathtt{g}}$ to be the canonical extension of
$\DSemtotal{g}$ as follows:
\begin{itemize}
	\item $\mathtt{g}_0$ is just the underlying partial (set) function of  $\mathtt{g}$;
	\item $\DSemipart{\mathtt{g}}{a} \coloneqq \DSemi{g}{a} $ if $a\in V$,
	and $\DSemipart{\mathtt{g}}{a} : 0\to \TagentS{a}{M} $ otherwise.
\end{itemize}

\subsection{CHAD-derivative as a structure-preserving Freyd category morphism}
The association
	\begin{equation} \label{eq:derivatives-partial-CHAD}
		\DSempartial{ }:  \PVMAN \to \PFam{\Vect ^\op}, \qquad  	 M \mapsto \left( M, \Tagen{M}   \right) , \qquad \mathtt{g} \mapsto 	\DSempartial{\mathtt{g}}
	\end{equation}
defines a functor. Indeed,  while identity is clearly preserved, we have that
$\DSempartial{\mathtt{h}}\circ \DSempartial{\mathtt{g}} = \DSempartial{\left( \mathtt{h}\circ\mathtt{g}\right)}$
for any pair of composable morphisms $\mathtt{h} : N\to K , \mathtt{g}: M\to N$ in $\PVMAN$,
since
\begin{itemize}
\item  $\mathtt{h}\circ\mathtt{g}$ is defined at $a\in M $ if and only if $\DSempartial{\mathtt{h}}\circ \DSempartial{\mathtt{g}}$ is defined at $a\in M $ if and only if $\DSempartial{\left( \mathtt{h}\circ\mathtt{g}\right)}$
is defined at $a\in M $;
\item denoting $\DSempartial{\mathtt{h}}\circ \DSempartial{\mathtt{g}} = \left( s, s' \right)$, we have that
\begin{itemize}
	\item by the chain rule, if $\mathtt{h}\circ\mathtt{g}$ is defined at $a\in M$,
$s'_a = \DSemipart{\mathtt{h}\circ\mathtt{g}}{a}  $;
\item if $\mathtt{h}\circ\mathtt{g}$ is undefined at $a\in M$, $s'_a : 0\to \TagentS{a}{M} $.
\end{itemize}
\end{itemize}
Moreover, by the above, it is straightforward to conclude that  $\DSempartial{ }$ is an $\wCpo$-enriched functor,
since it preserves the order, the colimits of $\ww$-chains, and the bottom morphisms. Indeed, on an open domain where a member of a chain is defined, its derivative agrees with that of every subsequent member and of their union, because differentiation is local. Therefore:
\begin{theorem}
\eqref{eq:derivatives-partial-CHAD}	is an $\wCpo$-enriched functor.
\end{theorem}

Finally, since $\DSempartial{ }$  is an extension of $\DSemtotal{}$, the square \eqref{diag:chad-derivative} is commutative, making the pair
\begin{equation} \label{eq:Partial-CHAD}
\DSemCHAD{}\coloneqq\left( \DSemtotal{},  \DSempartial{ }\right) : \FFMAN\to \FFFVECT
\end{equation}
  a Freyd category morphism.
\begin{equation} \label{diag:chad-derivative}
	\diag{freyd-category-morphism-CHAD}
\end{equation}

\begin{theorem}\label{theorem:SEMANTICAL-ITERATIVE-CHAD}
	The CHAD-derivative $\DSemCHAD{} =\left( \DSemtotal{},  \DSempartial{}\right) $
	defines an $\ww$-Freyd category morphism.
\end{theorem}

\subsection{What about iteration?}\label{susect:finally-iteration}
Raw \textit{iteration} constructs,\footnote{This is the usual notion of iteration~\cite{bloom1993iteration, DBLP:conf/lics/SimpsonP00}, but free of the usual equational rules, \textit{e.g.} \cite{2022arXiv221007724L}.} as understood here,
allow for the implementation of iterative algorithms with dynamic stopping criteria. A loop body either returns a result in $B$ or produces a new state in $A$ and continues. The variant type $B\sqcup A$ expresses this choice. A computation is undefined if some step is undefined or if no step returns a result.
The precise syntax is given in Section \ref{sec:correctness-iterative-CHAD}, while the usual
denotational semantics in terms of sets is given by the following: for each partially defined function
\begin{equation}\label{eq:fAtoAplusB}
 f : A\to B\sqcup A ,
\end{equation}
we compute the partial function $\iterationn f : A\to B $ which is obtained as follows:
firstly, we consider the function $ \cpairL \coproje{B}, f \cpairR : B\sqcup A\to B\sqcup A $,
where $\coproje{B} : B\to B\sqcup A $ is the coprojection (coproduct inclusion). Secondly, given $a\in A$,
we define $\iterationn f (a) =b $ if there are $b\in B$ and  $n\in\NN $ such that $ \cpairL \coproje{B}, f  \cpairR ^{(n)} \circ \coproje{A} (a) = \coproje{B}(b) $; otherwise $\iterationn f $ is not defined at $a$.
This is well-defined: the map $\cpairL\coproje{B},f\cpairR$ fixes every element in the $B$-summand. Once a result is reached, all later defined iterates therefore give the same result. This defines iteration in the category $\catPSet$ of sets and partially defined functions.

More generally, we start by giving a general definition of a category with iteration, and then
we establish this within the specific context of Freyd categories.

\begin{definition}[Raw iteration]\label{rem:functions-iteration-family-free}
An \textit{iterative category} is a pair $(\mathcal{C}, \iterationn)$, where $\mathcal{C}$ is a category  with finite coproducts and $\iterationn$ is a family of functions between hom-sets, called \textit{context-free iteration}.
\begin{equation} \label{eq:functions-iteration-family-free}
	\left(  \iterationn  ^{\left( A, B\right) }: \catC\left( A, B\sqcup A\right) \to \catC\left( A,  B\right) \right) _{\left( A, B\right)\in \obj{(\catC)}\times \obj{(\catC)} }
\end{equation}
An \textit{iterative category morphism} $\hat{F}: \left( \catC , \iterationn  \right) \to \left( \catC ', \iterationn  ' \right) $ is a  functor
\begin{equation}
	\hat{F}:  \catC  \to  \catC '
\end{equation}
that strictly preserves finite coproducts and iteration, meaning that, for each $\left(  A, B\right)\in\objects{(\catC)}\times \objects{(\catC)} $ and each $g \in \catC\left( A, B\sqcup A\right) $,
\begin{equation} \label{eq:preservation-iteration-omega-Freyd-category}
	\iterationn  ^{\left( \hat{F}(A), \hat{F}(B)\right) } \left( \hat{F}(g)\right) = \hat{F}\left( \iterationn ^{\left( A, B\right) } g\right),
\end{equation}
which is usually simply denoted by
$$\iterationn \left( \hat{F}(g)\right) = \hat{F}(\iterationn g) .$$
\end{definition}
\begin{definition}[Basic fixed point equation]\label{rem:basic-fixed-axiom-iteration}
An \textit{iterative category} $(\mathcal{C}, \iterationn)$ satisfies the \textit{basic fixed point equation} if the following equation is satisfied: for any morphism $ f\in \catC\left( A, B\sqcup A\right) $,
\begin{equation}\label{eq:iteration-with-one-equation}
	\cpairL  \ID _{B}, \iterationn ^{\left(A, B \right) } f\cpairR \circ  f = 	  \iterationn ^{ \left(A, B \right) } f .
\end{equation}
\end{definition}

\begin{remark}
	Iterative categories and iterative category morphisms form, of course, a category. Therefore, we have an induced notion of iterative category isomorphism.
\end{remark}

\begin{remark} \label{rem:basic-fixed-axiom-iteration-wcpo}
While we emphasize the adoption of \textit{raw iteration} as in Definition \ref{rem:functions-iteration-family-free}, it is worth noting that additional equational rules could be imposed. This flexibility arises because the concrete denotational semantics of our languages inherently satisfy the usual equational laws: including specifically the basic fixed point equation, established in \eqref{eq:iteration-with-one-equation}.

In particular, the fact that the $\wCpo$-enriched (concrete) models for dependently typed languages with iteration adhere to the equation presented in \eqref{eq:iteration-with-one-equation} is crucial for establishing Theorem \ref{theo:main-result-iterations-are-equivalent1}, which serves as the foundation for establishing a dependently typed language with iteration and, hence, deriving the syntactic (equational) iteration derivative in our target language. Consequently, readers have the choice to decide whether to require the iteration (in the syntax) to satisfy \eqref{eq:iteration-with-one-equation}.
\end{remark}

\begin{definition}[\iiterativeFn]\label{def:context-free-iterative-Freyd-category}
An \textit{\iterativeF} is a quintuple $\left( \catV , \catC , j, \otimes , \iterationn \right)$
	where $\left( \catV , \catC , j, \otimes \right) $ is a Freyd category and $\left( \catC , \iterationn\right) $ is an iterative category.

Furthermore, a Freyd category morphism $\left( F, \hat{F}\right): \left(\catV , \catC , j , \otimes \right)\to \left(\catV ', \catC ' , j' , \otimes ' \right) $ is an \textit{\iterativeF morphism }
if $\hat{F}: \left( \catC , \iterationn  \right) \to \left( \catC ', \iterationn  ' \right) $ is an
iterative category morphism.

The \iterativeFs and \iterativeF morphisms form a category: the
\iterativeF morphism composition is given by the pointwise composition.
\end{definition}

\begin{definition}[Iteration with context]\label{def:context-free-iteration-context-sensitive-iteration}
Given an \textit{\iterativeF} $\left( \catV , \catC , j, \otimes , \iterationn \right)$,
we define the family of functions
\begin{equation} \label{eq:functions-iteration-family-free-context}
	\left(  \iterationn  ^{\left( A, B\right) }_C: \catC\left(C\otimes A, B\sqcup A\right) \to \catC\left(C\otimes A,  B\right) \right) _{\left( C,  B, A\right)\in \obj{(\catC)}\times\obj{(\catC)}\times \obj{(\catC)} }
\end{equation}
by the following.

For each $\left( C, B, A\right)\in\objects{(\catV)}\times \objects{(\catV)}\times \objects{(\catV)} $ and each $g \in \catC\left( C\otimes A,  B\sqcup A\right) $,
denoting  by
$$\proj{C} : C\times A\to C , \proj{B}: C\times B \to B, \qquad\diagk{C\times A} : C\times A \to \left( C\times A\right)\times \left( C\times A\right)  $$
the respective projection morphisms and diagonal morphism in the cartesian category $\catV$, we set $\pproj{B}=j(\proj{B}):C\otimes B\to B$ and define
\begin{equation} \label{eq:functions-iteration-family-free-context-definition}
	\iterationn  ^{\left( A, B\right) }_C g \defeqq \left( \iterationn  ^{\left( C\otimes A, B\right) } \left( \cpairL  \pproj{B}, \ID _{C\otimes A}    \cpairR\circ \weirdcoherence\circ  \pairfreydL \proj{C}  , g\pairfreydR   \right) \right)
\end{equation}
in which
\begin{equation}
\weirdcoherence  :  C\otimes  \left( B\sqcup A\right)  \xto{\cong}     \left( C\otimes B\right) \sqcup \left( C\otimes A\right)
\end{equation}
is the isomorphism induced by the distribution of the action $\otimes : \catV \times \catC \to \catC $  over coproducts, and
\begin{equation}
	\pairfreydL  \proj{C}  , g\pairfreydR  \defeqq \left(  \proj{C}  \otimes g\right) \circ j\left( \diagk{C\times A}\right).
\end{equation}

We call \eqref{eq:functions-iteration-family-free-context}	 the \textit{context-sensitive iteration of $\left( \catV , \catC , j, \otimes , \iterationn \right)$}. %
\end{definition}

\begin{remark}
	It should be noted that an \textit{\iterativeF morphism } $\left( F, \hat{F}\right): \left(\catV , \catC , j , \otimes \right)\to \left(\catV ', \catC ' , j' , \otimes ' \right) $ preserves context-sensitive iteration; namely,
\begin{equation} \label{eq:preservation-iteration-omega-Freyd-category-contextsensitive}
	\iterationn  ^{\left( \hat{F}(A), \hat{F}(B)\right) }_{\hat{F}(C)} \left( \hat{F}(g)\right) = \hat{F}\left( \iterationn ^{\left( A, B\right) }_C g\right),
\end{equation}
for any triple $\left( A,B,C \right)\in \obj{(\catC)}\times \obj{(\catC)}\times \obj{(\catC)} $.
\end{remark}

Definition \ref{def:context-free-iteration-context-sensitive-iteration} shows how \iterativeFs
provide us with a context-sensitive iteration out of context-free iterations. The body used in \eqref{eq:functions-iteration-family-free-context-definition} has type
$C\otimes A\to B\sqcup(C\otimes A)$. It runs $g$ with the current state and the fixed context. On termination it discards the context; on continuation it pairs the unchanged context with the next state. Thus the construction is the semantic form of carrying the free variables of a loop as part of its state.

\subsection{Concrete models: enriched iteration}\label{susect:enrriched-finally-iteration}

We can recover the setting of our main example by specializing to $\wCpo$-enriched categories
and $\ww$-Freyd categories as follows.
Let $\catC $ be an $\wCpo$-enriched category with finite ($\wCpo$-)coproducts. Given a morphism
$f\in  \catC\left(A, B\sqcup A\right) $, we consider the \textit{colimit}, denoted by $\underline{\mathtt{f}}$, of the following $\ww$-chain
\begin{equation}\label{eq:colimit-iteration-diagram}
	\cpairL \ID _{B}, \bot  \cpairR\circ  \mathbf{g} \leq \cpairL \ID _{B}, \bot  \cpairR\circ \mathbf{g}^2\leq \cdots \leq \cpairL \ID _{B}, \bot  \cpairR\circ \mathbf{g}^n\leq \cdots
\end{equation}
in $\catC\left( B\sqcup A    ,   B  \right)$, where
\begin{equation}
	\mathbf{g}\coloneqq     \cpairL \coproje{B}, f \cpairR : B\sqcup A \to B\sqcup A .
\end{equation}
To see that \eqref{eq:colimit-iteration-diagram} is increasing, put $h=\cpairL\ID_B,\bot\cpairR$. By the coproduct equations,
\[
 h\circ\mathbf{g}=\cpairL\ID_B,h\circ f\cpairR,
 \qquad h=\cpairL\ID_B,\bot\cpairR\leq h\circ\mathbf{g}.
\]
Composing this inequality with $\mathbf{g}^n$ gives
$h\circ\mathbf{g}^n\leq h\circ\mathbf{g}^{n+1}$ for every $n$. The inequality uses the least morphism in the second component of $h$; it does not require $\mathbf{g}\leq\mathbf{g}^2$.

We, finally, define the context-free iteration by
\begin{equation}\label{eq:iteration-omega-chains}
	\iterationn  ^{\left( A, B\right) } f\coloneqq  \underline{\mathtt{f}}\circ\coproje{A} .
\end{equation}

\begin{theorem}\label{theo:wCpoenrichedcategory-underlyingiterative}
	Every $\wCpo$-enriched category $\catC $ with finite $\wCpo$-coproducts has a natural
	underlying iterative category
	$\left( \catC , \iterationn\right)$ once we define $\iterationn $ by \eqref{eq:iteration-omega-chains}. Moreover, the iterative category $\left( \catC , \iterationn\right)$  satisfies the basic fixed equation.

	Furthermore, if  an $\wCpo$-functor $\hat{F}: \catC\to \catC ' $ preserves finite $\wCpo$-coproducts,
	then $\hat{F}$ induces an iterative category morphism $$\left( \catC , \iterationn\right)\to \left( \catC ' , \iterationn '\right) $$ between the underlying iterative categories.
\end{theorem}
\begin{proof}
The preceding argument makes the defining chain well-defined. Write $q=\underline{\mathtt{f}}$ for its least upper bound. Continuity of composition and invariance of the least upper bound under removing a finite prefix give $q\circ\mathbf{g}=q$. Moreover, $q\circ\coproje{B}=\ID_B$. Consequently
$q=\cpairL\ID_B,\iterationn f\cpairR$, and restriction of $q\circ\mathbf{g}=q$ to the $A$-summand gives
$\cpairL\ID_B,\iterationn f\cpairR\circ f=\iterationn f$.

In the conditions above, $\hat{F}$ preserves iteration since it preserves finite $\wCpo$-coproducts, $\bot$ and the colimits of $\ww$-chains of morphisms.
\end{proof}

\begin{definition}[Enriched iteration]\label{def:wcpo-enriched-iteration-concrete}
Let $\catC $ be an $\wCpo$-enriched category with finite $\wCpo$-coproducts.
The iteration $\iterationn$ of the underlying iterative category 	$\left( \catC , \iterationn\right)$, as established in Theorem \ref{theo:wCpoenrichedcategory-underlyingiterative}, is called \textit{$\wCpo$-enriched iteration}.
\end{definition}

\begin{remark}\label{rem:basic-iterations-wcpo-set}
Since the categories $\catPSet$ and $\wCpo$ are $\wCpo $-enriched, we
have natural notions of iteration as defined in Theorem \ref{theo:wCpoenrichedcategory-underlyingiterative}.
Moreover the inclusion
$$\inCC{-} : \catPSet \to\wCpo, $$
which takes each set $A$ to the \wcpo $\inCC{A}$ defined by freely adding a bottom element to the corresponding discrete $\ww$-cpo  $A$ is an $\wCpo $-functor and, hence,
an iterative category morphism.
\end{remark}

\begin{corollary}\label{eq:natural-iteration-induced-by-omega-chains-Freyd-category}
	Every $\ww$-Freyd category $\left(\catV , \catC , j , \otimes \right)$ has an underlying
	\iterativeF\,   $\left( \catV , \catC , j, \otimes , \iterationn \right)$ once we endow it
	with the iterative category $\left( \catC , \iterationn\right) $ defined in Theorem \ref{theo:wCpoenrichedcategory-underlyingiterative}.

	Furthermore, any $\ww$-Freyd category morphism $\left( F, \hat{F}\right): \left(\catV , \catC , j , \otimes \right)\to \left(\catV ', \catC ' , j' , \otimes ' \right) $ is an \iterativeF morphism between the underlying \iterativeFsn{.}
\end{corollary}
\begin{proof}
It follows from Theorem \ref{theo:wCpoenrichedcategory-underlyingiterative}.
\end{proof}

\begin{remark}
Whenever we refer to an $\ww$-Freyd category $\left( \catV , \catC , j, \otimes \right)$, we consider the underlying \iterativeF $\left( \catV , \catC , j, \otimes , \iterationn \right) $ defined above. Furthermore, it should be noted that, as in the case of any \iterativeF, we have the (context-sensitive) iteration as in Definition \ref{def:context-free-iteration-context-sensitive-iteration}.
\end{remark}

\subsubsection{Concrete iterative CHAD}\label{sub:concrete-chad-omega-freyd}
By extending the CHAD-derivative functor to a Freyd category morphism $\DSemCHAD{}$, we have established how the CHAD-derivative interacts with computations involving non-termination and non-differentiability in our setting. More precisely, we have shown that $\DSemCHAD{}$ preserves the domain structure we defined -- specifically, the domain structure characterized by open sets. This preservation demonstrates that we can maintain our structure-preserving principle even in settings where totality does not hold.

Moreover, as $\FFMAN $ and $\FFFVECT$ are $\ww$-Freyd categories by Theorem \ref{theo:ww-Freyd-category-Manifolds} and Theorem \ref{theo:ww-Freyd-category-Fam},
we get that both are \iterativeFs  by Corollary~\ref{eq:natural-iteration-induced-by-omega-chains-Freyd-category}.
The iteration in both \iterativeFs  resembles that in $\catPSet $ described above. More precisely, in the case
of  $\PFam{\Vect ^\op} $, if
$$\left( V , f, f'\right) : \left( A, X\right) \to \left( B\sqcup A, \cpairL Y, X \cpairR\right) $$
is a morphism in $\PFam{\Vect ^\op} $, we have that $\iterationn \left( V , f, f'\right) =
\left( U, g, g'\right) $ where $(U,g) = \iterationn (V,f) $ in $\catPSet $
and, for each $u\in U$, let $n\geq 1$ be the first step at which the body returns a result in $B$. Thus, writing $a_0=u$, there are states $a_0,\ldots,a_{n-1}\in V$ such that
$f(a_m)=\coproje{A}(a_{m+1})$ for $m<n-1$, and $f(a_{n-1})=\coproje{B}(g(u))$. Then
$$g'_u \coloneqq f'_{a_0}\circ f'_{a_1}\circ\cdots\circ f'_{a_{n-1}}:Y(g(u))\to X(u).$$
The forward computation determines the finite sequence of states, and the cotangent is propagated backwards through precisely those steps. For a one-step terminating loop, this is just $f'_u$.

Furthermore, since $\DSemCHAD{} = \left( \DSemtotal{},  \DSempartial{ }\right) $ is an $\ww$-Freyd category morphism by Theorem \ref{theorem:SEMANTICAL-ITERATIVE-CHAD},
we get that the CHAD-derivative functor $\DSempartial{ }$  preserves iteration by Corollary \ref{eq:natural-iteration-induced-by-omega-chains-Freyd-category}.
In other words, we have shown that we can maintain our structure-preserving principle even in iterative setting, with
\begin{equation}
 \DSempartial{\left(\iterationn f\right)  } = \iterationn \DSempartial{\left( f\right)  } .
\end{equation}

\begin{theorem}\label{theo:CHAD-derivative-iterative-Freyd-category-morphism}
	$\DSemCHAD{} = \left( \DSemtotal{},  \DSempartial{ }\right) $  is an \iterativeF morphism.
\end{theorem}

\section{Fibred iteration via parameterized initial algebras}\label{sec:initial-algebras}
To study iteration constructs in a dependently typed language whose semantics supports the iterative category structure $\left( \PFam{\Vect^\op}, \iterationn\right)$ of our concrete model, we first establish an equational counterpart to the iteration construct. This counterpart provides us with a syntactic framework for realizing iteration within $\PFam{\Vect^\op}$ by means of initial algebra semantics.

We begin by demonstrating that the underlying $2$-functor of the suitable indexed category maps the iteration  $\iterationn$ in $\catPSet$ to the parameterized initial algebra operator (or $\mu$-operator) in $\CAT$. By exploiting this preservation property, we derive a purely equational characterization of iteration in $\PFam{\Vect^\op}$, which in turn enables a syntactic implementation of the derivative of functions defined in terms of iteration.

This result shows that whenever the target language supports appropriate (indexed) inductive types (of substitution functors), our framework can directly implement iteration. In other words, once the iteration construct and the non-termination effect are incorporated into the cartesian types of the target language of \cite{lucatellivakar2021chad}, all the necessary requirements for iterative CHAD are met. Indeed, as detailed in Section~\ref{sec:target-language}, our target language imposes even milder requirements.

Beyond solving the practical and theoretical problem of implementing the iterative CHAD, our observations about the target language also yield a \textit{novel contribution} to \textit{Programming Languages Theory}; namely, by introducing the notion of \textit{iteration-extensive indexed category}, we provide a principled (and practical) way of incorporating an iteration construct into a dependently typed language in a coherent manner.

The question can be expressed informally as follows. A loop either returns a result or produces a new state and continues. If the type of the auxiliary data depends on that state, we must also explain how the data is carried through the loop. Parameterized initial algebras provide the corresponding fold: one specifies what to do at a return and how to pass through one further step. We first establish this semantic principle, then show that it agrees with the usual iteration in our concrete models. Finally, iterative indexed categories retain exactly the folder operations that the target syntax needs.

\subsection{Parameterized initial algebras}\label{section:initial-algebra-semantics}

There is an extensive literature on initial algebra semantics (e.g., \cite{zbMATH03241329, zbMATH01850600, initialalgebrasPatricia}). Parameterized initial algebras and their functoriality are treated in \cite{lehmannSmyth1981, freyd1991algebraically} and \cite[Section~6.1]{fiore2004axiomatic}. For our specific context, we also refer the reader to \cite[Section~3]{lucatellivakar2021chad}, which provides an overview of the relevant background. Here, we use this established initial algebra semantics to study the particular functors arising from substitution along an iterative program. The point is to relate iteration in the base category to initial algebras in its fibres, and then to recover iteration in the op-Grothendieck construction.

Herein, we present a concise approach, by making the observations needed to frame our work related to
CHAD. We start by recalling the basic definition of \textit{parameterized initial algebra} in $\catCat $.

\begin{definition}[The category of $E$-algebras]\label{def:category-of-e-algebras}
	Let  $\catD$ be a category and $E : \catD\to \catD$ be an endofunctor.
	The category of $E$-algebras, denoted by $E\AAlg$, is defined as follows. The objects are pairs
	$(W, \zeta ) $ in which $W\in \catD $ and $ \zeta : E(W)\to W $ is a morphism of $\catD $.
	A morphism between $E$-algebras  $(W, \zeta ) $ and $(Y, \xi) $ is a morphism
	$g: W\to Y $ of $\catD $ such that
	\begin{equation}
		\begin{tikzcd}
			E(W) \arrow[rr, "E(  g)"] \arrow[swap,dd, "\zeta"] && E(Y) \arrow[dd, "\xi"] \\
			&&\\
			W \arrow[swap, rr, "g"] && Y
		\end{tikzcd}
	\end{equation}
	commutes. Assuming the existence of the initial $E$-algebra
	$\minAl{E}:E\left( \inAl{E}\right) \to\inAl{E}$ (an initial object in $E\AAlg$), we denote by
\begin{equation}\label{eq:basic-fold-semantics-appendix}
	\foldd{E, \left(Y, \xi\right)}: \inAl{E} \to Y,
\end{equation}
the unique morphism in $\catD $ such that
\begin{equation}
	\begin{tikzcd}
		E\left( \inAl{E}\right) \arrow[rrr, "{E\left(\foldd{E,\left(Y, \xi\right)}\right)}"] \arrow[swap,dd, "\minAl{E}"] &&& E(Y) \arrow[dd, "\xi"] \\
		&&&\\
		\inAl{E} \arrow[swap, rrr, "{\foldd{E,  \left(Y, \xi\right)}}"] &&& Y
	\end{tikzcd}
\end{equation}
commutes.  Whenever it is clear from the context, we denote $ \foldd{E,\left(Y, \xi\right)} $ by $\foldd{E,\xi} $.
\end{definition}

The forgetful functor $U:E\AAlg\to\catD$ creates colimits preserved
by $E$, and hence all absolute colimits. By Beck's monadicity
theorem, it is therefore monadic whenever it has a left
adjoint.\footnote{For a derivation of the classical form of
	Beck's monadicity theorem from Dubuc's adjoint triangle theorem,
	see \cite{dubuc1968adjoint}. More recent monadicity criteria,
	with an explicit discussion of the role of absolute colimits/Kan extensions,
	are developed in
	\cite[Section~2.1]{lucatelliNunes2021descent} and
	\cite[Corollary~5.11 and Remark~5.14]{lucatelliNunes2022semantic}.}

Given a functor
$	H : \catD '\times\catD \to \catD $ and an object
$X $ of $\catD ' $, we denote by $H^X $ the endofunctor
\begin{equation}\label{eq:basic-functor-inductive-types-restricted}
	H^X\defeqq H(X, -): \catD \to \catD .
\end{equation}
On morphisms, $H^X(g)=H(\ID_X,g)$. In this setting, if $\mu H^X$ exists for any object $X\in\catD'$ then their universal properties induce a functor $\inAl{H,\catD '}: \catD' \to \catD$, the \emph{carrier functor of the parameterized initial algebras}. The following proposition spells out its action.

\begin{proposition}[$\mu$-operator]\label{prop:general_parameterized_initial_algebras}
	Let $ H:\catD '\times \catD \to\catD $ be a functor. Assume that, for each object $X\in\catD '$, the functor $H^X = H(X,-) $ is such that
	$ \inAl{H ^X} $ exists. In this setting, the association
	\begin{eqnarray*}
		\inAl{H ,\catD '}:  \catD ' & \to     & \catD\\
		X            & \mapsto & \inAl{H ^X}\\
		\left( f: X\to Y \right) & \mapsto & \foldd{H^X, \minAl{{H^Y}} \circ  H\left(f, \ID_{\inAl{H ^Y}} \right)} .
	\end{eqnarray*}
defines the carrier functor of the parameterized initial algebras.
\end{proposition}
\begin{proof}
	See \cite{lehmannSmyth1981, freyd1991algebraically} and \cite[Section~6.1]{fiore2004axiomatic}; the formulation used here is also recalled in \cite[Section~3]{lucatellivakar2021chad}.
\end{proof}

Motivated by Proposition \ref{prop:general_parameterized_initial_algebras},
given a functor $ H:\catD '\times \catD \to\catD $,
we say that the parameterized initial algebra $\inAl{H ,\catD '}$ exists
if $H^X $ has initial algebra for any $X\in \catD '$.

\subsubsection{Fold for parameterized initial algebras}\label{subsubsection-useful-notation-fold-parameterized-algebras}
It is particularly useful to notice that, given any functor $H: \catD '\times\catD\to \catD $, an object $Y\in\catD '$ and a $H^Y$-algebra $(W, \zeta ) $, we want to consider the morphism
$\foldd{H^Y, (W, \zeta ) }$.
In order to be concise, we introduce the following notation.
\begin{equation}\label{eq:subsubsection-useful-notation-fold-parameterized-algebras}
\foldd{H, Y, \zeta }\defeqq \foldd{H, Y, (W, \zeta ) }\defeqq \foldd{H^Y, (W, \zeta ) } : \inAl{H, \catD'}\left( Y \right)\to W .
\end{equation}

\subsection{Parameterized initial algebras in terms of colimits}

Adámek's initial-chain construction \cite{adamek1974} computes the initial algebra of an endofunctor $E$ from an initial object and colimits of $\ww$-chains, provided that $E$ preserves the relevant colimit. We recall its parameterized form below; see also \cite[Lemma~4.4]{fioreSaville2017}.

\begin{lemma}\label{lem:adamek-computing-parameterized-initial-algebras}
	Let $\catD $ be a category with initial object $\initiall $ and colimits of $\ww$-chains. If $ H:\catD '\times \catD \to\catD $ is a functor such that, for any $X\in \catD '$,  $H\left( X, - \right) $ preserves colimits of $\ww $-chains, then	$\inAl{H ,\catD '}$ exists.

Furthermore, fixing $G\defeq \pairL \proj{\catD '}, H  \pairR : \catD'\times\catD\to\catD'\times\catD$, where $\proj{\catD'}$ is the first projection in $\CAT$,
denoting by $\bot : \catD ' \to \catD $ the functor constantly equal to $\initiall, $
and $\underline{\iota } $ the unique natural transformation whose first component is the identity,
$$\underline{\iota }: \pairL \ID_{\catD '}, \bot \pairR \to   	G\circ \pairL \ID_{\catD '}, \bot \pairR ,$$
with $\underline{\iota}_X=(\ID_X,i_X)$ for the unique map $i_X:\initiall\to H^X(\initiall)$, we can conclude that, if $\underline{H}$
 is the colimit of the $\ww$-chain of functors
\begin{equation}\label{eq:construction-parameterized-Adamek}
 \pairL \ID_{\catD '}, \bot \pairR \xto{\underline{\iota } }   	G\circ \pairL \ID_{\catD '}, \bot \pairR  \xto{G\left(\underline{\iota }\right) }  G ^2\circ \pairL \ID_{\catD '}, \bot \pairR \xto{G^2\left(\underline{\iota }\right) }  \cdots  \xto{G^{n-1}\left(\underline{\iota }\right) } G ^n\circ \pairL \ID_{\catD '}, \bot \pairR \xto{G^n\left(\underline{\iota }\right) } \cdots
\end{equation}
in the functor category $\catCat [\catD',\catD'\times\catD]$, then
\begin{equation}
	\inAl{H ,\catD '} \cong \pi_{\catD}\circ \underline{H}.
\end{equation}
\end{lemma}
\begin{proof}
Pointwise, for each $X\in \catD '$, we have that, indeed, $(\proj{\catD } \circ \underline{H})(X) $ is the colimit of the diagram
	\begin{equation}\label{eq:construction-parameterized-Adamek0}
		\initiall \xto{i_X }   	H^X (\initiall ) \xto{H^X\left(i_X\right) }  \left( H^X\right)^2  \left( \initiall \right) \xto{ \left( H^X\right)^2 \left(i_X\right) }  \cdots  \xto{ \left( H^X\right)^{n-1} \left(i_X\right) } \left( H^X\right)^n\left( \initiall \right) \xto{\left( H^X\right)^n\left(i_X\right) } \cdots
	\end{equation}
which is $\inAl{H^X}$ by Adámek's theorem \cite{adamek1974}. Naturality in $X$ follows from the universal property of the pointwise colimit, and agrees with Proposition~\ref{prop:general_parameterized_initial_algebras}.
\end{proof}

\subsection{Initial algebras of substitution functors}
A familiar use of (parameterized) initial algebras concerns functors built from standard type formers such as products, coproducts, and exponentials, yielding semantics for various inductive types. However, in the presence of dependent types, we gain a new class of syntactically definable functors arising from term substitution in types, often called \emph{change-of-base functors}.

To begin, observe that any endomorphism \(f: A \to A\) in \(\catC\) induces a functor
\[
\catL(f) : \catL(A) \to \catL(A).
\]
Since substitution preserves the chosen initial objects on the nose, its initial chain is constant at the initial object. More directly, we have the following.

\begin{lemma}\label{lem:trivial-initial-algebras-substitution-functors}
	Let \(\catL : \catC^{op} \to \catCat\) be an indexed category equipped with indexed initial objects.
	Then for any \(f : A \to A\) in \(\catC\), the initial \(\catL(f)\)-algebra is given by
	\[
	(\initiall, \id : \initiall \to \initiall).
	\]
\end{lemma}

On the other hand, when working in dependent type theory with sum types, we typically model it by an indexed category
\(\catL : \catC ^\op \to \CAT\)
that satisfies the extensivity property with respect to coproducts (a notion and perspective introduced in \cite[Definition~31]{lucatellivakar2021chad}). In this setting, a broader class of substitution functors arises. Specifically, each morphism \(f : A \to B \sqcup A\) induces a functor
\begin{equation}\label{eq:parameterized-initial-algebra}
	\catL(B) \times \catL(A)
	\xrightarrow{\cong}
	\catL(B \sqcup A)
	\xrightarrow{\catL(f)}
	\catL(A),
\end{equation}
whose parameterized initial algebras
\[
\catL(B) \;\longrightarrow\; \catL(A)
\]
can be non-trivial and play a central role in understanding iteration in the dependently typed setting.

Motivated by the concrete example given in~\ref{subsect:subst-concrete-Pset}, we analyze in more detail these parameterized initial algebras for the substitution functors of the form~\eqref{eq:parameterized-initial-algebra}, and their foundational connection with iteration, in Section~\ref{subsection:iteration-extensive-categories}.

\subsection{Parameterized initial algebras of substitution functors}\label{subsect:subst-concrete-Pset}
To lay the foundation for our observations, we again start by revisiting the elementary denotational semantics of iteration in terms of sets and partially defined functions between them, as in Section \ref{susect:finally-iteration}. The category $\catPSet$ is naturally endowed with a 2-categorical structure -- that is, the structure coming from the $\wCpo$-enrichment given by the domain order between functions.
As a consequence, for any partially defined function  $f : A\to  B\sqcup A$ and any set $K$, we have an induced functor
\begin{equation}
	\underline{\catPSet\left( f,K\right) } \defeqq \catPSet\left( f ,  K\right)\circ \cISO : \catPSet\left( B ,  K\right)\times \catPSet\left( A ,  K\right)\to \catPSet\left( A ,  K\right)
\end{equation}
where $\cISO$ is the isomorphism
\begin{equation}
\catPSet\left( B ,  K\right)\times \catPSet\left( A ,  K\right)\cong \catPSet\left( B\sqcup A ,  K\right), \qquad \left( t, r\right) \mapsto \cpairL t,r  \cpairR
\end{equation}
induced by the universal property of the coproduct $B\sqcup A$.
We claim that
\begin{equation}\label{eq:PSet-is-extensive-iteration}
  \catPSet\left(\iterationn f,  K\right) =  \inAl{\underline{\catPSet\left( f,K\right) }, \catPSet\left( B,  K\right)}
\end{equation}
where $\iterationn f$ is the function defined in Subsection~\ref{susect:finally-iteration}.

The Equation \eqref{eq:PSet-is-extensive-iteration} makes the indexed category
\begin{equation}\label{eq:PSET-Indexed-category}
\catPSet\left( - ,  K\right) : 	\catPSet ^\op \to \CAT
\end{equation}
into an \textit{iteration-extensive indexed category}, as introduced in \ref{subsection:iteration-extensive-categories} below.

\subsection{Revisiting extensivity}
In Section \ref{subsection:iteration-extensive-categories}, we introduce a principled notion of iteration on the op-Grothendieck construction for a model of our target language. To do so, we first require that this op-Grothendieck construction admits finite coproducts. Below, we recall how finite-coproduct-extensive indexed categories help us model sum types in our setting -- namely, how they ensure the existence of finite coproducts in the op-Grothendieck construction.

In \cite[Definition~31]{lucatellivakar2021chad}, we introduced the notion of a
\emph{binary-coproduct-extensive indexed category}
\[
\mathcal{L} : \catC^\op \to \CAT,
\]
which was referred to simply as an \emph{extensive indexed category} in that work. Motivated by this concept, we further developed the notion of extensive indexed categories with respect to a class of diagrams in  \cite[Section~6.7]{LV24b}; we recall this definition below.

\begin{definition}[Extensive indexed categories]\label{def:extensive-S}
	Let $\mathfrak{S}$ be a class of small categories, regarded as diagram shapes. We say that an indexed category $\catL : \catC ^\op \to \CAT $ is $\mathfrak{S}$-colimit-extensive if $\catC ^\op$ has, and $\catL $ preserves, the
	limits of diagrams $S\to \catC ^\op $ for $S\in\mathfrak{S}$.
	For instance:
	\begin{itemize}
		\item we say that the indexed category $\catL : \catC ^\op \to \CAT $ is \textit{binary-coproduct-extensive} if $\catC$ has binary coproducts and $\catL $ preserves binary products of $\catC^\op $;
		\item we say that the indexed category $\catL : \catC ^\op \to \CAT $ is \textit{finite-coproduct-extensive} if $\catC$ has finite coproducts and $\catL $ preserves finite products of $\catC^\op $.
	\end{itemize}
	\textit{Unlike in \cite{lucatellivakar2021chad}, throughout this work, we use the term ``extensive indexed category'' to mean what is more precisely a ``finite-coproduct-extensive indexed category'' as defined above.}
\end{definition}

Clearly, every finite-coproduct-extensive indexed category is a binary-coproduct-extensive indexed category. Moreover, we recall that, by \cite[Lemma~33]{lucatellivakar2021chad}, we have the following result:

\begin{lemma}\label{lem:it-was-already-proven-lemma}
	Let $\catL : \catC ^\op\to \CAT $ be an indexed category such that $\catC $ has finite coproducts. We have that $\catL $ is binary-coproduct-extensive and $\catL \left( W\right) $ is non-empty for some $W\in \catC $ if, and only if, $\catL $ is a finite-coproduct-extensive indexed category.
\end{lemma}
\begin{proof}
	By \cite[Lemma~33]{lucatellivakar2021chad}, if $\catL $ is a  binary-coproduct-extensive indexed category such that $\catL \left( W\right) $ is non-empty for some $W\in \catC $, then we get that $\catL $ preserves the terminal object of $\catC ^\op $; that is to say, $\catL\left( \initiall \right) \cong \terminall $. Therefore $\catL $ is finite-coproduct-extensive.

	Reciprocally, if $\catL $ is a finite-coproduct-extensive indexed category, then
	it is  binary-coproduct-extensive, and $\catL\left(\initiall\right)\cong \terminall $ is the terminal category and, hence, non-empty.
\end{proof}

\begin{remark}
In other words, Lemma \ref{lem:it-was-already-proven-lemma} states that, given any category $\catC $ with finite coproducts, the only binary-coproduct-extensive indexed category that is not finite-coproduct-extensive indexed is the indexed category $\emptyset : \catC ^\op \to \CAT $ constantly equal to the
empty category $\emptyset $.
\end{remark}

The notion of a \(\Sigma\)-bimodel for sum types in \cite[Definition~37]{lucatellivakar2021chad} also requires initial and terminal objects in every fibre. We recall it to distinguish this additional requirement from finite-coproduct extensivity.

\begin{definition}
	Let $\catL : \catC ^\op\to\CAT $ be a binary-coproduct-extensive indexed category.
	We say that $\catL $ is a \(\Sigma\)-bimodel for sum types if $\catC $ has initial
	object $\initiall $ and every fibre $\catL(W)$ has terminal and initial objects.
\end{definition}

\begin{lemma}\label{lemma49}
	Let $\catL : \catC ^\op\to \CAT $ be an indexed category.
	$\catL $ is a \(\Sigma\)-bimodel for sum types if, and only if, it is finite-coproduct-extensive and every fibre has initial and terminal objects.
\end{lemma}
\begin{proof}
	If $\catL $ is a \(\Sigma\)-bimodel for sum types, we get that $\catC$ has finite coproducts and $\catL \left( \initiall \right) $ is not empty (since it has a terminal object). Therefore, $\catL $ is a finite-coproduct-extensive indexed category by  Lemma \ref{lem:it-was-already-proven-lemma} .

	Conversely, finite-coproduct extensivity implies binary-coproduct extensivity and the existence of an initial object in the base. Together with the assumed initial and terminal objects in every fibre, this is precisely the definition of a \(\Sigma\)-bimodel for sum types.
\end{proof}

Finite-coproduct extensivity alone does not ensure these fibrewise objects: if $K$ is the discrete category on two objects, then $K^{(-)}:\Set^\op\to\CAT$ is finite-coproduct-extensive, but its fibre at a singleton is $K$, which has neither an initial nor a terminal object.

The appeal of these extensive notions lies in the properties they induce in the (op-)Grothendieck constructions, as demonstrated in our previous work \cite{lucatellivakar2021chad, LV24b}. To keep this discussion brief herein, we highlight only the following result:

\begin{theorem}\label{theo:L-finite-coproduct-extensive-indexed-category}
	Let $\catL :  \catC ^\op\to \CAT $ be a finite-coproduct-extensive indexed category. In this case, $\displaystyle\Sigma_{\catC } \catL ^\op$ has finite coproducts. More precisely:
	\begin{itemize}
	\item the initial object is given by $\left( \initiall , \terminall \right) $, where $\terminall $ is the only object of $\catL\left( \initiall \right) \cong \terminall $;
	\item the binary coproduct is given by
	$$ \left( B, Y\right) \sqcup \left( A, X\right)\cong \left( B\sqcup A, \cISO ^{\left( B, A\right) }\left( Y, X\right) \right) $$
	where $\cISO ^{\left( B, A\right) } :\catL\left( B\right) \times  \catL\left(A\right)  \xto{\cong} \catL\left( B\sqcup A\right) $ is the inverse of the comparison morphism.
	\end{itemize}
\end{theorem}
\begin{proof}
The initial object follows from $\catL(\initiall)\cong\terminall$. A map from $(B\sqcup A,\cISO^{(B,A)}(Y,X))$ to $(C,Z)$ consists of a base map $h=[h_B,h_A]$ and a fibre map $\catL(h)(Z)\to\cISO^{(B,A)}(Y,X)$. Under the extensive comparison, this fibre map is exactly a pair
\[
\catL(h_B)(Z)\to Y,\qquad \catL(h_A)(Z)\to X.
\]
These are precisely the data of maps from $(B,Y)$ and $(A,X)$ to $(C,Z)$, proving the coproduct universal property.
\end{proof}

We now introduce iteration into the op-Grothendieck construction of an extensive indexed category. As we explain in detail below, by strengthening our notion of extensivity, we can get this construction naturally, and ensure it is fibred. Moreover, under suitable hypotheses, it is the unique fibred iteration satisfying the fixed point equation.

\subsection{Iteration-extensive indexed categories}\label{subsection:iteration-extensive-categories}
We introduce below the notion of \textit{iteration-extensive indexed category} (Definition~\ref{eq:iteration-extensive-indexed-categories}), one of the novel contributions of the present paper.

This concept aims to provide the basic principle to answer the following general question: ``how can one incorporate
iteration in a dependently typed language in a principled, practical and coherent manner?''
We focus, however, on the question at hand, which has to do with implementing the CHAD derivative for programs involving iteration.

We show that the concrete semantics of the target language satisfies the property we call \emph{iteration extensivity}. This property allows us to define a fibred iteration  in
$\PFam{\Vect ^\op}$   purely in equational terms through its corresponding indexed category $\Vect_{\bot}^{\left( - \right)}: \catPSet ^\op \to \CAT $ (as defined in Eq.~\ref{def:Vectpartial}), without relying on the $\wCpo$-structure. This will give us a basic principle to define the syntax of our target language, and the implementation of our program transformation.

Motivated by Eq.~\eqref{eq:PSet-is-extensive-iteration} and the notions of extensive indexed categories, we introduce the following
notion -- which is our proposed guiding principle to introduce iteration in dependently typed languages.

\begin{definition}[Iteration-extensivity]\label{eq:iteration-extensive-indexed-categories}
An \textit{iteration-extensive indexed category} is a triple  $\left( \catC , \iterationn , \catL \right) $ where $\left( \catC , \iterationn \right) $
is an iterative category, $\catL:\catC^{op}\to\CAT$ is an extensive  indexed category, and, for any
morphism $f\in \catC\left( A , B\sqcup A \right) $,
\begin{equation}
\catL \left( \iterationn ^{(A,B)} f \right) \cong	\inAl{\catL ( f  )\circ \cISO ^{\left( B, A\right) }, \catL (B)  }
\end{equation}
where   $\cISO ^{\left( B, A\right) } :\catL\left( B\right) \times  \catL\left(A\right)  \xto{\cong} \catL\left( B\sqcup A\right) $ is the inverse of the comparison morphism.

When the iterative category
$ (\catC,\iterationn) $ is understood from the context, we omit it from the notation and simply say that the indexed category
$\catL : \catC ^\op \to \CAT $
is \emph{iteration-extensive} (or \emph{extensive for iteration}).
\end{definition}

The isomorphism in this definition is an isomorphism of functors $\catL(B)\to\catL(A)$. Its existence is the property of iteration-extensivity; no particular comparison is required here. When constructing a folder, we choose such an isomorphism and transport the initial algebra along it. We make this transport explicit below, since an isomorphism between carriers should be distinguished from compatibility with their algebra structures.

As mentioned above, for any object $K\in\catPSet$, Eq.~\eqref{eq:PSet-is-extensive-iteration}
shows that
$\catPSet\left( - ,  K\right) : \catPSet ^\op \to\CAT  $
is iteration-extensive w.r.t. the $\wCpo$-enriched iteration in $\catPSet$. More generally, by abuse of notation, given any $\wCpo$-category $\catC$, we can consider the indexed category
\begin{equation}
	\catC \left( -, K \right) : \catC ^\op \to \CAT
\end{equation}
where $\catC \left( A, K \right) $ is the $\ww$-complete partially ordered set of
morphisms $A\to K $ in $\catC $. In other words, $\catC \left( -, K \right)$
is actually obtained from composing the arrows
\begin{equation}
\catC ^\op \xto{\catC \left[ -, K \right]} \wCpo \to \CAT,
\end{equation}
where $\catC \left[ -, K \right] : \catC ^\op\to \wCpo $ is the $\wCpo$-enriched
representable functor, and $\wCpo \to \CAT$ is the inclusion.

\begin{theorem}[Iteration as a parameterized initial algebra] \label{theo:main-proof-iteration-initial-algebra-colimit}
	Let $\catC$ be an $\wCpo$-category with finite $\wCpo$-coproducts. We consider the underlying iterative category $\left( \catC, \iterationn\right) $ as defined
in Theorem \ref{theo:wCpoenrichedcategory-underlyingiterative}.
For any object $K$ of $\catC $, the triple $\left( \catC , \iterationn , \catC\left( -, K\right)  \right) $
is an iteration-extensive indexed category.

Moreover, for each pair $(A,B)\in \obj(\catC)\times \obj(\catC)$ and each morphism $f\in  \catC\left( A, B\sqcup A\right) $,
\begin{equation}
\iterationn ^{(A,B)} f : A\to B
\end{equation}
is the \textbf{only} morphism such that
\begin{equation}
\catC ( \iterationn ^{(A,B)} f , K  ) = 	\inAl{\catC ( f , K  )\circ \cISO^{\left( B, A\right) } , \catC ( B , K  )  }
\end{equation}
 for any object $K$ in $\catC $.
\end{theorem}
\begin{proof}
We give the types explicitly. Fix $A,B,K\in\catC$ and $f:A\to B\sqcup A$, and write
\[
\catD'=\catC(B,K),\qquad \catD=\catC(A,K),\qquad
H:\catD'\times\catD\to\catD,\quad H(t,r)=\cpairL t,r\cpairR\circ f.
\]
Here the homs are their ordered categories, and $H$ is a functor because composition and copairing are monotone. Its parameterized initial algebra therefore refers to initial algebras in these ordered categories, rather than in discrete hom-sets.

Let $g=\cpairL\coproje{B},f\cpairR:B\sqcup A\to B\sqcup A$ and $p=\cpairL\ID_B,\bot\cpairR:B\sqcup A\to B$. The chain defining iteration is $p\leq p\circ g\leq p\circ g^2\leq\cdots$; adjoining its first term does not change the colimit in Eq.~\eqref{eq:colimit-iteration-diagram}. Under the coproduct comparison, precomposition by $g$ is the functor
\[
G=\pairL\proj{\catD'},H\pairR:\catD'\times\catD\to\catD'\times\catD,
\]
and precomposition by $p$ is $J=\pairL\ID_{\catD'},\bot\pairR:\catD'\to\catD'\times\catD$. More precisely, the locally continuous representable induces the functor
\[
Q:\catC(B\sqcup A,B)\to\catCat(\catD',\catD'\times\catD),\qquad
Q(r)(t)=\bigl(t\circ r\circ\coproje{B},t\circ r\circ\coproje{A}\bigr).
\]
It preserves colimits of increasing $\ww$-chains, since these are computed pointwise and composition is continuous. Moreover, $Q(p)=J$ and $Q(r\circ g)=G\circ Q(r)$. Thus it takes the chain defining iteration to
\begin{equation}\label{eq:construction-parameterized-Adamek22}
J\xto{\underline{\iota}}G\circ J\xto{G(\underline{\iota})}G^2\circ J\longrightarrow\cdots .
\end{equation}
Writing $\underline{\mathtt f}=\colim_n p\circ g^n$, we obtain
\[
\catC(\iterationn f,K)=\proj{\catD}\circ Q(\underline{\mathtt f})
=\proj{\catD}\circ\colim_n G^n\circ J
=\inAl{H,\catD'}.
\]
The last equality follows from Lemma~\ref{lem:adamek-computing-parameterized-initial-algebras}: each $H(t,-)$ preserves increasing $\ww$-chain colimits. Since the categories involved are posets, the canonical isomorphisms are equalities. Finally, any other morphism with the stated property induces the same maps $\catC(B,K)\to\catC(A,K)$ for every $K$. Taking $K=B$ and evaluating at $\ID_B$ proves uniqueness, equivalently by the Yoneda Lemma.
\end{proof}
For $\catPSet$, no $\ww$-chain colimits are needed in the fibre category. Fix the iteration of Remark~\ref{rem:basic-iterations-wcpo-set}. For any category $K$ with a chosen initial object $0$, define
\begin{equation}
K_{\leastelement}^{(-)}:\catPSet^\op\to\CAT
\end{equation}
by $K_{\leastelement}^A=K^A$ and substitution along a partial function $u:A\to B$:
\[
\bigl(K_{\leastelement}^uY\bigr)(a)=
\begin{cases}
Y(u(a)),&u(a)\text{ is defined},\\
0,&u(a)\text{ is undefined}.
\end{cases}
\]
On family morphisms, reindexing selects the corresponding component where $u$ is defined and $\ID_0$ otherwise. Write $\cISO^{(B,A)}:K^B\times K^A\cong K^{B\sqcup A}$ for the inverse of the coproduct comparison.

\begin{theorem}\label{theo:main-theorem-consequence-for-Vect}
For every category $K$ with a chosen initial object, the triple
\begin{equation}
\left(\catPSet,\iterationn,K_{\leastelement}^{(-)}\right)
\end{equation}
is an iteration-extensive indexed category. For every partial function $f:A\to B\sqcup A$, the iteration
\begin{equation}
\iterationn^{(A,B)}f:A\to B
\end{equation}
is the unique partial function $u:A\to B$ such that, for every such $K$, there is a natural isomorphism of functors $K^B\to K^A$,
\begin{equation}
K_{\leastelement}^{u}\cong
\inAl{K_{\leastelement}^{f}\circ\cISO^{(B,A)},K^B}.
\end{equation}
\end{theorem}
\begin{proof}
The usual comparison $K^{B\sqcup A}\cong K^B\times K^A$, together with $K^\varnothing\cong\terminall$, shows that $K_{\leastelement}^{(-)}$ is extensive. Fix $f:A\to B\sqcup A$ and $Y\in K^B$, and write
\[
H_Y(X)=K_{\leastelement}^{f}\bigl(\cISO(Y,X)\bigr).
\]
Pointwise,
\[
H_Y(X)(a)=
\begin{cases}
0,&f(a)\text{ is undefined},\\
Y(b),&f(a)=\coproje{B}(b),\\
X(a'),&f(a)=\coproje{A}(a').
\end{cases}
\]
Let $I_Y=K_{\leastelement}^{\iterationn f}Y$. Thus $I_Y(a)=Y(b)$ when iteration from $a$ reaches $b\in B$ after finitely many steps, and $I_Y(a)=0$ otherwise. The evident componentwise identities give an algebra $H_Y(I_Y)\to I_Y$.

To see that it is initial, let $\alpha:H_Y(X)\to X$ be any algebra. If iteration from $a=a_0$ terminates through
\[
f(a_i)=\coproje{A}(a_{i+1})\quad(0\leq i<n),\qquad
f(a_n)=\coproje{B}(b),
\]
then the component of the unique algebra morphism $I_Y\to X$ at $a$ is
\[
Y(b)\xto{\alpha_{a_n}}X(a_n)
\xto{\alpha_{a_{n-1}}}\cdots
\xto{\alpha_{a_0}}X(a).
\]
If iteration from $a$ does not terminate, it is the unique map $0\to X(a)$. These components form an algebra morphism, and the algebra equations force exactly these components, so it is unique. The construction is natural in $Y$. Hence the carrier of the parameterized initial algebra is precisely $K_{\leastelement}^{\iterationn f}$, proving iteration-extensivity.

For the uniqueness of the partial function, suppose $u:A\to B$ has the displayed property for every such $K$. Take $K=\Set$ and, for each $b\in B$, the family $Y^b$ with $Y^b(b)=\terminall$ and $Y^b(b')=\varnothing$ for $b'\neq b$. Then $\bigl(K_{\leastelement}^{u}Y^b\bigr)(a)$ is a singleton exactly when $u(a)=b$, and is empty otherwise. The same characterization holds for $\iterationn f$, so the asserted natural isomorphism forces $u=\iterationn f$.
\end{proof}

\noindent\textbf{Specialization to vector spaces.}
Taking $K=\Vect$, we obtain the iteration-extensive indexed category
\begin{equation}
\left(\catPSet,\iterationn,
\Vect_{\leastelement}^{(-)}=\catCatb(\inCC{-},\Vect):\catPSet^\op\to\CAT\right).
\end{equation}
In particular,
\begin{equation}
\begin{aligned}
\Vect_{\leastelement}^{\left(\iterationn^{(A,B)}f\right)}
&\cong\inAl{\Vect_{\leastelement}^{f}\circ\cISO^{(B,A)},\Vect_{\leastelement}^B}\\
&=\inAl{\lhi{f},\Vect_{\leastelement}^{B}}.
\end{aligned}
\end{equation}
Here $\lhi{f}=\Vect_{\leastelement}^{f}\circ\cISO^{(B,A)}:
\Vect_{\leastelement}^B\times\Vect_{\leastelement}^A\to\Vect_{\leastelement}^A$ acts by
\begin{equation}\label{eq:fundamental-initial-algebra-iteration}
(Y,X)\longmapsto\cpairL Y,X\cpairR\circ\inCC{f}
\end{equation}
and
\begin{equation}
(\alpha:Y\to Y',\beta:X\to X')\longmapsto
\cpairL\alpha,\beta\cpairR\circ\inCC{f}:
\cpairL Y,X\cpairR\circ\inCC{f}\to\cpairL Y',X'\cpairR\circ\inCC{f}.
\end{equation}

\begin{remark}
	Colimits of $\ww$-chains give a natural iteration in $\catCatb$, up to natural isomorphism. For $K\in\catCatb$, the special case of Theorem~\ref{theo:main-theorem-consequence-for-Vect} can also be recovered from the Yoneda Lemma and the fact that the full inclusions $\catPSet\xto{\inCC{-}}\wCpo\to\catCatb$ preserve coproducts and colimits of $\ww$-chains of morphisms.
\end{remark}

\subsubsection{Iteration-extensive Freyd indexed categories}
We can finally define the concept of iteration-extensive Freyd indexed categories. We can think of them as models of languages with dependent types that depend on programs that may use iteration, with call-by-value semantics. The functor $j:\catV\to\catC$ separates values from computations, while $\catL:\catC^\op\to\CAT$ supplies the types and morphisms in each computational context. Thus the extra initial algebra structure concerns substitution along computations, including computations containing loops.

There is one further compatibility needed for the Freyd construction. Carrying a value through an effectful computation also carries its dependent data through that computation. To make this explicit, write the value projections in $\catC$ as $j(\proj A)$, and consider natural comparisons
\begin{equation}\label{eq:indexed-context-comparison}
\delta_{A,g}:\catL(\ID_A\otimes g)\circ\catL(j(\proj A^{A\times D}))
\Longrightarrow\catL(j(\proj A^{A\times C})),\qquad g:C\to D.
\end{equation}
A \emph{compatible action on dependent data} consists of such comparisons for which the following formula defines an action
\[
\hat\otimes:\Sigma_{\catV}(\catL\circ j^\op)^\op\times\Sigma_{\catC}\catL^\op
\longrightarrow\Sigma_{\catC}\catL^\op:
\]
\begin{equation}\label{eq:indexed-context-action}
\begin{split}
(f,f')\hat\otimes(g,g')
&=\bigl(f\otimes g,\\
&\quad(\catL(j(\proj A))(f')\circ\delta_{A,g,\catL(j(f))(Y)})
\sqcup\catL(j(\proj C))(g')\bigr),
\end{split}
\end{equation}
for $(f,f'):(A,X)\to(B,Y)$ in the value construction and $(g,g'):(C,Z)\to(D,W)$ in the computation construction. On objects the action is
\[
(A,X)\hat\otimes(C,Z)=
\bigl(A\otimes C,\catL(j(\proj A))(X)\sqcup\catL(j(\proj C))(Z)\bigr).
\]
Here we use the indexed coproduct comparisons and the second-projection identity for a Freyd action to identify the source of the second component. We require this action to preserve finite coproducts in its second argument, to extend the product on the value construction, and to have the unit and associativity isomorphisms induced by that product and the base action. Thus requiring a compatible action includes functoriality and these unit and associativity coherence conditions; equivalently, it specifies a lifting of the distributive Freyd action with the displayed object and morphism formulas. For pure $g$, $\delta_{A,g}$ is the identity after the projection identifications, so Eq.~\eqref{eq:indexed-context-action} reduces to the familiar coproduct $\catL(j(\proj A))(f')\sqcup\catL(j(\proj C))(g')$ of reindexed fibre maps. Morphisms of indexed Freyd structures preserve this action as well as their other displayed structure.

This compatibility cannot in general be replaced by an equality of the two functors in Eq.~\eqref{eq:indexed-context-comparison}: an effectful computation may fail to return even when the carried value does not depend on its result. In our partial-map models the comparison is simply restriction to the domain where the computation returns. We include the compatible action in the indexed Freyd structures below, suppressing its notation in their tuples.

\begin{definition}[Iteration-extensive Freyd indexed category]\label{def:Freyd-iteration=extensive=Freyd-category}
	An iteration-extensive Freyd indexed category consists of a compatible action on dependent data and a tuple	$\left( \catV , \catC , j,   \otimes , \iterationn , \catL \right)$ such that $\left( \catC , \iterationn , \catL \right)$  is an iteration-extensive indexed category in which $\catL $ has indexed finite coproducts, and $\left( \catV , \catC , j,   \otimes \right) $ is a distributive Freyd category.
\end{definition}

As a consequence of Lemma \ref{lem:op-Grothendieck-construction-of-iteration-extensive-categry} below, we see that iteration-extensive Freyd indexed categories have underlying iterative Freyd indexed categories. We postpone a detailed discussion of this fact to Sec.~\ref{subsection:iterative-freyd-categories-are-everywhere}.

\subsection{The fibred iteration induced by an iteration-extensive indexed category}\label{subsect:Grothendieck-iteration-extensive-indexed-category}

An \emph{iteration-extensive indexed category} induces an iteration on its (op-)Grothendieck construction: specifically, the op-Grothendieck construction of such an indexed category naturally becomes an iterative category when equipped with the \emph{container iteration} as established in Def.~\ref{def:op-Grothendieck-construction-of-iteration-extensive-categry} and Lemma~\ref{lem:op-Grothendieck-construction-of-iteration-extensive-categry} below.

In what follows, we show that the universal property of the initial algebra determines a unique fibred iteration satisfying its algebra equation. When this equation is the basic fixed point equation (see Definition~\ref{rem:basic-fixed-axiom-iteration}), it gives the corresponding uniqueness of fibred iteration.

More precisely, the base fixed point equation supplies a canonical identity algebra on the substituted result type. If this algebra is initial, its folder yields the unique fibred iteration satisfying the basic fixed point equation. Theorem~\ref{theo:iteration-extensive-indexed-category} makes the role of this algebra structure precise, and Theorem~\ref{theo:class-of-examples-of-iteration-extensive-categories} verifies its initiality for our enriched models.

We start by defining what we mean by \textit{fibred iteration}.

\begin{definition}[Fibred iteration]\label{def:fibred-iteration}
		Let $\left( \catC , \iterationn \right) $ be an iterative category and $\catL : \catC ^\op \to\CAT$  a finite-coproduct-extensive indexed category. If $\left( \Sigma_{\catC} \catL ^\op  , \underline{\iterationn}   \right) $ is an iterative category such that the associated fibration
		\begin{equation} \label{eq:basic-op-fibration-grothendieck-construction}
		P_{\catL} : \Sigma_{\catC} \catL ^\op \to \catC ,
		\end{equation}
		defined by $P_{\catL}\left( f,  f'\right) = f $, yields an iterative category morphism
		\begin{equation} \label{eq:fibration-iterative-categories}
			\left( \Sigma_{\catC} \catL ^\op , \underline{ \iterationn } \right)\to \left(  \catC , \iterationn\right)  ,
		\end{equation}
	we say that $\left( \Sigma_{\catC} \catL ^\op  , \underline{\iterationn}   \right) $ is a \textit{fibred iterative category} over  $\left( \catC , \iterationn\right) $.
	In the setting above, $\underline{ \iterationn }$ is a \textit{fibred iteration} for the triple $\left( \catC , \iterationn , \catL \right)$. The raw iteration $\underline{\iterationn}$ is chosen data. Being fibred is the property that this choice is preserved by $P_{\catL}$; in particular, different choices can give different fibred iterations over the same triple.
\end{definition}

\begin{definition}[Container iteration]\label{def:op-Grothendieck-construction-of-iteration-extensive-categry}
Let $\left(\catC,\iterationn,\catL\right)$ be an iteration-extensive indexed category. Write $H_f=\catL(f)\circ\cISO^{(B,A)}$ for $f:A\to B\sqcup A$, and choose an isomorphism
\[
\theta_f:\catL(\iterationn f)\xto{\cong}\inAl{H_f,\catL(B)}.
\]
For each $\left(f,f'\right):\left(A,X\right)\to\left(B,Y\right)\sqcup\left(A,X\right)$, we define
\begin{equation}\label{def:container-iteration}
\hat{\iterationn}^{\left((A,X),(B,Y)\right)}(f,f')
\defeqq\left(\iterationn^{(A,B)}f,\foldd{H_f,Y,(X,f')}\circ\theta_{f,Y}\right).
\end{equation}
The iteration $\hat{\iterationn}$ is called the \textit{container iteration} induced by the iteration-extensive indexed category $\left(\catC,\iterationn,\catL\right)$, with these choices of comparison. Equivalently, one may take $\catL(\iterationn f)(Y)$ itself as the initial-algebra carrier, transporting the algebra structure along $\theta_{f,Y}$; then the second component is simply its fold.
\end{definition}
\begin{remark}
	It should be noted that, in the context of Def.~\ref{def:op-Grothendieck-construction-of-iteration-extensive-categry}, $f'$ indeed defines an $\catL\left( f\right) \circ \cISO ^{\left( B, A \right)} \left( Y, -\right)$-algebra structure on $X$; namely,
	$$ f': \catL\left( f\right) \circ \cISO ^{\left( B, A \right)} \left( Y, X\right) \to X . $$
	Therefore Equation \eqref{def:container-iteration} establishes a well defined iterative category.
\end{remark}

We establish below that the container iteration $\hat{ \iterationn } $ for an iteration-extensive indexed category gives rise to an iterative category
$\left( \Sigma_{\catC} \catL ^\op , \hat{ \iterationn } \right)$.

\begin{lemma}[Container iterations are  iterations]\label{lem:op-Grothendieck-construction-of-iteration-extensive-categry}
	Every iteration-extensive indexed category $\left( \catC , \iterationn , \catL  \right) $ gives rise to an iterative category
	\begin{equation}\label{eq:iterative-category-out-of-iteration-extensive-Grothendieck}
		\left( \Sigma_{\catC} \catL ^\op , \hat{ \iterationn } \right),
	\end{equation}
	where  $\hat{ \iterationn } $ is the container iteration as in Definition \ref{def:op-Grothendieck-construction-of-iteration-extensive-categry}.
		Moreover, the container iteration $\hat{ \iterationn } $ is a fibred iteration for $\left( \catC , \iterationn , \catL  \right) $, in the sense of Def.~\ref{def:fibred-iteration}.
\end{lemma}
\begin{proof}
		Put $U=\catL(\iterationn f)(Y)$. The chosen comparison has type $\theta_{f,Y}:U\to\inAl{H_f^Y}$ and the fold has type $\inAl{H_f^Y}\to X$. Their composite therefore has exactly the type $U\to X$ required of a morphism over $\iterationn f$ in the op-Grothendieck construction. The projection sends this pair to $\iterationn f$, which proves the fibred property. No equation is required for a raw iteration.
\end{proof}

The initial algebra explains more than the existence of a correctly typed second component: it determines that component by a universal property. To state this without suppressing the comparison, write $U=\catL(\iterationn f)(Y)$ and transport the initial-algebra structure to $U$:
\[
\alpha_{f,Y}=\theta_{f,Y}^{-1}\circ\minAl{H_f^Y}\circ H_f(\ID_Y,\theta_{f,Y}):H_f(Y,U)\to U.
\]

\begin{theorem}[Universal property of the container iteration]\label{theo:iteration-extensive-indexed-category}
Let $\left(\catC,\iterationn,\catL\right)$ be an iteration-extensive indexed category, with the choices of Definition~\ref{def:op-Grothendieck-construction-of-iteration-extensive-categry}. The container iteration is the unique fibred iteration whose second component $k:U\to X$, for every $(f,f')$, satisfies
\[
k\circ\alpha_{f,Y}=f'\circ H_f(\ID_Y,k).
\]
If the base iteration satisfies the basic fixed point equation and the canonical algebra
\[
\left(\catL(\iterationn f)(Y),\ID\right)
\]
is initial for every $f,Y$, choose the comparisons to identify this algebra with the chosen initial algebra. With these choices, every fibred iteration $\tilde{\iterationn}$ satisfying the basic fixed point equation obeys
\begin{equation}
\tilde{\iterationn}=\hat{\iterationn}.
\end{equation}
\end{theorem}
\begin{proof}
The pair $(U,\alpha_{f,Y})$ is initial by transport of the initial $H_f^Y$-algebra. Its unique algebra morphism into $(X,f')$ is $k=\foldd{H_f,Y,f'}\circ\theta_{f,Y}$. This proves the first assertion, expressed by the commutativity of
\begin{equation}\label{eq:initial-algebra-diagram-for-iteration-translation}
\begin{tikzcd}[column sep=large]
H_f(Y,U)\arrow[r,"{H_f(\ID_Y,k)}"]\arrow[d,swap,"\alpha_{f,Y}"] & H_f(Y,X)\arrow[d,"f'"]\\
U\arrow[r,swap,"k"] & X .
\end{tikzcd}
\end{equation}

For the second assertion, put $u=\iterationn f$. The base fixed point equation gives $u=\cpairL\ID_B,u\cpairR\circ f$. Since substitution is strict and $\cISO^{(B,A)}$ is the inverse of the coproduct comparison, we have, as functors $\catL(B)\to\catL(A)$,
\[
H_f\circ\pairL\ID_{\catL(B)},\catL(u)\pairR
=\catL(f)\circ\catL(\cpairL\ID_B,u\cpairR)
=\catL(u).
\]
Thus $H_f(Y,U)=U$, which supplies the displayed identity algebra. Under the stated initiality assumption, its comparison with any chosen initial algebra is the unique algebra isomorphism. Consequently $\alpha_{f,Y}=\ID_U$, and the preceding square says exactly $k=f'\circ H_f(\ID_Y,k)$. Together with the base equation, this is
\begin{equation}\label{eq:iteration-with-one-equation-theorem}
\cpairL\ID_{(B,Y)},\hat{\iterationn}(f,f')\cpairR\circ(f,f')
=\hat{\iterationn}(f,f').
\end{equation}
Any fibred iteration satisfying this equation has the same first component $u$ and a second component making the same square commute. Initiality makes that second component unique.
\end{proof}

In the enriched models below, the initial-chain construction supplies the canonical identity algebra and proves its initiality (Theorem~\ref{theo:class-of-examples-of-iteration-extensive-categories}).

\begin{definition}
In the context of Theorem \ref{theo:iteration-extensive-indexed-category},
$\left( \Sigma_{\catC} \catL ^\op , \hat{ \iterationn } \right) $ is the \textit{op-Grothendieck construction of the  iteration-extensive indexed category $\left( \catC , \iterationn , \catL  \right) $}.
\end{definition}

\begin{theorem}[Grothendieck construction of iteration-extensive Freyd indexed categories]\label{theo:Freyd-goes-to-Freyd}
	The op-Grothendieck construction takes each iteration-extensive Freyd indexed category $\left( \catV , \catC , j,   \otimes , \iterationn , \catL \right)$ to an iterative Freyd category
	\begin{equation}
		\left( \Sigma_{\catV} \left( \catL\circ j^\op\right) ^\op , \Sigma_{\catC} \catL ^\op , \hat{j},   \hat{\otimes} , \hat{\iterationn}  \right)
	\end{equation}
where $\hat{j}\left( A, X\right) \defeqq \left( j(A), X\right)  $,
\begin{equation}
	\begin{aligned}
(f,f') \hat{\otimes} (g,g')&=\bigl(f\otimes g,\\
&\quad(\catL(j(\proj A))(f')\circ\delta_{A,g,\catL(j(f))(Y)})\sqcup\catL(j(\proj C))(g')\bigr)
\end{aligned}  ,
\end{equation}
and $\left( \Sigma_{\catC} \catL ^\op , \hat{ \iterationn } \right) $  is the op-Grothendieck construction of the  iteration-extensive indexed category $\left( \catC , \iterationn , \catL  \right) $.
\end{theorem}
\begin{proof}
	$\displaystyle\Sigma_{\catV } \left( \catL\circ j^\op\right)  ^\op$ is indeed
	a distributive category: products follow from the indexed finite coproducts, and coproducts from Theorem~\ref{theo:L-finite-coproduct-extensive-indexed-category}. Under the extensive comparison, the distributivity map is the base distributivity isomorphism with identity fibre components. The compatible action on dependent data gives the lifted Freyd action, as specified in Eq.~\eqref{eq:indexed-context-action}. Lemma~\ref{lem:op-Grothendieck-construction-of-iteration-extensive-categry} supplies the fibred iteration, so these data form the asserted iterative Freyd category.
\end{proof}

In Theorem \ref{theo:main-result-iterations-are-equivalent1}, we prove that the container iteration introduced above coincides with the usual iteration in the case of the {concrete models for dependently typed languages with iteration}, or, more precisely, $\wCpo$-enriched concrete models, like the case of the \textit{concrete semantics of our target language}. This result is foundational to our contribution, as it shows that our container iteration indeed implements the expected behavior inherited from the $\wCpo$-enrichment of the concrete semantics -- while not depending on that enrichment itself, given that the categorical semantics (and the syntax) is not $\wCpo$-enriched.

\subsection{Concrete container iterations are container iterations} \label{subsect:concrete-iterations-wcpo}
In what follows, we show that our definition of container iteration agrees with the usual notion of iteration via colimits in concrete (i.e., $\wCpo$-enriched) models. More precisely, we demonstrate that in the setting of $\wCpo$-enriched indexed categories, the concrete iteration -- defined as colimits of $\ww$-chains -- satisfies the property stated in Theorem \ref{theo:iteration-extensive-indexed-category}; namely, it is a fibred iteration satisfying the basic fixed point equation. Consequently, by the universal property of the container iteration, the two notions of iteration coincide.

In other words, the results of this section show that our definition of container iteration in
dependently typed theories with sum types indeed is coherent with the iteration of concrete models of dependently typed languages.

\subsubsection{Key ideas of the iteration-coincidence result}
We start by observing that the op-Grothendieck construction of any \(\wCpo\)-enriched indexed category \(\catL\) is itself naturally \(\wCpo\)-enriched. Consequently, on one hand, if \(\catL\) is an $\wCpo$-finite-coproduct-extensive indexed category (see Def.~\ref{def:enriched-model-for-sum-types}), then the op-Grothendieck construction \(\Sigma_{\catC}\,\catL^\op\) inherits an iteration, denoted herein by \(\iterationn_\ww\), from the \(\wCpo\)-enrichment (as established in  Theorem~\ref{theo:wCpoenrichedcategory-underlyingiterative}). This gives rise to an iterative category
$$
\bigl(\Sigma_{\catC}\,\catL^\op, \iterationn_\ww\bigr)
$$
that satisfies the basic fixed point equation.
We refer to \(\iterationn_\ww\) as the \emph{concrete $\wCpo$-container iteration} induced by the $\wCpo$-finite-coproduct-extensive indexed category $\left( \catC , \catL \right) $.

On the other hand, if \(\catL\) is again an \(\wCpo\)-enriched (coproduct-)extensive indexed category, then it provides an example of an iteration-extensive indexed category, thereby inducing a \emph{container iteration} \(\hat{\iterationn}\) on its op-Grothendieck construction, as in Definition \ref{def:op-Grothendieck-construction-of-iteration-extensive-categry}. In this setting, our main result, Theorem~\ref{theo:main-result-iterations-are-equivalent1}, shows that these two iterations coincide.

We start by establishing that the Grothendieck construction of an indexed $\wCpo$-enriched category is $\wCpo$-enriched.

\subsubsection{Basic definition: enriched indexed category}
The enrichment records how partial computations approximate one another. The fibres themselves need not be locally ordered. What matters is that an increasing chain of computations induces a colimiting chain of substitution functors; this is the structure used to compare the two iterations. Our use of ``$\wCpo$-enriched indexed category'' below refers to this particular $2$-functorial structure.

An \textit{$\wCpo$-enriched indexed category} herein is a pair $\left( \catC , \catL  \right) $ in which $\catC $ is an $\wCpo$-enriched category and $\catL : \catC ^\op \to \CAT $ is a $2$-functor such that (1) for any object $A$ of $\catC$, $\catL \left( A\right) $ has initial object and colimit of $\ww$-chains, (2) the $2$-functor $\catL $ locally preserves initial objects and colimits of $\ww$-chains of morphisms. More explicitly, this means that
\begin{itemize}
	\item $\catL : \catC ^\op \to \catCat $ is an indexed category;
	\item  for any object $A$ of $\catC$, $\catL \left( A\right) $ has initial object and colimits of $\ww$-chains;
	\item for each pair $\left( A, B \right) $ of objects in $\catC $, we
	have a functor
	\begin{equation}
		\catL _{A,B} : \catC\left( A, B \right)\to \catCat\left( \catL \left( B\right) , \catL \left( A \right) \right) ;
	\end{equation}
that preserves initial objects and colimits of $\ww$-chains, and satisfies the following explicit conditions;
\item for each pair
	$\left( f : A\to B , g: A\to B \right) $ of morphisms in $\catC$  satisfying  $f\leq g $ in $\catC$, we have a chosen
	natural transformation $\catL \left( f\leq g \right) : \catL \left( f \right)\to \catL \left( g \right) $, such that:
\begin{enumerate}
	\item $\catL \left( f\leq f \right) = \ID _{\catL \left( f\right) } $;
	\item $\catL \left( f\leq g \right)\circ \catL \left( h\leq f \right) = \catL \left( h\leq g \right)$;
	\item $\catL \left( f'\leq g' \right)\ast\catL \left( f\leq g \right) = \catL \left( f\circ f'\leq g\circ g' \right)$, where $\ast $ denotes the horizontal composition
	of natural transformations, with $f',g':D\to A$ and $f,g:A\to B$;
\end{enumerate}
	\item $\catL \left( \bot : A\to B \right)   $ is the functor $\catL \left( \bot \right)  : \catL \left(  B \right) \to\catL\left(  A \right)   $ that is constantly equal to the initial object $\initiall $ of $\catL\left( A\right) $;
	\item for any pair $\left( A, B\right) $ of objects in $\catC$ and any $\ww$-chain
\begin{equation}\label{eq:omega-chain-that-is-preserved}
\mathcal{G}\defeqq \left( g_0 \leq g_1\leq \cdots \leq g_n \cdots \right)
\end{equation}
of morphisms in $\catC\left( A,B \right) $, we have that
$$ \colim\left( \catL\circ \mathcal{G}\right) \cong \catL \left( \colim\left( \mathcal{G} \right) \right) : \catL \left( B\right) \to \catL \left( A \right)  .    $$
\end{itemize}

At this point, it is important to keep some examples in mind.
	We start by observing that:

\begin{lemma}[Representable functor]
	Let $\catC $ be any $\wCpo$-category. For any object $K$ of $\catC $, the pair $\left( \catC  , \catC\left( -, K \right) \right) $ is an $\wCpo$-enriched indexed category, where
	\begin{equation}\label{eq:basic-example-ofwcpo-indexed-category}
		\catC\left( -, K \right) : \catC ^\op \to \CAT
	\end{equation}
is defined by the composition of arrows
\begin{equation}
	\catC ^\op \xto{\catC \left[ -, K \right]} \wCpo \to \CAT ,
\end{equation}
in which $\catC \left[ -, K \right] : \catC ^\op\to \wCpo $ is the $\wCpo$-enriched
representable functor.
\end{lemma}
\begin{proof}
	It follows directly from the fact that $\catC $ is $\wCpo$-enriched, and the fact that
	the ordered homs have bottom elements and increasing $\ww$-chain colimits, with composition preserving these in each variable.
\end{proof}

Furthermore, it is clear that:
\begin{theorem}
Given an object $K$ of $\catCatb$, considering the inclusions
	\begin{equation}
		\inCC{-} : \catPSet\to \catCatb, \qquad\mbox{and}\qquad  \inCC{-} : \wCpo\to \catCatb
	\end{equation}
the pairs
\begin{equation}
\left( \wCpo , \catCatb\left( \inCC{-} , K \right) : \wCpo ^\op \to \CAT    \right), \quad
 \left( \catPSet, \catCatb\left( \inCC{-} , K \right): \catPSet ^\op \to \CAT    \right)
\end{equation}
are $\wCpo$-enriched indexed categories.
\end{theorem}
\begin{proof}
	Initial objects and colimits of $\ww$-chains in the displayed functor categories are computed pointwise. Precomposition with the inclusions sends bottom maps to the constantly initial functors, and increasing chains to their pointwise colimits. The required $2$-functoriality follows from precomposition.
\end{proof}

\subsubsection{Enriched Grothendieck construction of an enriched indexed category}
The op-Grothendieck construction of an $\wCpo$-enriched indexed category naturally inherits a fibred $\wCpo$-structure, which we establish in Lemma \ref{lem:inherited-wCpo-enrichment}.

\begin{lemma}\label{lem:inherited-wCpo-enrichment}
	Let $\left( \catC , \catL \right)$ be an $\wCpo$-enriched indexed category.
	The op-Grothendieck construction
$\displaystyle \Sigma_{\catC }\catL ^\op $
	naturally inherits an $\wCpo$-enrichment. Concretely, for two morphisms
	\begin{equation}
	\left( f, f'\right) , \left( g, g'\right) : \left( A, X \right) \to \left( B, Y \right)
\end{equation}
	in
$\displaystyle\Sigma_{\catC } \catL ^\op $
	we define
\begin{equation}
	\left( f, f'\right)\leq \left( g, g'\right)
\end{equation}
	if and only if $f \leq g$ in $\catC$, and the Diagram \eqref{eq:commutative-triangle-definition-wCpo} commutes in $\catL(A)$.
	\begin{equation}\label{eq:commutative-triangle-definition-wCpo}
		\diag{trianglewCPO}
	\end{equation}
	With this ordering, $\Sigma_{\catC}\catL^\op$ becomes an $\wCpo$-category, and the associated fibration, namely \eqref{eq:fibration-is-wCPO-enrichedd},	is an $\wCpo$-enriched functor.
	\begin{equation}\label{eq:fibration-is-wCPO-enrichedd}
	P_{\catL}:	\Sigma_{\catC} \catL^\op\to  \catC
	\end{equation}
\end{lemma}

\begin{proof}
	In order to verify that $\displaystyle\Sigma_{\catC } \catL ^\op $ endowed with the order structure defined above is an $\wCpo$-category we verify that, for each pair $\left( \left( A, X\right) , \left( B, Y\right) \right) $ of objects in  $\displaystyle\Sigma_{\catC } \catL ^\op$, we have that:
	\begin{itemize}
		\item the morphism defined by $\left( \bot , \iota_X \right) $, where $\iota _X : \catL\left( \bot \right)\left( Y\right)  = \initiall\to X $ is the unique morphism, is the initial object/bottom element of $\displaystyle\Sigma_{\catC } \catL ^\op \left( \left( A, X\right) , \left( B, Y\right) \right)  $;
		\item given an $\ww $-chain of morphisms
$$ \mathcal{G}\defeqq \left( \left( g_0 , g_0 '\right)  \leq \left( g_1 , g_1 '\right) \leq \cdots \leq \left( g_n , g_n '\right)\cdots \right) $$
in $\displaystyle\Sigma_{\catC } \catL ^\op \left( \left( A, X\right) , \left( B, Y\right) \right)  $, has a  colimit, which is defined by $\left( \colim g_t , g' \right) $
where $$g' : \catL \left(\colim g_t \right) \left( Y \right) \to X $$ is induced by the universal property of $$\catL \left(\colim\ g_t \right)\left( Y \right) \cong \colim\left( \catL \left( g_t \right)\left( Y \right)  \right) $$
and the morphisms $\left( g_t '\right) _{t\in\NN} $;
\item composition is an $\wCpo$-morphism. For example, for $h:D\to A$ and an increasing chain $g_n:A\to B$, the needed preservation after reindexing follows from
\[
\catL(h)\catL(\colim_n g_n)
=\catL\bigl((\colim_n g_n)\circ h\bigr)
\cong\colim_n\catL(g_n\circ h)
=\colim_n\catL(h)\catL(g_n).
\]
The other variable is treated by the same argument and the universal property of the displayed fibre colimits. Bottom maps are preserved because their images under $\catL$ are constantly initial. Separate continuity gives continuity on increasing chains of pairs, since the diagonal is cofinal in $\ww\times\ww$.
	\end{itemize}
Finally, with the above, we can also conclude that \eqref{eq:fibration-is-wCPO-enrichedd} is indeed
an $\wCpo$-functor.
\end{proof}

\begin{definition}[Op-Grothendieck construction]\label{def:op-Grothendieck-enrichment}
Let   $\left( \catC , \catL \right)$ be an $\wCpo$-enriched indexed category.
The op-Grothendieck construction $\displaystyle\Sigma^\op\left( \catC , \catL \right) \defeqq\Sigma_{\catC } \catL ^\op $  is, herein, considered to be endowed with the $\wCpo$-structure introduced in Lemma \ref{lem:inherited-wCpo-enrichment}. For structured indexed categories, $\Sigma^\op$ likewise denotes fibrewise dualization followed by the Grothendieck construction.
\end{definition}

\subsubsection{Enriched extensivity}
We introduce the $\wCpo$-enriched counterpart to the notion of (finite-coproduct-)extensive indexed category; namely:
\begin{definition}[Enriched extensivity]\label{def:enriched-model-for-sum-types}
	An $\wCpo$-enriched indexed category $\left( \catC, \mathcal{L} \right) $ is \textit{$\wCpo$-finite-coproduct-extensive} if $\catC $ has finite $\wCpo$-coproducts, and	$\mathcal{L} $ preserves the finite $\wCpo$-products of $\catC ^\op $.
\end{definition}

\begin{remark}
	It is not in the scope of the present work, but it is clear that
	given a class of diagrams $\mathfrak{S}$, we have the notion of $\wCpo$-enriched extensive
	indexed category w.r.t. $\mathfrak{S}$; namely, the enriched version of Def.~\ref{def:extensive-S}, \textit{e.g.}, \cite[Section~6.7]{LV24b}.
\end{remark}

We should observe that
our basic examples of $\wCpo$-enriched indexed categories are $\wCpo$-finite-coproduct-extensive. More precisely:
\begin{theorem}
	Let $\catC $  be an  $\wCpo$-enriched category with finite $\wCpo$-coproducts. For any $K\in \catC$, the representable $2$-functor
	$$\catC\left( -, K\right) : \catC ^\op \to \CAT$$
	makes the pair $\left( \catC, \catC\left( -, K\right)  \right) $ into a $\wCpo$-finite-coproduct-extensive indexed category.
\end{theorem}
\begin{proof}
	It follows directly from the universal property of $\wCpo$-coproducts.
\end{proof}

\begin{theorem}\label{theo:this-proves-the-case-of-our-concrete-model}
	Given an object $K$ of $\catCatb$, considering the inclusions
	\begin{equation}
		\inCC{-} : \catPSet\to \catCatb, \qquad\mbox{and}\qquad  \inCC{-} : \wCpo\to \catCatb
	\end{equation}
	the pairs
	\begin{equation}
		\left( \wCpo , \catCatb\left( \inCC{-} , K \right) : \wCpo ^\op \to \CAT    \right), \quad
		\left( \catPSet, \catCatb\left( \inCC{-} , K \right): \catPSet ^\op \to \CAT    \right)
	\end{equation}
	are $\wCpo$-finite-coproduct-extensive indexed categories.
\end{theorem}
\begin{proof}
	The preceding result gives the enriched indexed structure. The finite coproduct comparisons are isomorphisms by the coproduct universal property in $\catCatb$, since the inclusions $\inCC{-}$ preserve finite coproducts.
\end{proof}

As in the non-enriched setting, the appeal of the notion of  $\wCpo$-finite-coproduct-extensive indexed category is that it ensures the existence of finite $\wCpo$-coproducts in the Grothendieck constructions; namely:

\begin{theorem}\label{def:basic-wcpo-coproducts-enriched-containers}
		Let $\left( \catC , \mathcal{L}\right) $ be an $\wCpo$-finite-coproduct-extensive indexed category. In this setting, we have that $\displaystyle\Sigma_{\catC } \catL ^\op $ has finite $\wCpo$-coproducts. Moreover,
		the associated fibration functor \eqref{eq:fibration-is-wCPO-enrichedd} is an $\wCpo$-functor that preserves finite $\wCpo$-coproducts.
\end{theorem}
\begin{proof}
	Use the coproducts of Theorem~\ref{theo:L-finite-coproduct-extensive-indexed-category}. Their hom-set comparison is an order isomorphism: the triangle in Eq.~\eqref{eq:commutative-triangle-definition-wCpo} restricts to the corresponding triangle on each summand. It therefore preserves bottom elements and increasing $\ww$-chain suprema. The projection to the base preserves these coproducts by construction.
\end{proof}

Recall that, by Theorem \ref{theo:wCpoenrichedcategory-underlyingiterative}, every $\wCpo$-enriched category $\catC$ has an underlying iterative category $\left( \catC , \iterationn \right) $ provided that $\catC$ has finite $\wCpo$-coproducts. With that, we observe that:

\begin{theorem}[Concrete enriched container iteration]\label{theo:concrete-iteration-ww-container}
	Let $\left( \catC , \mathcal{L}\right) $ be an $\wCpo$-finite-coproduct-extensive indexed category. The $\wCpo$-enrichment of $\displaystyle\Sigma_{\catC } \catL ^\op $ yields an iterative category
	\begin{equation}\label{eq:OK-wcpo-coproducts-SIGMA}
		\bigl(\Sigma_{\catC}\,\catL^\op, \iterationn_\ww\bigr)
	\end{equation}
with the $\wCpo$-enriched iteration $\iterationn_\ww$ as
established in Def.~\ref{def:wcpo-enriched-iteration-concrete}.
Furthermore, by Theorem~\ref{theo:wCpoenrichedcategory-underlyingiterative},
$\iterationn_\ww$ satisfies the basic fixed point equation.
\end{theorem}
\begin{proof}
	Indeed, by Lemma \ref{lem:inherited-wCpo-enrichment} and Theorem \ref{def:basic-wcpo-coproducts-enriched-containers},
	$\bigl(\Sigma_{\catC}\,\catL^\op, \iterationn_\ww\bigr)$ is an $\wCpo$-category with $\wCpo$-coproducts. Hence, by Theorem~\ref{theo:wCpoenrichedcategory-underlyingiterative}, that underlying iterative category $\bigl(\Sigma_{\catC}\,\catL^\op, \iterationn_\ww\bigr)$
	satisfies the basic fixed point equation.
\end{proof}

\begin{definition}[Enriched container iteration]\label{def:concrete-iteration-ww-container}
	In the setting of Theorem \ref{theo:concrete-iteration-ww-container}, we refer to the iteration \(\iterationn_\ww\) as the \emph{concrete $\wCpo$-container iteration} induced by the $\wCpo$-finite-coproduct-extensive indexed category $\left( \catC , \catL \right) $.
\end{definition}

\subsubsection{Enriched extensive indexed categories are iteration-extensive}
The $\wCpo$-finite-coproduct-extensive indexed categories provide a class of examples of
iteration-extensive indexed categories. Consequently, every
$\wCpo$-finite-coproduct-extensive indexed category
 $\left( \catC , \catL \right)$ inherits a container iteration $\hat{\iterationn}$  from its underlying iteration-extensive indexed category by Lemma \ref{lem:op-Grothendieck-construction-of-iteration-extensive-categry} (see Definition \ref{def:op-Grothendieck-construction-of-iteration-extensive-categry}). More precisely:

\begin{theorem}[Underlying iteration-extensive indexed category]\label{theo:class-of-examples-of-iteration-extensive-categories}
	Every $\wCpo$-finite-coproduct-extensive indexed category $\left( \catC , \catL \right)$ has an underlying iteration-extensive indexed category $\left( \catC, \iterationn , \catL \right) $,  where $\iterationn $ is the iteration inherited by the $\wCpo$-structure in $\catC$ (as in Definition \ref{def:wcpo-enriched-iteration-concrete}). Moreover, the canonical identity algebra on $\catL(\iterationn f)(Y)$ is initial for every $f:A\to B\sqcup A$ and $Y\in\catL(B)$.
\end{theorem}
\begin{proof}
Let $f:A\to B\sqcup A$. Put $g=\cpairL\coproje{B},f\cpairR$ and $p=\cpairL\ID_B,\bot\cpairR$. By the definition of enriched iteration,
\begin{equation}\label{eq:iteration-omega-chains-again}
u\defeqq\iterationn^{(A,B)}f=\underline{\mathtt f}\circ\coproje{A},\qquad
\underline{\mathtt f}=\colim_n p\circ g^n,
\end{equation}
where the colimit is taken over
\begin{equation}\label{eq:colimit-iteration-diagram-again}
p\leq p\circ g\leq p\circ g^2\leq\cdots
\end{equation}
in $\catC(B\sqcup A,B)$. To see directly that the chain is increasing, $p$ and $p\circ g$ both restrict to $\ID_B$ on the first summand, while their restrictions to $A$ satisfy $\bot\leq p\circ f$. Monotonicity of composition gives all subsequent inequalities.

Equivalently, let $q_n=p\circ g^n\circ\coproje{A}:A\to B$. Then $q_0=\bot$, $q_{n+1}=\cpairL\ID_B,q_n\cpairR\circ f$, and $u=\colim_nq_n$. Fix $Y\in\catL(B)$ and write $H_f=\catL(f)\circ\cISO^{(B,A)}$. Strict $2$-functoriality and extensivity identify
\[
\catL(q_0)(Y)=\initiall,\qquad
\catL(q_{n+1})(Y)=H_f\bigl(Y,\catL(q_n)(Y)\bigr).
\]
The images of $q_n\leq q_{n+1}$ are exactly the connecting maps in the initial chain for $H_f(Y,-)$. Local preservation of colimits gives
\begin{equation}\label{eq:conclusion-main-theorem-concrete}
U\defeqq\catL(u)(Y)\cong\colim_n\catL(q_n)(Y).
\end{equation}
In the parameterized notation of Lemma~\ref{lem:adamek-computing-parameterized-initial-algebras}, this is the second component of the colimit of
\begin{equation}\label{eq:colimit-iteration-diagram-image-again}
J\xto{\underline{\iota}}G\circ J\xto{G(\underline{\iota})}G^2\circ J\longrightarrow\cdots,
\end{equation}
where $J=\pairL\ID_{\catL(B)},\bot\pairR:\catL(B)\to\catL(B)\times\catL(A)$ and $G=\pairL\proj{\catL(B)},H_f\pairR$ is an endofunctor of $\catL(B)\times\catL(A)$.

For completeness, the required preservation here concerns this particular initial chain. The fixed point equation $u=\cpairL\ID_B,u\cpairR\circ f$ gives $H_f(Y,U)=U$. Moreover, applying $H_f(Y,-)$ to the colimit cocone in Eq.~\eqref{eq:conclusion-main-theorem-concrete} yields the cocone obtained by applying $\catL$ to $q_{n+1}\leq u$. This follows from $2$-functoriality and the coproduct comparison. The latter cocone is colimiting by local preservation, since the shifted chain has the same supremum. Thus $H_f(Y,-)$ preserves the colimit of its initial chain, and Adámek's theorem \cite{adamek1974} makes $(U,\ID_U)$ initial. This also identifies its structure map: it is the identity carrying the shifted colimit cocone to itself.

All these constructions are natural in $Y$. Consequently
\[
\catL(\iterationn f)\cong\inAl{\catL(f)\circ\cISO^{(B,A)},\catL(B)},
\]
which proves iteration-extensivity and the additional assertion about the canonical algebra. We use these canonical initial algebras for the induced container iteration.
\end{proof}

Since, in the setting of Def.~\ref{def:concrete-iteration-ww-container}, every concrete $\wCpo$-container iteration satisfies the basic fixed point equation,  we get the following result by the universal property of container iterations (Def.~\ref{def:op-Grothendieck-construction-of-iteration-extensive-categry}) induced by iteration-extensive categories (see  Theorem \ref{theo:iteration-extensive-indexed-category} for the universal property).

\begin{corollary}\label{cor:unique-concrente-container-iteration}
	Let  $\left( \catC, \iterationn , \catL \right) $ be the underlying iteration-extensive indexed category of an $\wCpo$-finite-coproduct-extensive indexed category.

	The container iteration $\hat{\iterationn}$, using the canonical initial algebras, for the iteration-extensive indexed category $\left( \catC, \iterationn , \catL \right)$  is the only fibred iteration for  $\left( \catC, \iterationn , \catL \right) $ such that $\bigl(\Sigma_{\catC}\,\catL^\op, \hat{\iterationn}\bigr)$ satisfies the basic fixed point equation.
\end{corollary}
\begin{proof}
	Indeed, since by Theorem \ref{theo:wCpoenrichedcategory-underlyingiterative} $\left( \catC, \iterationn\right) $ satisfies the basic fixed point equation, the initiality assertion of Theorem~\ref{theo:class-of-examples-of-iteration-extensive-categories} lets us apply Theorem~\ref{theo:iteration-extensive-indexed-category}.
\end{proof}

\begin{theorem}\label{theo:main-result-iterations-are-equivalent1}
	Let  $\left( \catC, \iterationn , \catL \right) $ be the underlying iteration-extensive indexed category of an $\wCpo$-finite-coproduct-extensive indexed category $\left( \catC, \catL \right) $.

Denoting by \(\iterationn_\ww\) the \emph{concrete $\wCpo$-container iteration} induced by the $\wCpo$-finite-coproduct-extensive indexed category $\left( \catC , \catL \right) $ as in Def.~\ref{def:concrete-iteration-ww-container}, we have that the identity functor on $\displaystyle  \Sigma_{\catC}\,\catL^\op $ induces an iterative category isomorphism
\begin{equation}
		\bigl(\Sigma_{\catC}\,\catL^\op, \hat{\iterationn}\bigr)\to \bigl(\Sigma_{\catC}\,\catL^\op, \iterationn_\ww \bigr) ,
\end{equation}
where $\hat{\iterationn}$ is the container iteration using these canonical initial algebras for the iteration-extensive indexed category $\left( \catC, \iterationn , \catL \right) $.
\end{theorem}
\begin{proof}
	By Corollary \ref{cor:unique-concrente-container-iteration}, it is enough to verify that indeed \(\iterationn_\ww\)  is a fibred iteration for the iteration-extensive indexed category $\left( \catC, \iterationn , \catL \right) $ that makes $\bigl(\Sigma_{\catC}\,\catL^\op, \iterationn_\ww \bigr) $ into an iterative category satisfying the basic fixed point equation.

By Theorem \ref{def:basic-wcpo-coproducts-enriched-containers},
		\begin{equation}\label{eq:fibration-is-wCPO-enrichedd-proof}
		P_{\catL}:	\Sigma_{\catC} \catL^\op\to  \catC
	\end{equation}
is an $\wCpo$-functor that preserves $\wCpo$-coproducts. Hence, by Theorem \ref{theo:wCpoenrichedcategory-underlyingiterative}, we get that \eqref{eq:fibration-is-wCPO-enrichedd-proof} yields an iterative category morphism
\begin{equation} \label{eq:fibration-iterative-categories-proof}
	\left( \Sigma_{\catC} \catL ^\op , \iterationn _\ww  \right)\to \left(  \catC , \iterationn\right)  .
\end{equation}
This proves that $\iterationn_\ww$ is a fibred iteration for the iteration-extensive indexed category $\left( \catC, \iterationn , \catL \right) $. Since, by Theorem~\ref{theo:concrete-iteration-ww-container}, $\bigl(\Sigma_{\catC}\,\catL^\op, \iterationn_\ww \bigr) $ satisfies the
basic fixed point equation, this concludes our proof.
\end{proof}
That is, $\hat{\iterationn}$ and $\iterationn_\ww$ coincide.

\subsubsection{Enriched indexed Freyd categories}
We retain the enriched Freyd structure explicitly because it records sufficient conditions on a concrete target model for all the preceding constructions to agree. It packages the approximation order used for computations together with the indexed structure used for dependent types. In this way, checking a concrete model once supplies its iteration-extensive and raw iterative structures, as the following corollary records.
\begin{definition}[Enriched indexed Freyd category]
		An $\ww$-Freyd indexed category consists of a compatible action on dependent data, locally continuous in the computation argument, and a tuple	$\left( \catV , \catC , j,   \otimes ,  \catL \right)$ such that $\left( \catC , \catL \right)$  is an $\wCpo$-finite-coproduct-extensive indexed category, $\catL$ has indexed finite coproducts, and $\left( \catV , \catC , j,   \otimes \right) $ is an $\ww$-Freyd category.
\end{definition}
\begin{corollary}\label{cor:fundamental-for-concrete-semantics-Iguess}
	Every $\ww$-Freyd indexed category  $\left( \catV , \catC , j,   \otimes ,  \catL \right)$ has  an underlying iteration-extensive Freyd indexed category $\left( \catV , \catC , j,   \otimes , \iterationn,  \catL \right)$ where $\left( \catC , \iterationn \right) $ is the iterative category underlying the $\wCpo$-category (with $\wCpo$-coproducts) $\catC$.

	Moreover, every $\ww$-indexed Freyd category   $\left( \catV , \catC , j,   \otimes ,  \catL \right)$ is taken to an $\ww$-Freyd category
	\begin{equation}\label{eq:Grothendick-ww-Freyd}
	\displaystyle \Sigma^\op\left( \catV , \catC , j,   \otimes ,  \catL \right)  \defeqq	\left( \Sigma_{\catV} \left( \catL\circ j^\op\right) ^\op , \Sigma_{\catC} \catL ^\op , \hat{j},   \hat{\otimes} \right)
	\end{equation}
by the op-Grothendieck construction. The iterative Freyd category underlying \eqref{eq:Grothendick-ww-Freyd} is isomorphic to the iterative Freyd category
obtained by taking the op-Grothendieck construction of the underlying iteration-extensive indexed category $\left( \catV , \catC , j,   \otimes , \iterationn,  \catL \right)$ as in Theorem  \ref{theo:Freyd-goes-to-Freyd}.

Finally, in particular, the underlying iterative category	$\left( \Sigma_{\catC} \catL ^\op  , \hat{\iterationn}  \right) $ satisfies the basic fixed point equation.
\end{corollary}
\begin{proof}
	The underlying iteration-extensive indexed structure follows from Theorem~\ref{theo:class-of-examples-of-iteration-extensive-categories}. The indexed finite coproducts give the product on values, while the compatible action supplies its extension to computations; its assumed local continuity gives the enriched Freyd structure. The iteration $\hat{\iterationn} $ is the unique fibred iteration satisfying the basic fixed point equation, and coincides with the concrete $\wCpo$-container iteration $\iterationn_\ww$.
\end{proof}

\subsection{Iterative indexed categories}\label{syn:container-folder-iteration}
The container iteration defined in the op-Grothendieck construction
of an iteration-extensive indexed category
introduced in Def.~\ref{def:op-Grothendieck-construction-of-iteration-extensive-categry} is defined purely at the equational level -- in other words, it is a notion of iteration that is independent of any non-equational structure/convergence.

As demonstrated in the previous results, the concrete $\wCpo$-container iteration induced by a $\wCpo$-finite-coproduct-extensive indexed category $\left( \catC, \catL \right) $ coincides precisely with the container iteration in the op-Grothendieck construction induced by the underlying iteration-extensive indexed category. In Section~\ref{subsect:revisiting-iteration-PFAMVECT} below, we illustrate explicitly that this is the setting of $\PFam{\Vect^\op}$. This result lays a foundational path toward establishing a concrete categorical semantics for the target language of our automatic differentiation macro, CHAD.

We now specify the categorical structure presented by the iteration and fold operations of our target language. The syntax itself is given in Section~\ref{sec:target-language}; its universal property uses the iterative Freyd indexed categories introduced below. Initiality supplies folders in the concrete semantics, but the syntax only needs the folders themselves. This distinction allows us to define the program transformation without requiring an $\wCpo$-completion of the syntactic categories.

These raw iteration and folder operations suffice to construct CHAD. Additional iteration equations require corresponding equations on the folders; Theorem~\ref{theo:iteration-extensive-indexed-category} gives the compatibility for the basic fixed point equation.

\begin{definition} \label{eq:iterative-free-indexed}
	A \textit{(raw) iterative indexed category} is \textit{herein} a quadruple
\begin{equation} 	\label{eq:quadruble-for-free}
	\left(\catC ,  \iterationn, \catL,  \folding{-,-} \right)
\end{equation}
	  where $\left( \catC , \iterationn\right) $ is an iterative category,  $\catL : \catC ^\op \to \CAT $ is a  finite-coproduct-extensive indexed category, and, for each morphism $f\in \catC\left( A,  B\sqcup A\right)   $ and each pair $\left( X, Y\right) \in \catL\left( A\right) \times\catL\left( B\right) $,    we have a function
	\begin{eqnarray*}
		\foldingg{f, -}{X,Y}: & \catL\left( A \right)\left( \catL\left( f\right)\circ \cISO ^{\left( B, A\right) } \left( Y, X \right), X   \right) &    \to \catL\left( A \right)\left(  \catL\left( \iterationn f\right)\left( Y \right)  , X \right) \\
		& g &\mapsto \folding{f, g} ,
	\end{eqnarray*}
	where $\cISO ^{\left( B, A\right) } :\catL\left( B\right) \times  \catL\left(A\right)  \xto{\cong} \catL\left( B\sqcup A\right) $ is the inverse of the comparison morphism.

The functions $ \folding{-,-} $ are called \textit{\foldersname}.
\end{definition}

\begin{remark}\label{rem:equivalent-definition-of-iterative-indexed-category}
	Definition \ref{eq:iterative-free-indexed} could be rephrased as follows:
		a \textit{(reverse) (raw) iterative indexed category} is \textit{herein} a quadruple \eqref{eq:quadruble-for-free} where $\left( \catC , \iterationn\right) $ is an iterative category,  $\catL : \catC ^\op \to \CAT $ is a  finite-coproduct-extensive indexed category, and, for each morphism $\left( f, f'\right) : \left( A, X\right) \to \left( B, Y\right) \sqcup \left( A, X\right) $ in $\displaystyle\Sigma_{\catC } \catL ^\op $, we have a fixed morphism
		\begin{equation}
			\folding{f, f'} : \catL \left( \iterationn f\right) (Y) \to X .
		\end{equation}

	The above shows how the \foldersname{}  are precisely the structure needed to define
	fibred iterations in the op-Grothendieck construction, as precisely established in Lemma \ref{lem:bijection-iterative-indexed-categories-Grothendieck}. This is confirmed by Theorem \ref{th:iterative-Grothendieck-iteration-free-container}.
\end{remark}

Motivated by the results we have on iteration-extensive indexed categories, we observe that:
\begin{theorem}[Iterative Grothendieck construction]\label{th:iterative-Grothendieck-iteration-free-container}
Let $\left(\catC ,  \iterationn , \catL,  \folding{-,-} \right) $ be an iterative indexed category.
The op-Grothendieck construction on $\catL$ inherits an iteration structure, giving rise to an iterative category
\begin{equation} \label{eq:iterative-Grothendieckconstruction}
 \left(  \displaystyle\Sigma_{\catC } \catL ^\op , \iterationg  \right)
\end{equation}
where, for each morphism
$$\left( f , f'\right) : \left( A, X\right) \to \left( B, Y \right)\sqcup \left( A, X \right)\cong \left( B\sqcup A, \cISO ^{\left( B, A\right) }\left( Y, X\right)\right)   $$
in $\displaystyle\Sigma_{\catC } \catL ^\op$, we define
\begin{equation}
	\iterationg  \left( f , f'\right) = \left( \iterationn f , \folding{f, f'}\right).
\end{equation}
Furthermore, $\iterationg$ is a fibred iteration for the triple $\left(\catC ,  \iterationn , \catL\right) $ (see Def.~\ref{def:fibred-iteration}).
For short, we say that $\left(  \displaystyle\Sigma_{\catC } \catL ^\op , \iterationg  \right)$ is the op-Grothendieck construction of the iterative indexed category $\left(\catC,\iterationn,\catL,\folding{-,-}\right)$.
\end{theorem}
\begin{proof}
	Since our definition of iteration is free of equations, we only need to see that, indeed, $\left( \iterationn f , \folding{f, f'}\right)$
	is a morphism $\left( A, X\right) \to \left( B, Y\right) $ in $\displaystyle\Sigma_{\catC } \catL ^\op$, which follows directly from the definitions.
\end{proof}

\begin{definition}[Raw container iteration]\label{def:syntactic-free-container-iteration}
	In the setting of Theorem \ref{th:iterative-Grothendieck-iteration-free-container}, the iteration
	$\iterationg  $ is called the \textit{raw container iteration}.
\end{definition}

\subsection{Iterative Freyd  indexed categories}
We now define an indexed Freyd category that suits our setting. More structured versions of this definition are certainly possible, but we choose the simpler formulation below to avoid overshadowing the paper’s main focus. The reader is invited to take a look, for instance, at Section~\ref{sec:target-language} for more details on the categorical semantics of the target language.

\begin{definition}[Iterative Freyd indexed categories]\label{def:iterative-Feyd-indexed-category}
	An iterative Freyd indexed category consists of a compatible action on dependent data and a tuple  $\left(\catL, \catV , \catC , j,   \otimes , \iterationn , \folding{-,-} \right) $ where $\left(\catC,\iterationn,\catL,\folding{-,-}\right)$  is an iterative indexed category such that
	$\catL $ has indexed finite coproducts, and
	$\left( \catV , \catC , j,  \otimes  \right) $ is a (distributive) Freyd category.
\end{definition}

\begin{theorem}\label{theo:iterative-Feyd-indexed-category}
	Let  $\left(\catL, \catV , \catC , j,   \otimes , \iterationn , \folding{-,-} \right) $  be an iterative Freyd indexed category. The op-Grothendieck constructions on $\catL $ and on $\catL\circ j^\op : \catV ^\op \to \CAT $
	induce an iterative Freyd category
	\begin{equation}
	\Sigma^\op{\left( \catL, \catV , \catC , j,   \otimes , \iterationn , \folding{-,-} \right)  }  \defeqq 	\left(  \displaystyle\Sigma_{\catV } \left( \catL\circ j^\op\right)  ^\op ,       \displaystyle\Sigma_{\catC } \catL ^\op , j_{\Sigma},  \otimesg, \iterationg   \right)
	\end{equation}
where  $\left(  \displaystyle\Sigma_{\catC } \catL ^\op , \iterationg  \right)$
is the iterative category as defined in Eq.~\eqref{eq:iterative-Grothendieckconstruction},
$j_\Sigma \left( A, X\right) \defeqq \left( j(A), X\right)$, and
\begin{equation}
	\begin{aligned}
(f,f') \otimesg (g,g')&=\bigl(f\otimes g,\\
&\quad(\catL(j(\proj A))(f')\circ\delta_{A,g,\catL(j(f))(Y)})\sqcup\catL(j(\proj C))(g')\bigr)
\end{aligned}  .
\end{equation}
\end{theorem}
\begin{proof}
	$\displaystyle\Sigma_{\catV } \left( \catL\circ j^\op\right)  ^\op$ is indeed
	a distributive category: products follow from the indexed finite coproducts, and coproducts from Theorem~\ref{theo:L-finite-coproduct-extensive-indexed-category}. Under the extensive comparison, the distributivity map is the base distributivity isomorphism with identity fibre components. The compatible action on dependent data gives $\otimesg$ by Eq.~\eqref{eq:indexed-context-action}. Theorem~\ref{th:iterative-Grothendieck-iteration-free-container} supplies the iteration from the chosen folders. Together these give the asserted iterative Freyd category.
\end{proof}

\subsection{Iteration-extensive indexed categories are iterative}\label{subsection:iterative-freyd-categories-are-everywhere}

We end this section observing that, indeed, the concrete semantics yields an iterative (Freyd) indexed category and, hence, it fits the framework established in Section~\ref{syn:container-folder-iteration}. We start by observing that every concept of indexed categories we introduced that yields iterative categories through the op-Grothendieck construction naturally provides examples of iterative indexed categories. More precisely:

\begin{lemma}\label{lem:bijection-iterative-indexed-categories-Grothendieck}
	Let $\left( \catC , \iterationn   \right) $ be an iterative category and $\catL : \catC ^\op \to\CAT$ a finite-coproduct-extensive indexed category. There is a bijection between fibred iterations for the
	triple $\left( \catC , \iterationn , \catL    \right) $ and iterative indexed categories  $	\left(\catC ,  \iterationn, \catL,  \folding{-,-} \right)$.
\end{lemma}
\begin{proof}
Every iterative indexed category yields a fibred iteration for the triple $\left( \catC , \iterationn , \catL    \right) $ by Theorem \ref{th:iterative-Grothendieck-iteration-free-container}.
Reciprocally, if $\underline{\iterationn} $ is a fibred iteration for  $\left( \catC , \iterationn , \catL    \right) $, then we can define the following. For each morphism
$$\left( f , f'\right) : \left( A, X\right) \to \left( B, Y \right)\sqcup \left( A, X \right)\cong \left( B\sqcup A, \cISO ^{\left( B, A\right) }\left( Y, X\right)\right)   $$
in $\displaystyle\Sigma_{\catC } \catL ^\op$,
by denoting $\left( g_{\iterationn (f, f') }, g'_{\iterationn (f, f') }\right)\defeqq \underline{\iterationn}\left( f, f'\right)  $, we establish
\begin{equation}
	\folding{f, f'} \defeqq  g'_{\iterationn (f, f')}.
\end{equation}
This yields an iterative indexed category (by Remark \ref{rem:equivalent-definition-of-iterative-indexed-category}).
It is clear to see that the constructions above are inverse to each other.
\end{proof}

\begin{theorem}\label{the:obvious-syntactic-underlying}
	Every iteration-extensive indexed category $\left(\catC,\iterationn,\catL\right)$ has an underlying iterative indexed category by defining, for each morphism
	$$\left( f , f'\right) : \left( A, X\right) \to \left( B, Y \right)\sqcup \left( A, X \right)\cong \left( B\sqcup A, \cISO ^{\left( B, A\right) }\left( Y, X\right)\right)   $$
	in $\displaystyle\Sigma_{\catC } \catL ^\op$,
	\begin{equation}
		\folding{f, f'} \defeqq  \foldd{\left( \catL (f)\circ \cISO ^{\left( B, A \right)} \right) ,  Y , \left( X , f'\right)  }\circ\theta_{f,Y}.
	\end{equation}
	Moreover, the container iteration $\hat{\iterationn }$ (Def.~\ref{def:op-Grothendieck-construction-of-iteration-extensive-categry}) induced by the iteration-extensive indexed category $\left(\catC,\iterationn,\catL\right)$
	coincides with the raw container iteration $\iterationg$ (Def.~\ref{def:syntactic-free-container-iteration}) induced by the underlying iterative indexed category.
\end{theorem}
\begin{proof}
	The proof of the first statement is straightforward, as it only consists of verifying that, indeed, any iteration in the Grothendieck construction of a finite-coproduct-extensive indexed category would induce an iterative indexed category.

	The second statement follows directly from the respective definitions.
\end{proof}

\begin{corollary}
	Every $\wCpo$-finite-coproduct-extensive indexed category has an underlying iterative indexed category. Moreover, the $\wCpo$-container iterations coincide with the raw container iterations induced by the underlying iterative indexed category.
\end{corollary}
\begin{proof}
	We can easily check this fact directly, or we can conclude the result by Theorem \ref{the:obvious-syntactic-underlying} since every $\wCpo$-finite-coproduct-extensive indexed category has an underlying iteration-extensive indexed category, and the container iteration coincides with the $\wCpo$-container iteration.
\end{proof}

We can, then, extend the observations above to Freyd categorical versions. More precisely:

\begin{corollary}
	Every iteration-extensive Freyd indexed category
	$\left( \catV , \catC , j,   \otimes , \iterationn , \catL \right)$ has an underlying iterative Freyd indexed category $\left(\catV , \catC , j,   \otimes , \iterationn , \catL,  \folding{-,-} \right) $,
	The induced iterative Freyd categories by the op-Grothendieck constructions (as established in Theorem \ref{theo:Freyd-goes-to-Freyd} and \ref{theo:iterative-Feyd-indexed-category}) coincide.
\end{corollary}

\begin{corollary}
		Every $\ww$-Freyd indexed category has an underlying iterative Freyd indexed category.
\end{corollary}

\begin{remark}[Biadditive types]
We can add one aspect to the categorical semantics of the fragment of the target language we care about; namely, it is biadditive. More precisely:
\begin{definition}[Biadditive iterative Freyd indexed categories]
	A \textit{biadditive iterative Freyd indexed category} is an iterative Freyd indexed category  $\left( \catV , \catC , j,   \otimes , \iterationn , \catL, \folding{-,-} \right) $ where
    $\catL $ has indexed biproducts.
\end{definition}
\end{remark}

\subsection{\texorpdfstring{\for{toc}{Concrete target semantics as an op-Grothendieck construction}\except{toc}{The concrete semantics of our target language: $\PFam{\Vect^\op}$ as an op-Grothendieck construction}}{Concrete target semantics as an op-Grothendieck construction}}\label{subsect:revisiting-iteration-PFAMVECT}
We now revisit iteration in the concrete semantics of our target language drawing on the concepts introduced in this section.
We start by recalling that, since \(\PFam{\Vect^\op}\) is an \(\wCpo\)-enriched category with finite $\wCpo$-coproducts, Theorem~\ref{theo:wCpoenrichedcategory-underlyingiterative} establishes that it forms an underlying iterative category, denoted by $\left( \PFam{\Vect^\op}, \iterationn\right) $.
This is ultimately the structure we want to consider in (the op-Grothendieck construction of) our concrete model.

 In order to frame this iteration in our established setting, we start by observing that
the indexed category
\begin{equation}\label{eq:Vect_leastelement}
	\Vect _{\leastelement} ^{\left( -\right) } : \catPSet^\op \to\CAT ,
\end{equation}
as defined in Equation \eqref{def:Vectpartial}, satisfies the following:
\begin{theorem}\label{theo:concrente-model-still}
	The pair $\left( \catPSet, \Vect _{\leastelement} ^{\left( -\right) } \right) $ is an $\wCpo$-finite-coproduct extensive indexed category. Moreover,
	we have an $\wCpo$-enriched isomorphism
	\begin{equation} \label{eq:view-op-Grothendieck-construction-partial}
		\displaystyle \PFam{\Vect^\op} \cong \displaystyle\Sigma^\op\left( \catPSet, \Vect _{\leastelement} ^{\left( -\right) } \right)
	\end{equation}
between the op-Grothendieck construction $\Sigma^\op\left( \catPSet, \Vect _{\leastelement} ^{\left( -\right) } \right) $, as established in Definition~\ref{def:op-Grothendieck-enrichment}, and  $\PFam{\Vect^\op}$, as established in Subsect.~\ref{subsect:concrete-semantics-FamVectop} (more precisely, Subsect.~\ref{def:wcpoFAMVECTop}).

In particular, the $\wCpo$-container iteration induced by   $\left( \catPSet, \Vect _{\leastelement} ^{\left( -\right) } \right) $  is the same as the iteration induced by the $\wCpo$-enrichment in
$\PFam{\Vect^\op} $ as established in Subsect.~\ref{sub:concrete-chad-omega-freyd}.
\end{theorem}
\begin{proof}
This is a particular case of Theorem \ref{theo:this-proves-the-case-of-our-concrete-model}, while the isomorphism is easily obtained by observing that Subsect.~\ref{subsect:opGrothendieck-construction-PFAM} indeed is an $\wCpo$-enriched isomorphism.
\end{proof}

The compatible action in this concrete case is worth making explicit. For a set $A$, a partial function $g:C\rightharpoonup D$, and a family $U\in\Vect_\bot^{A}$, the component of $\delta_{A,g,U}$ at $(a,c)$ is
\[
\begin{cases}
\ID_{U(a)}:U(a)\to U(a),&\text{if }g(c)\text{ is defined},\\
0:0\to U(a),&\text{otherwise}.
\end{cases}
\]
These are the components induced by the inequality between the partial first projection after $\ID_A\otimes g$ and the total first projection. The $2$-functoriality of $\Vect_\bot^{(-)}$ gives naturality and compatibility with composition. Formula~\eqref{eq:indexed-context-action} then agrees with the ordinary action on partial families: where $g$ is defined it takes the coproduct of the two fibre maps, and elsewhere its domain fibre is zero. This identifies the action with the one on $\PFam{\Vect^\op}$; its unit, associativity, distributivity, and continuity follow from the corresponding operations on partial families. Thus the compatible enriched action required above is present in the concrete semantics.

\begin{definition}[Concrete Freyd indexed category]\label{def:the-real-semantics-of-the-target-language}
 We can establish the $\ww$-Freyd indexed category $\concINDEX \defeqq \left( \Set,\catPSet , j, \otimes , \Vect _{\leastelement} ^{\left( -\right) } \right) $ where $\left( \Set,\catPSet , j, \otimes \right) $ is the usual $\ww$-Freyd category, and  $\left( \catPSet, \Vect _{\leastelement} ^{\left( -\right) } \right) $ is the  $\wCpo$-finite-coproduct extensive indexed category established in Theorem \ref{theo:concrente-model-still}.
\end{definition}

\begin{corollary}
	The $\ww$-Freyd category $\Sigma^\op\left( \concINDEX\right) $ obtained by the op-Grothendieck construction of the $\ww$-Freyd indexed category  $\left( \Set,\catPSet , j, \otimes , \Vect _{\leastelement} ^{\left( -\right) } \right) $ (as established in Corollary~\ref{cor:fundamental-for-concrete-semantics-Iguess}) is isomorphic to the $\ww$-Freyd category $\FFFVECT$, as defined in Theorem \ref{theo:ww-Freyd-category-Fam} (Eq.~\refeq{eq:FFVECT}).
\end{corollary}

\part{Iterative CHAD}\label{part:iterative-chad}

We now return to the original problem: transforming a program with loops into a program that returns its primal result together with its backward map. The source program uses variants to decide whether a loop returns or continues. The target program follows this computation and uses the folders developed in Part~\ref{part:iteration} for the backward component, with its dependent cotangent types.

We use the preceding categorical development to obtain this program transformation. Sections~\ref{sec:source-language} and~\ref{sec:target-language} give the source and target languages, including the typing and equational rules used in their categorical interpretations. We then specify their concrete semantics and derive iterative CHAD from the universal property of the source language in Section~\ref{sec:correctness-iterative-CHAD}. The correctness proof follows the same principle as in the total setting: translating a program and then interpreting it agrees with first interpreting it and then taking its CHAD-derivative, now also for programs containing loops.

\section{Source language as a standard call-by-value language}\label{sec:source-language}
\label{ap:source-language}
In this section, we present a basic call-by-value language that serves as the foundation for our source language. Depending on the semantics, we may allow non-termination (resulting in a partial language) -- with the evaluation strategy being call-by-value -- or disallow it (resulting in a total language where evaluation strategy does not matter). We give the syntax explicitly so that the categorical constructions of Part II can now be read as operations on programs.

Our language is built upon ground types $\reals ^n$  ($n\in\NN$), with sets of primitive operations $\op\in\Op _{\mathbf{m}}^\mathbf{n} $, for each $(k_1, k_2)\in \NN\times \NN $ and each $\left( \mathbf{n}, \mathbf{m} \right) = \left( \left( n_1, \ldots , n_{k_1} \right), \left( m_1, \ldots , m_{k_2} \right)\right) \in \NN ^{k_1}\times\NN ^{k_2}  $.
The type $\reals^n$ is a ground type for a fixed-length array. Its interpretation is isomorphic to the product of $n$ copies of $\RR$, but we keep arrays and products syntactically distinct so that array primitives can be specified directly.
The types $\ty{1},\ty{2},\ty{3}$, and computations $\trm{1},\trm{2},\trm{3}$ of our language are as follows.\\
\begin{figure}[!ht]
	\framebox{\begin{minipage}{0.98\linewidth}
\small\begin{syntax}
    \ty{1}, \ty{2}, \ty{3} & \gdefinedby & & \syncat{types}                          \\
    &\gor& \reals ^n                     & \synname{real arrays}\\
    &&&\\
    \trm{1}, \trm{2}, \trm{3} & \gdefinedby & & \syncat{computations}                          \\
    &\gor & \op(\trm{1}_1,\ldots,\trm{1}_n) & \synname{operation}\\
    &\gor& \var{1},\var{2},\var{3}                      & \synname{variables}\\
    &\gor & \letin{\var{1}}{\trm{1}}{\trm{2}} & \synname{sequencing}\\
	&\gor& \tIni[1]{\trm{1}} \gor\cdots\gor   \tIni[n]{\trm{1}} & \synname{sum inclusions}\\
  \end{syntax}%
~
 \begin{syntax}
	&\gor\quad\, & \Unit  \gor  \ty{1}_1 \t*\cdots\t* \ty{1}_n & \synname{products}\\
    &\gor & \Init  \gor \ty{1}_1 \t+\cdots\t+ \ty{1}_n  & \synname{sums}\\
  &&&\\
  &\gor & \lrvMatch{\trm{3}}{\begin{array}{l}\;\;\tIni[1]\var{1}_1\To\trm{1}_1\\
    \mid \cdots\\
\gor \tIni[n]\var{1}_n\To \trm{1}_n\end{array}}\hspace{-15pt}\; & \synname{sum match}\\
	&\gor\quad\, &  \tTuple{\trm{1}_1,\ldots,\trm{1}_n} & \synname{tuples}\\
	&\gor\quad\, & \tMatch{\trm{2}}{\var{1}_1,\ldots,\var{1}_n}{\trm{1}}\hspace{-10pt}\; & \synname{prod. match}\\
	&\gor&\tIt{\var{1}}{\trm{1}}\hspace{-10pt}\; & \synname{iteration}\\
\end{syntax}
\end{minipage}}
\caption{\label{fig:sl-terms-types-kinds} A grammar for the types and terms of the source language.}
\end{figure}\\
We use the following syntactic sugar for the $i$-th projection out of a tuple:
$$
\tPrj[i]\trm{1}\defeq \tMatch{\trm{1}}{\var{1}_1,\ldots,\var{1}_n}{\var{1}_i}.
$$

The computations of our source language are typed according to the rules of Fig.~\ref{fig:types1}.
We consider the standard call-by-value $\beta\eta$-equational theory of \cite{moggi1988computational}
for our language, including the usual sequencing and commuting equations, which we list in Fig. \ref{fig:beta-eta1}. The equations preserve call-by-value evaluation: substituting a value is harmless, whereas a computation must be sequenced before its result is used. In particular, the unit and associativity equations for let-bindings are part of the equational theory.
To present this equational theory, we distinguish a subset of computations that we call \emph{(complex) values} $\val{1},\val{2},\val{3}$, in the sense of \cite[Chapter~3]{levy2012call}, to consist of those computations built without iteration, using only primitives designated as total.
We fix the designated total primitives when specifying the language; numerical constants can be included among them.

\begin{figure}[!ht]
	\fbox{\parbox{0.98\linewidth}{\begin{minipage}{\linewidth}\noindent\[
  \begin{array}{c}
   \inferrule{
	\set{\Ginf {\trm{1}_j} \reals ^ {n_j}}_{j=1}^{k_1}\quad \left((k_1,k_2)\in\NN^2, \, \left( \mathbf{n}, \mathbf{m} \right)\in \NN ^{k_1}\times \NN^{k_2},  \,\op\in\Op^{\mathbf{n}}_{\mathbf{m}}\right)
}{
	\Ginf {\op(\trm{1}_1,\ldots,\trm{1}_{k_1})}{\reals ^{m_1}\t+ \cdots\t+  \reals ^{m_{k_2}} }
}\\ \\
\inferrule{
  ((\var{1} : \ty{1}) \in \ctx)
}{
  \Ginf {\var{1}}{\ty{1}}
}
\quad
\inferrule{
  \Ginf{\trm{1}}{\ty{2}}\quad
  \Ginf[,\var{1}:\ty{2}]{\trm{2}}{\ty{1}}
}{
  \Ginf{\letin{\var{1}}{\trm{1}}{\trm{2}}}{\ty{1}}
}\quad
   \inferrule{\Ginf {\trm{1}} {\ty{1}_i}}
   {\Ginf{\tIni[i]\trm{1}}\ty{1}_1\t+\cdots \t+\ty{1}_n}(i=1,\ldots,n)
   \\\\
   \inferrule{\Ginf{\trm{3}}{\ty{2}_1\t+\cdots\t+\ty{2}_n}\quad
   \Ginf[,\var{1}_1:\ty{2}_1]{\trm{1}_1}{\ty{1}}\quad\cdots\quad
   \Ginf[,\var{1}_n:\ty{2}_n]{\trm{1}_n}{\ty{1}}
   }{\Ginf{\vMatch{\trm{3}}{\tIni[1]\var{1}_1\To\trm{1}_1\mid\cdots\mid \tIni[n]\var{1}_n\To \trm{1}_n}}{\ty{1}}}
  \\
  \\
   \inferrule{\Ginf {\trm{1}_1} {\ty{1}_1}\quad\cdots\quad \Ginf {\trm{1}_n} {\ty{1}_n}}
   {\Ginf {\tTuple{\trm{1}_1,\ldots,\trm{1}_n}} {\ty{1}_1\t*\cdots\t*\ty{1}_n}}
   \quad
   \inferrule{
   \Ginf {\trm{3}}{\ty{1}_1\t*\cdots\t*\ty{1}_n}\quad
   \Ginf[{,\var{1}_1:\ty{1}_1,\ldots,\var{1}_n:\ty{1}_n}] {\trm{1}}{\ty{2}}}
   {\Ginf{\tMatch{\trm{3}}{\var{1}_1,\ldots,\var{1}_n}{\trm{1}}}{\ty{2}}}
\end{array}
\]
\end{minipage}}}
	\caption{Typing rules for our standard call-by-value language constructed over the ground types $\reals ^n $ and the primitive operations in $\Op$.
		\label{fig:types1}}
\end{figure}

\begin{figure}[!ht]
	\fbox{\parbox{0.98\linewidth}{\begin{minipage}{\linewidth}\noindent
				\[
\begin{array}{l}
    \inferrule{\freeGinf[\var{1}:{\ty{2}}] {\trm{1}}{\ty{1}\t+\ty{2}}
    }{\freeGinf[\var{1}:{\ty{2}}] {\tIt{\var{1}}{\trm{1}}}{\ty{1}}}
\end{array}
\]\end{minipage}}}
	\caption{Typing rules for iteration on top of our standard call-by-value language.}
\end{figure}

\begin{figure}[!ht]
	\fbox{\parbox{0.97\linewidth}{\scalebox{0.85}{\begin{minipage}{\linewidth}\noindent\[
\begin{array}{l}
    \letin{\var{1}}{\val{1}}{\trm{1}}=\subst{\trm{1}}{\sfor{\var{1}}{\val{1}}}
    \qquad \letin{\var{1}}{\trm{1}}{\var{1}}=\trm{1}
    \\[4pt]
    \letin{\var{2}}{(\letin{\var{1}}{\trm{1}}{\trm{2}})}{\trm{3}}
    =\letin{\var{1}}{\trm{1}}{(\letin{\var{2}}{\trm{2}}{\trm{3}})}\\
    \vMatch{\tIni[i]\trm{2}}{\tIni[1]\var{1}_1\To\trm{1}_1\mid\cdots\mid\tIni[n]\var{1}_n\To\trm{1}_n}=
    \letin{\var{1}_i}{\trm{2}}{\trm{1}_i} \\[4pt]
    \letin{\var{3}}{\trm{2}}{\trm{1}}\freeeq{\var{1}_1,\ldots,\var{1}_n}\lrvMatch{\trm{2}}
    { \begin{array}{l}\;\,\,\tIni[1]\var{1}_1\To\subst{\trm{1}}{\sfor{\var{3}}{\tIni[1]\var{1}_1}}\gor\cdots \\ \gor \tIni[n]\var{1}_n\To\subst{\trm{1}}{\sfor{\var{3}}{\tIni[n]\var{1}_n}}\end{array}
    }
    \\
    \tMatch{\tTuple{\val{1}_1,\ldots,\val{1}_n}}{\var{1}_1,\ldots,\var{1}_n}{\trm{1}}=\subst{\trm{1}}{\sfor{\var{1}_1}{\val{1}_1},\ldots,\sfor{\var{1}_n}{\val{1}_n}} \\[4pt]
    \subst{\trm{1}}{\sfor{\var{3}}{\val{1}}}\freeeq{\var{1}_1,\ldots,\var{1}_n}\tMatch{\val{1}}{\var{1}_1,\ldots,\var{1}_n}{\subst{\trm{1}}{\sfor{\var{3}}{\tTuple{\var{1}_1,\ldots,\var{1}_n}}}} \\
\letin{\var{1}_1}{\trm{1}_1}{\cdots\letin{\var{1}_n}{\trm{1}_n}{\tTuple{\var{1}_1,\ldots,\var{1}_n}}}\freeeq{\var{1}_1,\ldots,\var{1}_n}\tTuple{\trm{1}_1,\ldots,\trm{1}_n}\\
\letin{\var{1}}{\trm{1}}{\tMatch{\var{1}}{\var{2}_1,\ldots,\var{2}_n}{\trm{3}}} \freeeq{\var{1}} \tMatch{\trm{1}}{\var{2}_1,\ldots,\var{2}_n}{\trm{3}}\\
\letin{\var{1}_1}{\trm{1}_1}{\cdots\letin{\var{1}_n}{\trm{1}_n}{\op({\var{1}_1,\ldots,\var{1}_n})}}\freeeq{\var{1}_1,\ldots,\var{1}_n}\op({\trm{1}_1,\ldots,\trm{1}_n})
\end{array}
\]
\end{minipage}}}
	}\caption{\label{fig:beta-eta1} Basic $\beta\eta$-equational theory for  our language.
		We write $\freeeq{\var{1}_1,\ldots,\var{1}_n}$ to indicate that the variables are fresh in the left hand side.
		In the associativity rule for let-bindings, $\var{1}$ may not be free in $\trm{3}$.
		Equations hold on pairs of computations of the same type.
		Note that we do not include equations for iteration, as they are not needed for our development.}
\end{figure}

Observe that a more general iteration construct with typing rule
\[
\frac{\Gamma,\var{1}:\ty{2}\vdash \trm{1}:\ty{1}\t+\ty{2}}{
	\Gamma,\var{1}:\ty{2}\vdash \tIt[\Gamma]{\var{1}}{\trm{1}}:\ty{1}
}
\]
 is derivable. Indeed, for $\Gamma=\var{2}_1:\ty{3}_1,\ldots,\var{2}_n:\ty{3}_n$, we define $\tIt[\Gamma]{\var{1}}{\trm{1}}$ as
\[
\begin{array}{l}
\letin{\var{3}}{\tTuple{\var{2}_1,\ldots,\var{2}_n,\var{1}}}{}\\
\tIt{\var{3}}{\tMatch{\var{3}}{\var{2}_1,\ldots,\var{2}_n,\var{1}}{\vMatch{\trm{1}}{\tIni[1]u\To \tIni[1]u\mid \tIni[2]v\To \tIni[2]\tTuple{\var{2}_1,\ldots,\var{2}_n,v}}}}
\end{array}
\]

The left summand contains a completed result, while the right summand contains the next state. The construction above carries the unchanged context alongside that state. Thus a loop with parameters needs no additional primitive iteration construct.

Following the same viewpoint as in \cite{vakar2020denotational, 2022arXiv221007724L, 2022arXiv221008530L}, we do not impose additional equations on the iteration construct, unlike in \cite{bloom1993iteration, goncharov2015unguarded}. This leads us to what we refer to as \textit{raw iteration}. We have chosen to work at this level of generality because no equations are needed to establish the correctness proof of CHAD.

We can think of this syntax as the freely generated iterative Freyd category with
$\Syn = \left( \SynV , \SynC, \SynJ , \SynTT, \Synit \right) $
on our total and partial operations $\op$.
Concretely,
\begin{itemize}
	\item $\SynV$ and $\SynC$ both have types $\ty{1},\ty{2},\ty{3},\ldots$ as objects;
	\item morphisms $\ty{1}\to\ty{2}$ are (complex) values $\var{1}:\ty{1}\vdash \val{1}:\ty{2}$ (in the case of $\SynV$) or computations $\var{1}:\ty{1}\vdash \trm{1}:\ty{2}$ (in the case of $\SynC$) modulo $\alpha$-renaming of bound variables and $\beta\eta$-equivalence;
	\item identities are the equivalence class $[\var{1}:\ty{1}\vdash \var{1}:\ty{1}]$;
	\item composition of $[\var{1}:\ty{1}\vdash \trm{1}:\ty{2}]$ and $[\var{2}:\ty{2}\vdash \trm{2}:\ty{3}]$ is
	given by $[\var{1}:\ty{1}\vdash \letin{\var{2}}{\trm{1}}{\trm{2}}:\ty{3}]$;
	\item $\SynJ$ is the inclusion of complex values in the larger set of computations;
	\item $\SynTT([\var{1}:\ty{1}\vdash \val{1}:\ty{1}'], [\var{2}:\ty{2}\vdash \trm{1}:\ty{2}'])=
	[\var{3}:\ty{1}\t* \ty{2}\vdash \tMatch{\var{3}}{\var{1},\var{2}}{\tTuple{\val{1},\trm{1}}}:\ty{1}'\t* \ty{2}']$;
	\item $\Synit([\var{1}:\ty{2}\vdash \trm{1}:\ty{1}\t+ \ty{2}])=[\var{1}:\ty{2}\vdash\tIt{\var{1}}{\trm{1}}:\ty{1}]$.
\end{itemize}
$\Syn$ has the following universal property:
\begin{quote}
	Given any \iterativeF $\left( \catV , \catC , j, \otimes , \iterationn\right)   $, for each pair
	$	\left( K , \mathtt{h}   \right) $
	where $K = \left(  K_ n \right) _ {n\in\NN} $ is a family of objects in $\catC $ and
	$\mathtt{h} = \left( h_\op \right) _{\op\in\Op} $ is a family of morphisms of the prescribed types
	$h_\op:K_{n_1}\times\cdots\times K_{n_{k_1}}\to K_{m_1}\sqcup\cdots\sqcup K_{m_{k_2}}$
	in $\catC$ (with a chosen value morphism $h^v_\op$ satisfying $j(h^v_\op)=h_\op$ for each designated total primitive), there is a unique \iterativeF morphism
	\begin{equation*}
		\left( H, \hat{H}\right)  : \Syn \to \left( \catV , \catC , j, \otimes , \iterationn\right)
	\end{equation*}
	such that $\hat{H} (\reals ^n ) = K _n $ and $\hat{H} (\op ) = h_\op $ for any $n\in\NN$ and any primitive operation $\op\in\Op$, and $H(\op)=h^v_\op$ for each designated total primitive.
\end{quote}

\section{Linear \texorpdfstring{$\lambda$}{lambda}-calculus as the target language}\label{sec:target-language}
\providecommand{\csubst}[2]{#1\langle #2\rangle}
\providecommand{\tContext}[2]{\mathbf{ctx}_{#1}(#2)}

We describe the target language for our AD transformation. A transformed program computes its original result and a linear map that sends a cotangent at that result back to a cotangent at the input. The linear types therefore depend on Cartesian values: for a variant type, the selected branch determines the cotangent type. The new ingredient for iteration is the folder that composes these linear maps along a terminating loop. Dependent pairs and linear-function types allow us to return the primal and its backward map together. The indexed categorical presentation records the two components of these target programs.
Cartesian contexts are dependent telescopes. Cartesian dependent types admit ordinary value substitution $\subst{\ty{2}}{\sfor{\var{1}}{\val{1}}}$, which acts throughout the subsequent context. Linear types also admit reindexing along computations, written $\csubst{\cty{1}}{\sfor{\var{1}}{\trm{1}}}$; this is a formal type constructor, not capture-avoiding substitution. For example, reindexing the constant family $\RR$ along a nowhere-defined computation $\Omega:\Unit\to\Unit$ in $\Vect_{\leastelement}^{(-)}$ gives the zero family. Thus computational reindexing retains the computation even when $\var{1}$ does not occur in $\cty{1}$.

For a computation in a context $\Gamma$, computational reindexing includes the context: it is reindexing along the computation that returns the unchanged context together with the result. For values it agrees with ordinary substitution. Computational reindexing has the identity, composition, functoriality, and biproduct equations of a strict indexed category. These equations, coproduct extensivity, and the coherent context comparisons of Eq.~\eqref{eq:indexed-context-comparison} are part of the syntax. We write $\tContext{\var{1}\leftarrow\trm{1}}{\lvar}$ for the corresponding map
$\csubst{\cty{1}}{\sfor{\var{1}}{\trm{1}}}\to\cty{1}$ when $\cty{1}$ is a type in the unchanged context. Its concrete interpretation is the identity where the computation returns and the unique map from the zero object otherwise. This context map is obtained from $\delta_{\Gamma,\trm{1}}$ by reindexing along the value diagonal $\Gamma\to\Gamma\times\Gamma$. The comparisons satisfy the unit, composition, and distributive action coherences of Eq.~\eqref{eq:indexed-context-action}.

The formation rule and equations for computational reindexing appear in Figures~\ref{fig:tl-type-system1} and~\ref{fig:tl-equations}. In particular, it agrees with ordinary substitution on values, both on linear types and on linear terms. Cartesian arguments of primitive linear operations, and the function in a linear application, are values.

Figure~\ref{fig:tl-type-system2} gives dependent elimination for Cartesian sums and pairs: each pattern substitutes through the remaining context and result type. In the target they replace the nondependent elimination rules of Figure~\ref{fig:types1}; those rules follow by sequencing and value substitution. For Cartesian computations, matching a computation abbreviates sequencing followed by matching its returned value; the overall result type must be well formed before that value is bound. For instance,
\[
\tMatch{\trm{1}}{\var{1},\var{2}}{\trm{2}}
\defeq
\letin{\var{3}}{\trm{1}}{\tMatch{\var{3}}{\var{1},\var{2}}{\trm{2}}},
\]
with $\var{3}$ fresh. Sum matching is treated in the same way. All displayed contexts and types are assumed well formed, and value substitution acts on the whole judgement.

The terms and types are presented in Figure \ref{fig:tl-terms-types-kinds}.
We shade out the constructors $\tRoll\trm{1}$ for inductive types, as we only need the corresponding eliminators, our \foldersname{}.
A stronger presentation may include roll constructors with the computation and uniqueness rules for initial algebras.

\begin{figure}[!ht]
	\framebox{\begin{minipage}{0.98\linewidth}
			\begin{syntax}
    \cty{1}, \cty{2}, \cty{3} & \gdefinedby & &  \syncat{linear types}\\
  &\gor & \creals^n & \synname{real array}\\
  & \gor & \lUnit & \synname{linear zero type}\\
  & \gor & \cty{1}_1\t*\cdots \t* \cty{1}_n & \synname{linear biproduct}\\
  & \gor & \csubst{\cty{1}}{\sfor{\var{1}}{\trm{1}}} & \synname{computational reindexing}\\
  & \gor & \vMatch{\trm{1}}{\tIni[1]\var{1}_1\To\cty{1}_1\mid\cdots\mid \tIni[n] \var{1}_n\To\cty{1}_n} & \synname{case distinction}\\
  &&&\\
  \ty{1}, \ty{2}, \ty{3} & \gdefinedby & & \syncat{Cartesian types}               \\
  && \ldots                      & \synname{as in Fig. \ref{fig:sl-terms-types-kinds}}\\
  &\gor\quad\, & \cty{1}\multimap \cty{2} & \synname{linear function}\\
  &\gor & \Sigma \var{1}:\ty{1} .\ty{2} & \synname{dependent pair}\\
&&&\\
  \trm{1}, \trm{2}, \trm{3} & \gdefinedby  && \syncat{computations}             \\
&& \ldots                      & \synname{as in Fig. \ref{fig:sl-terms-types-kinds}} \\
&\gor & \lop(\trm{1}_1,\ldots,\trm{1}_k;\trm{2}) & \synname{linear operation}\\
& \gor & \tContext{\var{1}\leftarrow\trm{1}}{\lvar} & \synname{context comparison}\\
& \gor & \lvar & \synname{linear identifier}\\
& \gor & \letin{\lvar}{\trm{1}}{\trm{2}} & \synname{linear let-binding}\\
& \gor & \tPrj[i]\trm{1} & \synname{projection}\\
&\gor & \lfun{\lvar}{\trm{1}}\,\gor  \lapp{\trm{1}}{\trm{2}}  & \synname{abstraction/application}\\
&\gor & \tZero\,\gor \trm{1}+\trm{2} & \synname{monoid structure}\\
&\color{gray}\gor & \color{gray}\tRoll\trm{1} & \color{gray}\synname{inductive type introduction}\\
&\gor & \tFold{\trm{1}}{\lvar}{\trm{2}} & \synname{inductive type elimination}
\end{syntax}

	\end{minipage}}
    \caption{\label{fig:tl-terms-types-kinds} A grammar for the types and terms of the target language,
    extending that of Fig. \ref{fig:sl-terms-types-kinds}.
    }
\end{figure}

These terms are typed according to the rules of Figures \ref{fig:tl-type-system1} and \ref{fig:tl-type-system2}.
In a judgement $\Gamma;\lvar:\cty{1}\vdash\trm{1}:\cty{2}$, the semicolon separates the Cartesian context $\Gamma$ from the single linear identifier $\lvar$. The Cartesian variables specify the point at which we differentiate and may occur in the types. The linear identifier carries a cotangent, and terms must preserve its zero and addition. Here linearity means additivity, rather than a requirement that the identifier occur exactly once.
We think of Cartesian types as denoting (families of) sets and of linear types as denoting (families of) commutative monoids. Linear abstractions are Cartesian values. In particular, a linear abstraction containing a computation is a value, whereas evaluating a computation of linear-function type may diverge. The function $\eta$-equation below is consequently restricted to values.
\begin{figure}[!ht]
	\framebox{\begin{minipage}{0.98\linewidth}\noindent\hspace{-24pt}\[
  \begin{array}{c}
    \inferrule{\Gamma\vdash\trm{1}:\ty{1}\quad
      \Gamma,\var{1}:\ty{1}\vdash\cty{1}:\ltype}
    {\Gamma\vdash\csubst{\cty{1}}{\sfor{\var{1}}{\trm{1}}}:\ltype}
    \qquad
    \inferrule{\Gamma\vdash\trm{1}:\ty{1}\quad\Gamma\vdash\cty{1}:\ltype}
    {\Gamma;\lvar:\csubst{\cty{1}}{\sfor{\var{1}}{\trm{1}}}\vdash\tContext{\var{1}\leftarrow\trm{1}}{\lvar}:\cty{1}}
    \\[8pt]
    \inferrule{~}{\Ginf[;\lvar:\cty{1}]{\lvar}{\cty{1}}}
    \quad\!
    \inferrule{\Ginf[;\lvar:\cty{1}]{\trm{1}}{\cty{2}}\quad \Ginf[{;\lvar:\cty{2}}]{\trm{2}}{\cty{3}}}
    {\Ginf[;\lvar:\cty{1}]{\letin{\lvar}{\trm{1}}{\trm{2}}}{\cty{3}}}
    \quad \!
    \inferrule{\Ginf{\trm{1}}{\ty{1}}\quad
    \Gamma,\var{1}:\ty{1};\lvar:\cty{2}\vdash \trm{2}:\cty{3}}
    {\Ginf[{;\lvar:\csubst{\cty{2}}{\sfor{\var{1}}{\trm{1}}}}]{\letin{\var{1}}{\trm{1}}{\trm{2}}}{\csubst{\cty{3}}{\sfor{\var{1}}{\trm{1}}}}}
    \\
    \\
    \inferrule{
        \set{\Ginf {\val{1}_i} {\reals^{n_i}}}_{i=1}^k
        \quad
        \Ginf[;\lvar:\cty{1}]{\trm{2}}{\LDomain{\lop}}\quad
         (\lop\in\LOp^{n_1,\ldots,n_k;n'_1,\ldots,n'_l}_{m_1,\ldots,m_r})
      }{
        \Ginf[;\lvar:\cty{1}] {\lop(\val{1}_1,\ldots,\val{1}_k;\trm{2})}{\LCDomain{\lop}}
      }\\
      \\
    \inferrule{\Ginf[;\lvar:\cty{1}]{\trm{1}_1}{\cty{1}_1}\quad \cdots\quad
    \Ginf[;\lvar:\cty{1}]{\trm{1}_n}{\cty{1}_n}
    }
    {
        \Ginf[;\lvar:\cty{1}]{\tTuple{\trm{1}_1,\ldots,\trm{1}_n}}{\cty{1}_1\t*\cdots\t* \cty{1}_n}
    ~
    }
    \quad
    \inferrule{\Ginf[;\lvar:\cty{1}]{\trm{1}}{\cty{2}_1\t*\cdots\t*\cty{2}_n}}{\Ginf[;\lvar:\cty{1}]{\tPrj[i]\trm{1}}{\cty{2}_i}}(i=1,\ldots,n)
    \\[8pt]
\inferrule{\Ginf[;\lvar:\cty{1}]{\trm{1}}{\cty{2}}}{\Ginf{\lfun{\lvar}\trm{1}}{\cty{1}\multimap \cty{2}}} \quad
\inferrule{\Ginf{\val{1}}{\cty{3}\multimap\cty{2}}\quad
\Ginf[;\lvar:\cty{1}]{\trm{2}}{\cty{3}}}{\Ginf[;\lvar:\cty{1}]{\lapp{\val{1}}{\trm{2}}}{\cty{2}}}
\\[8pt]
\inferrule{~}{\Ginf[;\lvar:\cty{1}]{\tZero}{\cty{2}}}\quad
\inferrule{\Ginf[;\lvar:\cty{1}]{\trm{1}}{\cty{2}}\quad\Ginf[;\lvar:\cty{1}]{\trm{2}}{\cty{2}} }
{\Ginf[;\lvar:\cty{1}]{\trm{1}+\trm{2}}{\cty{2}}}\\
\\
\inferrule{
\Ginf{\trm{1}}{\ty{1}_1\t+ \cdots\t+ \ty{1}_n}\quad
\set{{\Gamma,\var{1}_i:\ty{1}_i;\lvar:\subst{\cty{2}}{\sfor{\var{1}}{\tIni[i]\var{1}_i}}}\vdash {\trm{3}_i}:{\subst{\cty{3}}{\sfor{\var{1}}{\tIni[i]\var{1}_i}}}}_{1\leq i\leq n}
}
{
\begin{array}{l}
\Ginf[;\lvar:\csubst{\cty{2}}{\sfor{\var{1}}{\trm{1}}}]{\vMatch{\trm{1}}{\tIni[1]\var{1}_1\To \trm{3}_1\mid\cdots\mid\tIni[n]\var{1}_n\To \trm{3}_n }}{\csubst{\cty{3}}{\sfor{\var{1}}{\trm{1}}}}
\end{array}
}
\end{array}
\]
\end{minipage}}
    \caption{Linear typing and computational reindexing for the target language. The last rule is derived from value case analysis, linear abstraction/application, and computational sequencing; $\Gamma,x:\ty{1}_1\t+\cdots\t+\ty{1}_n\vdash\cty{2},\cty{3}:\ltype$ is understood.\label{fig:tl-type-system1}}\;
\end{figure}
\begin{figure}[!ht]
	\framebox{\begin{minipage}{0.98\linewidth}\noindent\hspace{-24pt}\[
  \begin{array}{c}
    \inferrule{\Ginf{\val{1}}{\ty{1}}\quad
    \Gamma,\var{1}:\ty{1}\vdash \trm{2}:\ty{3}}
    {\Ginf{\letin{\var{1}}{\val{1}}{\trm{2}}}{\subst{\ty{3}}{\sfor{\var{1}}{\val{1}}}}}
    \qquad
    \inferrule{\Ginf{\val{1}}{\ty{1}}\quad
    \Ginf{\trm{2}}{\subst{\ty{2}}{\sfor{\var{1}}{\val{1}}}}}
    {\Ginf{\tPair{\val{1}}{\trm{2}}}{\Sigma\var{1}:\ty{1}.\ty{2}}}
    \\\\
    \inferrule{
      \{\Gamma,\var{1}_i:\ty{1}_i,\subst{\Delta}{\sfor{\var{3}}{\tIni[i]\var{1}_i}}\vdash\trm{1}_i:\subst{\ty{3}}{\sfor{\var{3}}{\tIni[i]\var{1}_i}}\}_{1\leq i\leq n}}
    {\Gamma,\var{3}:\ty{1}_1\t+\cdots\t+\ty{1}_n,\Delta\vdash
      \vMatch{\var{3}}{\tIni[1]\var{1}_1\To\trm{1}_1\mid\cdots\mid\tIni[n]\var{1}_n\To\trm{1}_n}:\ty{3}}
    \\\\
    \inferrule{\Gamma,\var{1}:\ty{1},\var{2}:\ty{2},\subst{\Delta}{\sfor{\var{3}}{\tTuple{\var{1},\var{2}}}}\vdash\trm{1}:\subst{\ty{3}}{\sfor{\var{3}}{\tTuple{\var{1},\var{2}}}}}
    {\Gamma,\var{3}:\Sigma\var{1}:\ty{1}.\ty{2},\Delta\vdash
      \tMatch{\var{3}}{\var{1},\var{2}}{\trm{1}}:\ty{3}}
    \\\\
    \inferrule{
      \Gamma\vdash\trm{1}:\ty{1}\quad
      \Gamma,\var{1}:\ty{1}\vdash\cty{1}:\ltype\quad
      \Gamma\vdash\cty{2}:\ltype\\
      \Gamma;\lvar:\csubst{\cty{1}}{\sfor{\var{1}}{\trm{1}}}\vdash\trm{2}:\cty{2}}
    {\Gamma\vdash\tPair{\trm{1}}{\lfun{\lvar}{\trm{2}}}:
      \Sigma\var{1}:\ty{1}.(\cty{1}\multimap\cty{2})}
  \end{array}
\]
\end{minipage}}
    \caption{Cartesian dependent typing and sequential pairing. In the elimination rules, substitution acts on the remaining context as well as the result type. The eliminators also apply to values; the pair rule includes ordinary products (and their $n$-ary form) with independent component types. Cartesian computations use the sequencing rule of Fig.~\ref{fig:types1} with result type well formed in the original context.\label{fig:tl-type-system2}}\;
\end{figure}

The last rule of Fig.~\ref{fig:tl-type-system1} is derived: its conclusion abbreviates
\[
\letin{z}{\trm{1}}{\bigl(\vMatch{z}{\tIni[i]x_i\To\lfun{w}r_i[w/\lvar]}_{i=1}^{n}\bigr)\mathbin{\bullet}\lvar}.
\]
Value case analysis gives the parenthesized value type $\cty{2}[z/x]\multimap\cty{3}[z/x]$; computational sequencing then gives the displayed reindexed input and output types. Here $z,w$ are fresh.

The last rule of Fig.~\ref{fig:tl-type-system2} makes explicit the dependent pairing used by CHAD. The first component is evaluated before its backward map is returned. When $\trm{1}$ returns $\var{1}$, the computationally reindexed input type of $\trm{2}$ is the fibre $\cty{1}$ at that returned value. More precisely, over the graph $\{(\gamma,p)\mid\semt{\trm{1}}(\gamma)=p\}$, the family $\csubst{\cty{1}}{\sfor{\var{1}}{\trm{1}}}$ is the family $\cty{1}$ at $p$. The linear abstraction therefore has the required type in this successful context. The pair is undefined exactly when its first component is undefined. This is also the pairing rule used for the iteration clause: after the primal loop returns, its folder gives a linear map from the cotangent fibre at the returned result.

\begin{figure}[!ht]
	\framebox{\begin{minipage}{0.98\linewidth}\noindent\hspace{-24pt}\[
  \begin{array}{c}
\color{black!65}
\inferrule{
\var{2}:\ty{1}\vdash\underline{\ty{1}}:\ltype\qquad
\var{1}:\ty{2}\vdash \trm{1}:\ty{1}\t+ \ty{2}\\
\Gamma,\var{1}:\ty{2}; \lvar: \underline{\ty{3}}\vdash
{\trm{2}}: \vMatch{\trm{1}}{
  \tInl \var{2}\To\underline{\ty{1}}
  \mid
\tInr\var{1}\To \csubst{\underline{\ty{1}}}{\sfor{\var{2}}{
\tIt{\var{1}}{\trm{1}}}}}
}
{\Gamma,\var{1}:\ty{2}; \lvar: \underline{\ty{3}}\vdash
\tRoll
\trm{2}:\csubst{\underline{\ty{1}}}{\sfor{\var{2}}{
\tIt{\var{1}}{\trm{1}}}}}
\\\\
\color{black}
\inferrule{
\var{2}:\ty{1}\vdash\underline{\ty{1}}:\ltype\qquad
\var{1}:\ty{2}\vdash \trm{1}:\ty{1}\t+ \ty{2}\\
\var{1}:\ty{2};\lvar:\underline{\ty{3}}\vdash \trm{2}:
\csubst{\underline{\ty{1}}}{\sfor{\var{2}}{
\tIt{\var{1}}{\trm{1}}}}
\\
\var{1}:\ty{2};\lvar:
\vMatch{\trm{1}}{ \tInl\var{2}\To \underline{\ty{1}}\mid \tInr\var{1}\To \underline{\ty{3}'}}\vdash
\trm{3}:\underline{\ty{3}'}
}
{%
\var{1}:\ty{2};\lvar:\underline{\ty{3}}\vdash \tFold
{\trm{2}}{\lvar}{\trm{3}} :\underline{\ty{3}'}}
\end{array}
\]
\end{minipage}}
	\caption{Typing rules for \foldersname{}  for the target language  on top of the rules of
		\ref{fig:types1}, \ref{fig:tl-type-system2} and \ref{fig:tl-type-system1}.
        \label{fig:tl-type-system0}}\;
\end{figure}

Fig.  \ref{fig:beta-eta1} and \ref{fig:tl-equations} display the equational theory we consider for the terms and
types, which we call $(\alpha)\beta\eta+$-equivalence, together with the context-action equations specified above. In particular, a computational reindexing is not removed merely because its bound variable is absent from the type.

\begin{figure}[!ht]
	\framebox{\begin{minipage}{0.98\linewidth}\hspace{-24pt} \[\begin{array}{l}
    \letin{\lvar}{\trm{1}}{\trm{2}}=\subst{\trm{2}}{\sfor{\lvar}{\trm{1}}}
    \\[5pt]
    \lapp{(\lfun{\lvar}{\trm{1}})}{\trm{2}} = \subst{\trm{1}}{\sfor{\lvar}{\trm{2}}} \!\!\qquad\qquad\qquad\qquad\qquad\qquad \val{1} = \lfun{\lvar}{\lapp{\val{1}}{\lvar}}
    \\[5pt]
    \trm{1} + \tZero = \trm{1}\qquad \tZero + \trm{1} = \trm{1}\qquad\qquad\qquad\qquad\qquad (\trm{1}+\trm{2})+\trm{3}=\trm{1}+(\trm{2}+\trm{3})\qquad
    \trm{1} + \trm{2} = \trm{2} + \trm{1}\\[6pt]
    (\Ginf[;\lvar:\cty{1}]{\trm{1}}{\cty{2}})\;\textnormal{ implies }\; \subst{\trm{1}}{\sfor{\lvar}{\tZero}}=\tZero\qquad \qquad\quad\! (\Ginf[;\lvar:\cty{1}]{\trm{1}}{\cty{2}})\;\textnormal{ implies }\;
    \subst{\trm{1}}{\sfor{\lvar}{\trm{2}+\trm{3}}}=\subst{\trm{1}}{\sfor{\lvar}{\trm{2}}}+\subst{\trm{1}}{\sfor{\lvar}{\trm{3}}}\\[6pt]
    \vMatch{\tIni[i]\trm{2}}{\tIni[1]\var{1}_1\To\cty{1}_1\mid\cdots\mid \tIni[n]\var{1}_n\To\cty{1}_n}=\csubst{\cty{1}_i}{\sfor{\var{1}_i}{\trm{2}}}
    \\[6pt]
    \csubst{\cty{1}}{\sfor{\var{2}}{\trm{2}}}\freeeq{\var{1}_1,\ldots,\var{1}_n}
\vMatch {\trm{2}}  {
        \tIni[1]{\var{1}_1}\To{\subst{\cty{1}}{\sfor{\var{2}}{\tIni[1]{\var{1}_1}}}}
\vor \cdots
\vor  \tIni[n]{\var{1}_n}\To{\subst{\cty{1}}{\sfor{\var{2}}{\tIni[n]{\var{1}_n}}}}
}
\\[6pt]
\tPrj[i]\tTuple{\trm{1}_1,\ldots,\trm{1}_n} = \trm{1}_i
\qquad\qquad\qquad\qquad\qquad\quad
\trm{1} = \tTuple{\tPrj[1]\trm{1},\ldots,\tPrj[n]\trm{1}}

\\[6pt]\hline\\[-4pt]
\subst{\bigl(\vMatch{\var{3}}{\tIni[1]\var{1}_1\To\trm{1}_1\mid\cdots\mid\tIni[n]\var{1}_n\To\trm{1}_n}\bigr)}{\sfor{\var{3}}{\tIni[j]\var{1}_j}}=\trm{1}_j\\[6pt]
\trm{3}=\vMatch{\var{3}}{\tIni[1]\var{1}_1\To\subst{\trm{3}}{\sfor{\var{3}}{\tIni[1]\var{1}_1}}\mid\cdots\mid\tIni[n]\var{1}_n\To\subst{\trm{3}}{\sfor{\var{3}}{\tIni[n]\var{1}_n}}}\\[6pt]
\subst{\bigl(\tMatch{\var{3}}{\var{1},\var{2}}{\trm{1}}\bigr)}{\sfor{\var{3}}{\tTuple{\var{1},\var{2}}}}=\trm{1}\qquad
\trm{3}=\tMatch{\var{3}}{\var{1},\var{2}}{\subst{\trm{3}}{\sfor{\var{3}}{\tTuple{\var{1},\var{2}}}}}
\\[6pt]\hline\\[-4pt]
\csubst{\cty{1}}{\sfor{\var{2}}{\letin{\var{1}}{\trm{1}}{\trm{2}}}}
    =\csubst{\bigl(\csubst{\cty{1}}{\sfor{\var{2}}{\trm{2}}}\bigr)}{\sfor{\var{1}}{\trm{1}}}\qquad
\csubst{\cty{1}}{\sfor{\var{1}}{\val{1}}}=\subst{\cty{1}}{\sfor{\var{1}}{\val{1}}}
    \qquad(\val{1}\text{ a Cartesian value})\\[6pt]
\csubst{(\cty{1}_1\t*\cdots\t*\cty{1}_n)}{\sfor{\var{1}}{\trm{1}}}
    =\csubst{\cty{1}_1}{\sfor{\var{1}}{\trm{1}}}\t*\cdots\t*\csubst{\cty{1}_n}{\sfor{\var{1}}{\trm{1}}}\\[6pt]
\letin{\var{1}}{\trm{1}}{\lvar}=\lvar
    \qquad\qquad
    \letin{\var{1}}{\trm{1}}{\tZero}=\tZero\\[6pt]
\letin{\var{1}}{\trm{1}}{(\letin{\lvar}{\trm{2}}{\trm{3}})}
    =\letin{\lvar}{(\letin{\var{1}}{\trm{1}}{\trm{2}})}{(\letin{\var{1}}{\trm{1}}{\trm{3}})}\\[6pt]
\letin{\var{1}}{\trm{1}}{(\trm{2}+\trm{3})}
    =(\letin{\var{1}}{\trm{1}}{\trm{2}})+(\letin{\var{1}}{\trm{1}}{\trm{3}})\\[6pt]
\letin{\var{1}}{\trm{1}}{\tTuple{\trm{2}_1,\ldots,\trm{2}_n}}
    =\tTuple{\letin{\var{1}}{\trm{1}}{\trm{2}_1},\ldots,\letin{\var{1}}{\trm{1}}{\trm{2}_n}}
\end{array}\]

	\end{minipage}}
	\caption{Target equations, in addition to Fig.~\ref{fig:beta-eta1}. The blocks give linear $\beta\eta$ and biadditivity, dependent pattern matching, and computational reindexing. Function $\eta$ is restricted to values. The last block's term equations are in linear judgements, with reindexed input and result types. Products and tuples include arity zero. The value-substitution and associativity equations for Cartesian \textbf{let} in Fig.~\ref{fig:beta-eta1} apply also in linear judgements. All equations are well typed, modulo $\alpha$-renaming and with fresh binders. No equations are imposed on iteration or folders.\label{fig:tl-equations}}
\end{figure}

We find it helpful to think of the target language from a categorical point of view as an iterative Freyd indexed category in the sense of Definition~\ref{def:iterative-Feyd-indexed-category}
$\LSyn:\CSynC^{op}\to\catCat$, concretely:
\begin{itemize}
\item similarly to Section~\ref{ap:source-language}, we have an iterative Freyd category
\[\CSyn=(\CSynV,\CSynC,\CSynJ,\CSynTT,\CSynit).\]
Its objects are the closed Cartesian types of the target language, and its morphisms are equivalence classes of values and computations, respectively. The inclusion of source syntax induces an iterative Freyd category morphism $\Syn\to\CSyn$;
\item a (strict) indexed category $\LSyn=\LSynC:\CSynC^{op}\to\catCat$
(from which we can also define  $\LSynV:\CSynV^{op}\to\catCat$
as $\LSynC\circ \CSynJ^{op}$) where the objects of $\LSyn(\ty{1})$ are equivalence classes
$[\cty{2}]$
of linear (dependent) types $\var{1}:\ty{1}\vdash \cty{2}$, with ground linear types specified in the terminal value context and weakened along value projections, and morphisms $[\trm{1}]:[\cty{2}]\to[\cty{3}]$ are equivalence classes
of
computations $\var{1}:\ty{1};\lvar:\cty{2}\vdash \trm{1}:\cty{3}$;
identities in $\LSyn(\ty{1})$ are given by the equivalence class of $\var{1}:\ty{1};\lvar:\cty{2}\vdash \lvar:\cty{2}$
and composition of $[\trm{1}]:[\cty{2}_1]\to[\cty{2}_2]$ and $[\trm{2}]:[\cty{2}_2]\to[\cty{2}_3]$ is given by
the equivalence class of $\var{1}:\ty{1};\lvar:\cty{2}_1\vdash \letin{\lvar}{\trm{1}}{\trm{2}}:\cty{2}_3$;
finally, given $[\trm{1}]:[\cty{2}]\to[\cty{3}]$ in $\LSyn(\ty{1})$ represented by
$\var{1}:\ty{1};\lvar:\cty{2}\vdash \trm{1}:\cty{3}$
and $[\trm{2}]:[\ty{1}']\to [\ty{1}]$ in $\CSynC$,
we define the change of base $\LSyn([\trm{2}])([\trm{1}])$ as the morphism
$[\csubst{\cty{2}}{\sfor{\var{1}}{\trm{2}}}]\to [\csubst{\cty{3}}{\sfor{\var{1}}{\trm{2}}}]$
in $\LSyn(\ty{1}')$ given by the equivalence class of $\var{1}':\ty{1}';\lvar:\csubst{\cty{2}}{\sfor{\var{1}}{\trm{2}}}\vdash \letin{\var{1}}{\trm{2}}{\trm{1}}:\csubst{\cty{3}}{\sfor{\var{1}}{\trm{2}}}$;
\item we observe ($\dagger$) that $\LSyn$ has finite indexed biproducts (hence is enriched over commutative monoids), that $\LSyn$ is extensive\footnote{See \cite{lucatellivakar2021chad} or \cite{LV24b} for more details on extensive indexed categories.} in the sense that $\LSyn([\ty{1}_1\sqcup \cdots\sqcup \ty{1}_n])\cong \LSyn(\ty{1}_1)\times \cdots\times \LSyn(\ty{1}_n)$ and that $\LSyn$ is an \emph{iterative indexed category} in the sense that we have iterative folders\footnote{
Adding roll constructors with both the computation and uniqueness rules for initial algebras instead requires that $\LSyn$ is iteration-extensive in the sense that
given $\var{1}:\ty{2}\vdash \trm{1}:\ty{1}\sqcup\ty{2}$, the roll constructors turn $\LSyn(\tIt{\var{1}}{\trm{1}})$ into the carrier of the parameterized initial algebras of the functor $\LSyn(\trm{1}):\LSyn(\ty{1})\times\LSyn(\ty{2})\cong \LSyn(\ty{1}\sqcup \ty{2})\to \LSyn(\ty{2})$.
With the canonical identity-algebra compatibility of Theorem~\ref{theo:iteration-extensive-indexed-category} as well, we can also demand the basic fixed point equation for iteration in the source language, since the CHAD code transformation will then respect it.
};
\item similarly, the Cartesian dependent types and values form a (strict) indexed category $\CSyn':\CSynV^{op}\to\catCat$ where the objects of $\CSyn'(\ty{1})$ are $(\alpha)\beta\eta+$-equivalence classes
$[\ty{2}]$
of Cartesian (dependent) types $\var{1}:\ty{1}\vdash \ty{2}$ and morphisms $[\val{1}]:[\ty{2}]\to[\ty{3}]$ are  $(\alpha)\beta\eta+$-equivalence classes
of
values $\var{1}:\ty{1},\var{2}:\ty{2}\vdash \val{1}:\ty{3}$;
identities in $\CSyn'(\ty{1})$ are given by the equivalence class of $\var{1}:\ty{1},\var{2}:\ty{2}\vdash \var{2}:\ty{2}$
and composition of $[\val{1}]:[\ty{2}_1]\to[\ty{2}_2]$ and $[\val{2}]:[\ty{2}_2]\to[\ty{2}_3]$ is given by
the equivalence class of $\var{1}:\ty{1},\var{2}:\ty{2}_1\vdash \letin{\var{2}}{\val{1}}{\val{2}}:\ty{2}_3$;
finally, given $[\val{1}]:[\ty{2}]\to[\ty{3}]$ in $\CSyn'(\ty{1})$ represented by
$\var{1}:\ty{1},\var{2}:\ty{2}\vdash \val{1}:\ty{3}$
and $[\val{2}]:[\ty{1}']\to [\ty{1}]$ in $\CSynV$,
we define the change of base $\CSyn'([\val{2}])([\val{1}])$ as the morphism
$[\subst{\ty{2}}{\sfor{\var{1}}{\val{2}}}]\to [\subst{\ty{3}}{\sfor{\var{1}}{\val{2}}}]$
in $\CSyn'(\ty{1}')$ given by the equivalence class of $\var{1}':\ty{1}',\var{2}:\subst{\ty{2}}{\sfor{\var{1}}{\val{2}}}\vdash \letin{\var{1}}{\val{2}}{\val{1}}:\subst{\ty{3}}{\sfor{\var{1}}{\val{2}}}$;
\item we observe ($\ddagger$) that $\CSyn'$ forms a model of dependent type theory (in the sense that it satisfies full, faithful, democratic comprehension -- see \cite{vakar2017search,LV24b}) with strong $\Sigma$-types (in the sense that the display maps/dependent projections are closed under composition) and that the linear hom-types represent linear maps by Cartesian values:
$\LSyn(\ty{1})([\cty{2}],[\cty{3}])\cong \CSyn'(\ty{1})(\Unit, [\cty{2}\multimap \cty{3}])$. This is the value-level correspondence given by linear abstraction and application; arbitrary computations of linear-function type may also diverge.
\end{itemize}
The types and terms, modulo the stated equations, form the initial model of the target theory on the chosen primitives. Here an interpretation preserves the full typed structure, including dependent pairs, linear-function values, computational reindexing, and the context comparisons. Its underlying linear indexed category is a biadditive iterative Freyd indexed category. Thus an interpretation of the target theory induces the indexed functor needed below.
Crucially, we now have the following when we take the op-Grothendieck construction of the indexed categories. This is the point where the folders in the target language become iteration on pairs consisting of a primal and its cotangent type.
\begin{lemma}
$\big(\Sigma_{\CSynV}\LSynV^{\op}, \Sigma_{\CSynC}\LSynC^{\op},\Sigma_{\CSynV}\LSynV^{\op}\to \Sigma_{\CSynC}\LSynC^{\op},\otimes_\Sigma,$\\$(\tIt{}{},\tFold{}{}{})\big)$ forms an iterative Freyd category.
\end{lemma}
\begin{proof}
This is Theorem~\ref{theo:iterative-Feyd-indexed-category} applied to the syntactic indexed category just described.
\end{proof}

\section{The iterative CHAD code transformation and its correctness}\label{sec:correctness-iterative-CHAD}

In this section, we establish our main result concerning the correctness of iterative CHAD. By introducing iteration constructs on top of the source language, we extend CHAD while keeping
the structure-preservation principle as long as we also add \foldersname{}   to our target language. More precisely, CHAD is still defined as a structure preserving
transformation between the source language and the op-Grothendieck construction of the target language concerning the respective indexed categories of biadditive types.

\subsection{Source language}\label{section:the-source-language-for-iterative-CHAD}
We consider a call-by-value functional programming language
with variant, product types, non-termination, and an iteration construct. Like in the case of
total CHAD (see, for instance, Subsect.~\ref{sec:source-language-total}), our language is constructed over ground types $\reals ^n $
for arrays of real numbers with static length $n$ ($n\in\NN $).
In fact, the source language we consider below is effectively the source language we
considered for total CHAD (see Section \ref{section:denotational-correctness-of-CHAD-total} and Section \ref{sec:source-language}) with iteration constructs on top of it,
and with a larger set of primitive operations, still denoted by $$\Op =\bigcup _{\left( k_1, k_2\right)\in\NN \times \NN }\bigcup _{\left( \mathbf{n}, \mathbf{m}\right)  \in \NN ^{k_1} \times \NN ^{k_2}} \Op  ^{\mathbf{n}}_{\mathbf{m}} ,$$  allowing them to implement partially defined functions.
More precisely, if $\mathbf{n} = \left( n_1 , \ldots , n_{k_1}\right)  $ and $\mathbf{m} = \left( m_1 , \ldots , m_{k_2}\right)  $,
a primitive operation  $\op\in \Op  ^\mathbf{n}_ \mathbf{m} $ may, now, implement a partially defined differentiable function, that is to say, a morphism
\begin{equation}
\sem{\op}\coloneqq \left( U_\op , \sempp{\op } \right) : M\to N
\end{equation}
of $\PVMAN$. Besides the examples of primitive operations we discussed
in Subsect.~\ref{sec:source-language-total} (which are totally defined), the reader can keep the following examples of primitive operations corresponding to partially defined functions in mind:
\begin{itemize}
	\item the norm operation $\normRR{n} \in \Op ^n _1 $ that intends to implement the function
	\begin{equation}
		\sem{\normRR{n}} \coloneqq \left( \RR^n -\left\{ 0 \right\}, 	\sempp{\normRR{n}}  \right)  : \RR ^n \to \RR
	\end{equation}
which is not defined in $0$ (the singularity) and is defined by $\sempp{\normRR{n}} (x) = || x ||$ for non-zero  $n$-dimensional vectors $x$;
\item the reciprocal operation $\recipp \in \Op ^1 _1$ that intends to implement the function
	\begin{equation}
	\sem{\recipp } \coloneqq \left( \RR -\left\{ 0 \right\},  	\sempp{\recipp } \right)  : \RR  \to \RR
\end{equation}
which is not defined in $0$  and is defined by $\sempp{\recipp } (x) = 1/x$ otherwise;
		\item the normalization operation $\orientationF{n} \in \Op^{n}_n$ that should implement
	\begin{equation}
		\sem{\orientationF{n}} = \left( \RR ^n-\left\{ 0 \right\} , \sempp{\orientationF{n}}\right)  : \RR ^n \to \RR ^n
	\end{equation}
	where $\sempp{\orientationF{n}}\left( x \right)= x/|x| $ for  $x\in\RR ^n-\left\{ 0 \right\} $;
	\item the sign function $\signR \in \Op^{1}_{(1,1)}$ that should implement
\begin{equation}
\sem{\signR} = \left( \RR _-\cup \RR _+ , \sempp{\signR}\right)  : \RR\to \RR _0\sqcup \RR _1
\end{equation}
	where $\sempp{\signR}\left( x \right)= \coproje{\RR _0} (x) $ if  $x\in\RR _+ $, and $\sempp{\signR}\left( x \right)= \coproje{\RR _1} (x) $ if  $x\in\RR _- $;
\item deciders which are generic operations that have coproducts in the codomain. For instance,
for each $a\in \RR $,
the operation $\deciderr{1}{(1,1)}{a} \in\Op _{(1,1)}	^1 $ that implements
\begin{equation}
	\sem{\deciderr{1}{(1,1)}{a}} = \left( \RR\setminus\{a\} , \sempp{\deciderr{1}{(1,1)}{a}} \right)  : \RR ^1\to \RR _1\sqcup \RR _2
\end{equation}
	which is not defined at $a\in\RR$, 	$\sempp{\deciderr{1}{(1,1)}{a}}(x) = \coproje{1}(x)\in\RR_1 $ if $x>a$,
	and $\sempp{\deciderr{1}{(1,1)}{a}}(x) = \coproje{2}(x)\in\RR_2 $ if $x < a$.
\end{itemize}

We establish, then, the source language to be a \textit{standard call-by-value language
with tuples, variant types, and iteration} constructed over our ground types $\reals ^n $ and our primitive operations in $\Op$, with the typing rule for iteration shown in Fig.~\ref{fig:typestermrecursion-iteration} on top of the language established in Section~\ref{sec:source-language} (Fig.~\ref{fig:types1}). Following the same principle of \cite{vakar2020denotational, 2022arXiv221007724L, 2022arXiv221008530L}, we do not impose further equations (\textit{e.g.}, \cite{bloom1993iteration, goncharov2015unguarded}) on our iteration construct  because it is unnecessary for our developments.
Observe that a more general iteration construct with typing rule
\[
\frac{\Gamma,\var{1}:\ty{2}\vdash \trm{1}:\ty{1}\t+\ty{2}}{
	\Gamma,\var{1}:\ty{2}\vdash \tIt[\Gamma]{\var{1}}{\trm{1}}:\ty{1}
}
\]
 is derivable. Indeed, for $\Gamma=\var{2}_1:\ty{3}_1,\ldots,\var{2}_n:\ty{3}_n$, we define $\tIt[\Gamma]{\var{1}}{\trm{1}}$ as
\[
\begin{array}{l}
\letin{\var{3}}{\tTuple{\var{2}_1,\ldots,\var{2}_n,\var{1}}}{}\\
\tIt{\var{3}}{\tMatch{\var{3}}{\var{2}_1,\ldots,\var{2}_n,\var{1}}{\vMatch{\trm{1}}{\tIni[1]u\To \tIni[1]u\mid \tIni[2]v\To \tIni[2]\tTuple{\var{2}_1,\ldots,\var{2}_n,v}}}}
\end{array}
\]

\begin{figure}[!ht]
	\fbox{\parbox{0.98\linewidth}{\begin{minipage}{\linewidth}\noindent
		\[
\begin{array}{l}
    \inferrule{\freeGinf[\var{1}:{\ty{2}}] {\trm{1}}{\ty{1}\t+\ty{2}}
    }{\freeGinf[\var{1}:{\ty{2}}] {\tIt{\var{1}}{\trm{1}}}{\ty{1}}}
\end{array}
\]\end{minipage}}}
	\caption{Typing rules for iteration, after appropriately adding ${\tIt{\var{1}}{\trm{1}}}$ to our grammar of computations. See Section \ref{sec:source-language} for more details.
		\label{fig:typestermrecursion-iteration}}
\end{figure}

The primitive operations above show why the partial setting is useful even before a loop is introduced: reciprocal and normalization are differentiable only on their open domains. Iteration adds another source of partiality, since a sequence of individually defined steps need not terminate.

\subsection{Categorical semantics of the source language and its universal property}
Very much like the total setting discussed in Section~\ref{section:denotational-correctness-of-CHAD-total},
we can frame the primitive types and operations into a structured (poly)graph/computad,  and take the
\iterativeF $\Syn = \left( \SynV , \SynC, \SynJ , \SynTT, \Synit \right) $ freely generated on it. Following this  path, we end up with the following syntactic categorical model.

\begin{theorem}[Universal Property of the Source Language]\label{theo:universal-property-source-Freyd-language}
	The \iterativeF $\Syn = \left( \SynV , \SynC, \SynJ , \SynTT, \Synit \right) $ corresponding to our source call-by-value language has the following universal property. Given any \iterativeF $\left( \catV , \catC , j, \otimes , \iterationn\right)   $, for each pair
	$	\left( K , \mathtt{h}   \right) $
	where $K = \left(  K_ n \right) _ {n\in\NN} $ is a family of objects in $\catC $ and
	$\mathtt{h} = \left( h_\op \right) _{\op\in\Op} $ is a consistent family of morphisms
	in $\catC$ (meaning that $h_\op:K_{n_1}\times\cdots\times K_{n_{k_1}}\to K_{m_1}\sqcup\cdots\sqcup K_{m_{k_2}}$ for $\op\in\Op^{\mathbf{n}}_{\mathbf{m}}$, with a chosen value morphism $h^v_\op$ satisfying $j(h^v_\op)=h_\op$ for each designated total primitive), there is a unique \iterativeF morphism
	\begin{equation}
		\left( H, \hat{H}\right)  : \Syn \to \left( \catV , \catC , j, \otimes , \iterationn\right)
	\end{equation}
	such that $\hat{H} (\reals ^n ) = K _n $ and $\hat{H} (\op ) = h_\op $ for any $n\in\NN$ and any primitive operation $\op\in\Op$, and $H(\op)=h^v_\op$ for each designated total primitive.
\end{theorem}

Making use of the universal property above and the fact that $\FFMAN = \left( \VMAN, \PVMAN , \pman , \otimes\right) $ is an $\ww$-Freyd category (hence an \iterativeFn), we can easily define our concrete semantics of the source language in terms of partially defined differentiable functions, morphisms of $\PVMAN $.
More precisely, by the universal property of $\Syn = \left( \SynV , \SynC, \SynJ , \SynTT, \Synit \right) $ , we can define the concrete semantics as an \iterativeF morphism
\begin{equation}\label{eq:sematics-FUNCTOR-iterativeF}
	\sem{-}: \Syn\to\FFMAN ,
\end{equation}
where $\FFMAN$ is the $\ww$-Freyd category defined in Eq.~\eqref{eq:freyd-category-source-language-semantics}.

\begin{theorem}[Semantics iterative Freyd morphism]\label{the:semantics-source-language-iterative-Freyd-category-morphism}
	There is only one \iterativeF morphism \eqref{eq:sematics-FUNCTOR-iterativeF}
	\begin{equation}\label{eq:semantics-sourrce-language-partial}
		\sem{-}: \Syn \to \FFMAN
	\end{equation}
	such that, for each
	$n\in \NN $, $\sem{\reals ^n} = \RR ^n $ and, for each $\left( \mathbf{n}, \mathbf{m}\right) \in \NN ^{k_1}\times \NN ^{k_2} $,
	$\sem{\op} $ is the morphism
	$ \RR^{n_1}\times \cdots \times \RR ^{n_{k_1}}\to\RR^{m_1}\sqcup \cdots \sqcup \RR ^{m_{k_2}} $
	of $\PVMAN $
	that $\op $ intends to implement.

	The iterative Freyd category morphism \eqref{eq:semantics-sourrce-language-partial}  gives the concrete semantics of the source language.
\end{theorem}
\begin{proof}
	Since $\FFMAN = \left( \VMAN, \PVMAN , \pman , \otimes\right) $ is an $\ww$-Freyd category, it is an
	\iterativeFn{} by Corollary~\ref{eq:natural-iteration-induced-by-omega-chains-Freyd-category}. Hence, the result follows from the universal property of $\Syn$.
\end{proof}

The universal property separates the choices made for primitives from the treatment of language structure. Once primitives have been interpreted, the meaning of sequencing, case distinction, and iteration is forced by preservation of that structure. We now arrange for the derivative target to have the same structure.

\subsection{Target language and its categorical semantics}
Our target language is defined following the same principle of Section \ref{sec:target-language-total}; namely,  we use a variant of the dependently typed enriched effect calculus~\citep[Chapter 5]{vakar2017search} that
extends our call-by-value source language established in Section \ref{section:the-source-language-for-iterative-CHAD} with suitable biadditive dependent types for typing derivatives and with \textit{primitive linear operations}
$\lop\in\LOp^{\mathbf{r};\mathbf{n}}_{\mathbf{m}}$ as building blocks for constructing those derivatives.
We refer the reader to Section \ref{sec:target-language} for the detailed grammar and typing rules of our language.

To serve as a practical target language for the automatic derivatives of all programs from the source language, we make the following assumption:
for each $\left( k_1, k_2\right) \in\NN^2 $, each $\left( \mathbf{n}, \mathbf{m}\right) \in \NN ^{k_1}\times \NN ^{k_2} $, and each $\op\in \Op_{\mathbf{m}}^{\mathbf{n}}$, there is a chosen program
\begin{equation*}
\begin{array}{l}
x_1:\reals^{n_1},\ldots,x_{k_1}:\reals^{n_{k_1}}\\
\quad;\lvar:
\vMatch{\op(x_1,\ldots,x_{k_1})}{\tIni[1] \_\To \tangentreals^{m_{1}}\mid\cdots\mid \tIni[k_2] \_\to
\tangentreals^{m_{k_2}}}\\
\quad\vdash D\op(x_1,\ldots,x_{k_1})(\lvar):\tangentreals^{n_1}\times\cdots\times \tangentreals^{n_{k_1}}
\end{array}
\end{equation*}
that intends to implement a function that gives the (transpose) derivative of $\op$.

Being a proper extension of our (call-by-value) source language established in Section \ref{section:the-source-language-for-iterative-CHAD}, our target language, established in  Section \ref{sec:target-language}, already allows for non-termination (and partially defined functions as semantics of
primitive operations). In order to implement our macro,  we add our novel \foldersname{}     on top of our language, following the principle of our categorical semantics -- namely, the \textit{iterative indexed categories} or, more precisely, the \textit{iterative Freyd indexed categories } as introduced in Subsect.~\ref{syn:container-folder-iteration}.
We show their typing rules in Fig. \ref{fig:tl-type-system00}. The grey rule is the optional roll constructor: the transformation only requires the folder elimination rule.
\begin{figure}[!ht]
	\framebox{\begin{minipage}{0.98\linewidth}\noindent\hspace{-24pt}\[
  \begin{array}{c}
\color{black!65}
\inferrule{
\var{2}:\ty{1}\vdash\underline{\ty{1}}:\ltype\qquad
\var{1}:\ty{2}\vdash \trm{1}:\ty{1}\t+ \ty{2}\\
\Gamma,\var{1}:\ty{2}; \lvar: \underline{\ty{3}}\vdash
{\trm{2}}: \vMatch{\trm{1}}{
  \tInl \var{2}\To\underline{\ty{1}}
  \mid
\tInr\var{1}\To \csubst{\underline{\ty{1}}}{\sfor{\var{2}}{
\tIt{\var{1}}{\trm{1}}}}}
}
{\Gamma,\var{1}:\ty{2}; \lvar: \underline{\ty{3}}\vdash
\tRoll
\trm{2}:\csubst{\underline{\ty{1}}}{\sfor{\var{2}}{
\tIt{\var{1}}{\trm{1}}}}}
\\\\
\color{black}
\inferrule{
\var{2}:\ty{1}\vdash\underline{\ty{1}}:\ltype\qquad
\var{1}:\ty{2}\vdash \trm{1}:\ty{1}\t+ \ty{2}\\
\var{1}:\ty{2};\lvar:\underline{\ty{3}}\vdash \trm{2}:
\csubst{\underline{\ty{1}}}{\sfor{\var{2}}{
\tIt{\var{1}}{\trm{1}}}}
\\
\var{1}:\ty{2};\lvar:
\vMatch{\trm{1}}{ \tInl\var{2}\To \underline{\ty{1}}\mid \tInr\var{1}\To \underline{\ty{3}'}}\vdash
\trm{3}:\underline{\ty{3}'}
}
{%
\var{1}:\ty{2};\lvar:\underline{\ty{3}}\vdash \tFold
{\trm{2}}{\lvar}{\trm{3}} :\underline{\ty{3}'}}
\end{array}
\]
\end{minipage}}
	\caption{Typing rules for \foldersname{}  for the target language  on top of the rules of
		\ref{fig:types1}, \ref{fig:tl-type-system2} and \ref{fig:tl-type-system1}.\label{fig:tl-type-system00}}\;
\end{figure}

Similarly to the case of total CHAD (see Subsect.~\ref{sec:target-language-total}), the syntactic model of our target theory has an underlying biadditive iterative Freyd indexed category
\[\LSynTF = \left( \LSyn , \CSynV , \CSynC , \CSynj ,   \otimes , \CSynit, \foldingS{-,-} \right)\]
of linear types over the Cartesian target syntax (see Section \ref{sec:target-language} for more details on the target language). The important aspect of this categorical structure is that we can define the concrete semantics as follows.

\Needspace{7\baselineskip}
\begin{theorem}[Syntactic categorical model]\label{theo:syntactic-categorical-semantics-clearly}
	The initial model of the target theory on our operations has an underlying biadditive iterative Freyd indexed category $\LSynTF = \left( \LSyn , \CSynV , \CSynC , \CSynj ,   \otimes , \CSynit, \foldingS{-,-} \right)$. Its initiality concerns interpretations of the full target theory, including its dependent pairs and linear-function types, as described in Section~\ref{sec:target-language}.
\end{theorem}

\begin{corollary}\label{coro:Iterative-Freyd-category-sematics}
The op-Grothendieck construction $\Sigma^\op\left( \LSynTF \right) $ gives an iterative Freyd category that we denote by
\begin{align}\label{eq:Grothendick-ww-Freydd}
\TPSYN &\defeqq \Sigma^\op\left(\LSynTF\right)\\
&=\Sigma^\op\left(\LSyn,\CSynV,\CSynC,\CSynj,\otimes,\CSynit,\foldingS{-,-}\right)\\
&=\left(\Sigma_{\CSynV}\left(\LSyn\circ\CSynj^\op\right)^\op,
\Sigma_{\CSynC}\LSyn^\op,\CSynj,\otimes_{T\Sigma},\TLSynit\right).
\end{align}
The iteration $\TLSynit$ is defined as follows: for each morphism
$$\left( f , f'\right) : \left( A, X\right) \to \left( B, Y \right)\sqcup \left( A, X \right)\cong \left( B\sqcup A, \cISO ^{\left( B, A\right) }\left( Y, X\right)\right)   $$
in $\displaystyle \displaystyle\Sigma_{\CSynC } \LSyn ^\op \left( \left( A, X\right), \left( B, Y\right)\sqcup\left( A, X\right) \right) $, we have
\begin{equation}
	\TLSynit  \left( f , f'\right) = \left( \iterationn f , \foldingS{f, f'}\right).
\end{equation}
\end{corollary}
\begin{proof}
	Since $\LSynTF = \left( \LSyn , \CSynV , \CSynC , \CSynj ,   \otimes , \CSynit, \foldingS{-,-} \right) $ is an iterative Freyd indexed category, the result follows from Theorem \ref{theo:iterative-Feyd-indexed-category}.
\end{proof}

The usual forgetful functors $\VMAN\to \Set $ and $\PVMAN\to\catPSet $ induce
a Freyd category morphism $\forgettingagain : \FFMAN\to \FFSET $, where $\FFSET \coloneqq \left( \Set , \catPSet , j, \otimes  \right) $ is the canonical $\ww$-Freyd category obtained by the usual inclusion $j : \Set\to\catPSet$.

For the purpose of stating Theorem~\ref{theo:concrete-target-language-partially-defined-functions}, we write $\sem{-}_0 : \CSyn \to \FFSET$ for the Cartesian interpretation of the target language. Its restriction to source syntax is $\forgettingagain\circ\sem{-}$, where $\sem{-}$ is the source interpretation of Eq.~\eqref{eq:sematics-FUNCTOR-iterativeF}. Cartesian dependent types are interpreted as families of sets, with ordinary reindexing along values; $\Sigma$-types are their dependent sums. Linear types are families of vector spaces, and $\cty{1}\multimap\cty{2}$ is the family of sets of linear maps between their fibres. Values denote total functions, computations denote partial functions, and computational reindexing of linear types is given by $\Vect_{\leastelement}^{(-)}$. These clauses define the Cartesian and linear interpretations together by induction on the syntax. In particular, sequential dependent pairing returns the primal and the corresponding linear map on its domain of definition, as explained after Fig.~\ref{fig:tl-type-system2}.

\begin{theorem}[Concrete categorical semantics of the target language as an indexed functor]\label{theo:concrete-target-language-partially-defined-functions}
Let $\concINDEX$ be the biadditive iterative Freyd indexed category established in Definition~\ref{def:the-real-semantics-of-the-target-language}. The Cartesian interpretation
\begin{equation}
	\sem{-}_0: \CSyn\to\FFSET
\end{equation}
together with the interpretations $\tangentreals^n\mapsto\overline{\RR^n}$ of ground linear types and the chosen interpretations of all primitive linear operations, determines an interpretation of the full target theory. Its underlying biadditive iterative Freyd indexed category morphism is\footnote{See \cite[Subsect.~6.9]{lucatellivakar2021chad} for indexed functors. The underlying indexed morphism preserves the iterative Freyd structure, indexed biproducts, coproduct extensivity, the chosen iterative folders, and the compatible context action.}
\begin{equation} \label{eq:concrete-semantics-of-target-language-partial-CHAD-indexed-functor}
\semt{-}^{\mathfrak{I}} : \LSynTF\to \concINDEX
\end{equation}
over the Cartesian interpretation $\sem{-}_0$, with underlying natural transformation
\begin{equation}\label{eq:semantics-of-target-natural-transformation}
	\semt{-}^{\mathfrak{I}} : \LSyn\to \Vect _{\leastelement} ^{\left( \sem{-}_0\right) }
\end{equation}
The natural transformation satisfies the following: for each $k_1\in \NN $ and each $\mathbf{n} = \left( n_1, \ldots , n_{k_1} \right)\in\NN ^{k_1} $,  the component
\begin{equation}
\semt{-}^{\mathfrak{I}}_{\reals^{n_1}\times \cdots \times \reals^{n_{k_1}}} : \LSyn\left(\reals^{n_1}\times \cdots \times \reals^{n_{k_1}}  \right)\to  \Vect _{\leastelement}  ^{\left( \sem{\reals^{n_1}}_0  \times\cdots\times \sem{\reals^{n_{k_1}}}_0  \right)}
\end{equation}
of the natural transformation \eqref{eq:semantics-of-target-natural-transformation}
at   $\reals^{n_1}\times \cdots \times \reals^{n_{k_1}} $ is such that, for each primitive $\op\in \Op^{\mathbf{n}}_\mathbf{m}$,
$$\semt{D\op}^{\mathfrak{I}}_{\mathbf{n} }\in\Vect _{\leastelement} ^{\left( \sem{\reals^{n_1}}_0  \times\cdots\times \sem{\reals^{n_{k_1}}}_0  \right)} \left( \semtlu{\reals_{\mathbf{m}} } \circ  \semt{\op}, \semtl{\reals^{\mathbf{n}}}  \right) , $$ where
$\semt{D\op}^{\mathfrak{I}}_{\mathbf{n} } $ is the family of linear transformations obtained by interpreting the chosen program $D\op$.
\end{theorem}
\begin{proof}
	Interpretation follows the type and term constructors. The indexed equations, biproducts, context comparisons, and folders are interpreted by the corresponding structure of $\concINDEX$. The Cartesian dependent sums and linear-function values are interpreted by dependent sums of sets and fibrewise linear maps, so their value substitution and $\beta\eta$-equations hold. Dependent sum and pair elimination reindex the entire context along the corresponding value patterns; the decompositions of dependent sums of sets validate their computation and uniqueness equations. The sequential dependent-pair rule is sound because computational reindexing agrees with the returned fibre on the graph of its primal computation. These interpretations therefore respect the full target theory and induce the displayed indexed morphism. Uniqueness as an interpretation of that theory follows by induction on the syntax.
\end{proof}

The op-Grothendieck constructions of $\LSynTF$ and $\concINDEX$, together with \eqref{eq:concrete-semantics-of-target-language-partial-CHAD-indexed-functor},
induce an iterative Freyd category morphism between the corresponding op-Grothendieck constructions. This defines the most important aspect of the semantics of the target language to our correctness proof.
More precisely:

\begin{corollary}[Container semantics of the target language]\label{theo:concrete-target-language-partially-defined-functions-containers}
		The op-Grothendieck construction functor takes \eqref{eq:concrete-semantics-of-target-language-partial-CHAD-indexed-functor} to
	an iterative Freyd category morphism
\begin{equation}\label{eq:Freyd-category-morphism-iteration}
	\semt{-} : \TPSYN   \to \FFFVECT
\end{equation}
	where $ \TPSYN  \defeqq \Sigma^\op\left( \LSynTF \right) $ corresponds to the category of containers over the target language, as in Corollary \ref{coro:Iterative-Freyd-category-sematics}.	 In particular, for each primitive operation $\op\in\Op^{\mathbf{n}}_{\mathbf{m}}$ and its chosen backward program $D\op$, we have that
	 \begin{equation}
	 	\semt{\left( \op , D\op\right)} = \left( \sem{\op}_0, \semt{D\op}^{\mathfrak{I}}_{\mathbf{n}}\right)
	 \end{equation}
\end{corollary}
\begin{proof}
	It follows from Theorem \ref{theo:iterative-Feyd-indexed-category} and Theorem \ref{th:iterative-Grothendieck-iteration-free-container}.
\end{proof}

\subsection{Iterative CHAD: the AD macro and its categorical semantics}\label{sec:MACRO}

We finally can extend the macro, defined in \cite{lucatellivakar2021chad} for total languages and discussed in Subsection~\ref{sub:correctness-macro-total}. This is still defined in the structure-preserving
manner described, resulting in the following
definition for iteration, which basically is given by the principle of our categorical semantics introduced in Subsect.~\ref{syn:container-folder-iteration}.

That is to say, after implementing the derivatives of the primitive operations following the same principle of total CHAD, we define CHAD as the only \textit{structure-preserving transformation (iterative Freyd category morphism)} between the source language and the op-Grothendieck construction (category of containers) obtained from the target language that extends the implementation on the primitive operations. This allows for a straightforward correctness proof (presented in Theorem \ref{theo:fundamental-correctness-total-partial-language}).

The earlier treatment of iteration and recursion in \cite{2022arXiv221007724L} uses a dual-numbers transformation and a logical-relations correctness argument. Here we extend the dependent cotangent transformation of expressive total CHAD: iteration is lifted to its op-Grothendieck target by the folders developed in Part II, and correctness follows from the first-order concrete derivative functor.

We first state the universal property that determines the transformation. As in total CHAD, correctness is relative to the chosen implementations of the derivatives of primitives: we assume that $\semt{(\op,D\op)}=\DSemCHAD{\sem{\op}}$ for every primitive operation $\op$. This assumption includes agreement of the domains of definition.
\begin{theorem}\label{theo:macro-iterative-Freyd-category-morphism}
	There is only one iterative Freyd category morphism
	\begin{equation}
		\DSynrevt =\left( 	\DSynrevt ^v, 	\DSynrevt ^c \right) :  \Syn \to \TPSYN
	\end{equation}
such that  for each $n\in\NN $,
\begin{equation}
	\DSynrevt ^v \left( \reals ^n \right) = \left( \reals ^n , \tangentreals^n\right),
\end{equation}
and, for each $\op\in\Op$,   $\DSynrevt \left(\op \right) $ implements the CHAD-derivative
of the primitive operation $\op\in\Op$ of the source language, that is to say,
\begin{equation}
	\DSynrevt ^c  \left(\op \right) = \left( \op, D\op\right) \quad\mbox{ and consequently }\quad \semt{\DSynrevt \left(\op \right)} = \DSemCHAD{\sem{\op}} ,
\end{equation}
where
\begin{equation}
	\DSemCHAD{}\coloneqq\left( \DSemtotal{},  \DSempartial{ }\right) : \FFMAN\to \FFFVECT
\end{equation}
 is the \textit{CHAD-derivative} $\ww$-Freyd category morphism defined in \eqref{eq:Partial-CHAD}.
The split transformation $\DSynrevt$ has the paired macro $\DSynmacrot{-}$ below as its concrete representation, as proved in Lemma~\ref{lem:paired-CHAD-representation}.
\end{theorem}
\begin{proof}
Since $\TPSYN$ is an iterative Freyd category, the universal property of $\Syn$ in Theorem~\ref{theo:universal-property-source-Freyd-language} gives existence and uniqueness of $\DSynrevt$. The type and term clauses below spell out this split construction and its paired implementation.
\end{proof}

We write
\begin{itemize}
    \item given $\Gamma;\lvar:\cty{1}\vdash \trm{1}:\cty{2}_i$, we write the $i$-th coprojection\\ $\Gamma;\lvar:\cty{1}\vdash
    \tCoProj{i}(\trm{1})\defeq \tTuple{\tZero,\ldots,\tZero,\trm{1},\tZero,\ldots,\tZero}
    :{\cty{2}_1}\t*{\cdots}\t*{\cty{2}_n}$;
    \item for a list $\var{1}_1,\ldots,\var{1}_n$ of distinct identifiers, we write $\idx{\var{1}_i}{\var{1}_1,\ldots, \var{1}_n}\defeq i$ for the index of the identifier $\var{1}_i$ in this list.
\end{itemize}

We define for each type $\ty{1}$ of the source language:
\begin{itemize}
    \item a cartesian type $\Dsynrevarg{\ty{1}}_1$ of reverse-mode primals;
    \item a linear type $\Dsynrevarg{\ty{1}}_2$ (with free term variable $\pvar$) of reverse-mode cotangents. The variable $\pvar:\Dsynrevarg{\ty{1}}_1$ denotes the primal point at which the cotangent is taken.
\end{itemize}
We extend $\Dsynrevarg{-}$ to act on typing contexts
$\Gamma=\var{1}_1:\ty{1}_1,\ldots,\var{1}_n:\ty{1}_n$ as
\[\begin{array}{lll}
    \Dsynrevarg{\Gamma}_1\defeq \var{1}_1:\Dsynrevarg{\ty{1}_1}_1,\ldots,\var{1}_n:\Dsynrevarg{\ty{1}_n}_1\qquad&\text{(a cartesian typing context)}\\
    \Dsynrevarg{\Gamma}_2\defeq {\subst{\Dsynrevarg{\ty{1}_1}_2}{\sfor{\pvar}{\var{1}_1}}}\t*{\cdots}\t*{\subst{\Dsynrevarg{\ty{1}_n}_2}{\sfor{\pvar}{\var{1}_n}}}\qquad&\text{(a linear type)}.\end{array}
\]

In $\Dsynrevarg{\Gamma}_2$, the substitution $\pvar:=\var{1}_i$ selects the cotangent type at the value of the $i$-th variable. We use the canonical reassociations of finite products; the cotangent type of an extended context is grouped as $\Dsynrevarg{\Gamma}_2\t*\subst{\Dsynrevarg{\ty{1}}_2}{\sfor{\pvar}{\var{1}}}$. Empty linear products and sums of linear maps denote the zero type and the zero map, respectively.

Similarly, we define for each term
$\Gamma\vdash\trm{1}:\ty{1}$ of the source language:
\begin{itemize}
\item a term $\Dsynrevarg{\Gamma}_1\vdash \Dsynrevarg[\vGamma]{\trm{1}}_1:\Dsynrevarg{\ty{1}}_1$ that represents the reverse-mode primal computation associated with $\trm{1}$;
\item a term $\Dsynrevarg{\Gamma}_1;\lvar:\csubst{\Dsynrevarg{\ty{1}}_2}{\sfor{\pvar}{\Dsynrevarg[\vGamma]{\trm{1}}_1}}\vdash \Dsynrevarg[\vGamma]{\trm{1}}_2:\Dsynrevarg{\Gamma}_2$ that represents the reverse-mode cotangent computation associated with $\trm{1}$.
\end{itemize}
Here, $\vGamma$ is the list of identifiers that occur in $\Gamma$ (that is, $\overline{\var{1}_1:\ty{1}_1,\ldots,\var{1}_n:\ty{1}_n}\defeq \var{1}_1,\ldots,\var{1}_n$).
Then, the idea is that $(\Dsynrevarg{-}_1, \Dsynrevarg{-}_2)$ defines the uniquely defined structure preserving functor from the source language to the op-Grothendieck construction of the target language that sends each operation $\op$ to its chosen transposed derivative $D\op$.

For the sake of efficiency, we often tuple up the primal and cotangent code transformations, so we can share repeated subcomputations between them.
That means that we send a source language term $\trm{1}$ typed according to $\Gamma\vdash\trm{1}:\ty{1}$ to a term of the target language typed as follows:
\[
\begin{array}{l}
    \Dsynrevarg{\Gamma}_1\vdash \Dsynrevarg[\vGamma]{\trm{1}}:\Sigma{\pvar:\Dsynrevarg{\ty{1}}_1}.{\Dsynrevarg{\ty{1}}_2\multimap \Dsynrevarg{\Gamma}_2}.
\end{array}\]
In the concrete semantics, the paired presentation agrees with
\[
\tPair{\Dsynrevarg[\vGamma]{\trm{1}}_1}{\lfun{\lvar}{\Dsynrevarg[\vGamma]{\trm{1}}_2}}.
\]
This agreement is established in Lemma~\ref{lem:paired-CHAD-representation}.
Observe that we can perform this tupling because the target language supports $\Sigma$-types (dependent sum types) and $\multimap$-types (types of homomorphisms, in the style of the enriched effect calculus), as described in Section~\ref{sec:target-language}. The dependent pairing rule evaluates the primal first. At its returned value, the computationally reindexed input of the backward term is the corresponding cotangent fibre. Thus the pair returns exactly when the primal returns.

Concretely, a morphism $(f,f'):(A,X)\to(B,Y)$ of the op-Grothendieck construction determines the partial dependent section
\[
 a\longmapsto (f(a),f'_a)\in\sum_{b\in B}\Vect(Y(b),X(a)),
 \qquad a\in\operatorname{dom}(f).
\]
Conversely, such a partial section determines $f$ and $f'$ on its domain; outside that domain, $f'_a:0\to X(a)$ is uniquely determined. These constructions are inverse. This is exactly the interpretation of dependent pairing and linear abstraction in our model, so tupling preserves both equality and the domain of definition.

In the displayed clauses, matching a computation abbreviates sequencing followed by value pattern matching, as in Section~\ref{sec:target-language}. Thus each intermediate computation is evaluated once, and the returned value determines the types of its accompanying backward maps. The clauses use computational reindexing, the coherent context comparisons, and canonical reassociations of products implicitly. In particular, a computation that returns an unchanged context still restricts the available cotangent to its domain of definition; the context comparison maps it back to the original context fibre.

The split transformation is defined on source terms modulo $\beta\eta$ by Theorem~\ref{theo:macro-iterative-Freyd-category-morphism}. For the paired macro, Lemma~\ref{lem:paired-CHAD-representation} consequently gives
\[
\trm{1}\beeq\trm{2}\quad\Longrightarrow\quad
\semt{\Dsynrevarg[\vGamma]{\trm{1}}}
=\semt{\Dsynrevarg[\vGamma]{\trm{2}}}.
\]

We then define our reverse-mode AD code transformation as follows on types
\begin{align*}
    &\Dsynrevarg{\reals^n}_1 \defeq \reals^n\\
    &\Dsynrevarg{\ty{1}_1\t*\cdots\t*\ty{1}_n}_1 \defeq \Dsynrevarg{\ty{1}_1}_1\t*\cdots\t*\Dsynrevarg{\ty{1}_n}_1
    \\
    &\Dsynrevarg{{\ty{1}_1\t+\cdots\t+ \ty{1}_n}}_1 \defeq {\Dsynrevarg{\ty{1}_1}_1\t+ \cdots \t+\Dsynrevarg{\ty{1}_n}_1}
    \\
    &\\
    & \Dsynrevarg{\reals^n}_2 \defeq \creals^n\\
    & {\Dsynrevarg{\ty{1}_1\t*\cdots\t*\ty{1}_n}_2} \defeq  \subst{\Dsynrevarg{\ty{1}_1}_2}{\sfor{\pvar}{\tPrj[1]\,\pvar}}\t*\cdots\t*\subst{\Dsynrevarg{\ty{1}_n}_2}{\sfor{\pvar}{\tPrj[n]\,\pvar}}\\
    &\Dsynrevarg{{\ty{1}_1\t+\cdots\t+ \ty{1}_n}}_2\defeq
    \vMatch{\pvar}{\tIni[1]\pvar\To \Dsynrevarg{\ty{1}_1}_2\mid\cdots\mid
    \tIni[n]\pvar\To\Dsynrevarg{\ty{1}_n}_2}
    \end{align*}
The coproduct clause selects the cotangent type of the branch actually returned by the primal. For example, at $\tIni[1]x:\reals\t+\reals^2$ the cotangent is a real number, whereas at $\tIni[2]y$ it is a two-dimensional vector. This dependent clause, like the product and primitive-operation clauses, is already present in expressive total CHAD~\cite{lucatellivakar2021chad}. The new clause below is the clause for iteration: its primal iterates, and its cotangent uses an iterative folder.

We define the transformation on terms as follows.
\begin{flalign*}
&\Dsynrevarg[\vGamma]{\op(\trm{1}_1,\ldots,\trm{1}_k)} \defeq  \ndots\ndots\cdot\cdot\cdot&& \pletin{\var{1}_1}{\var{1}_1'}{\Dsynrevarg[\vGamma]{\trm{1}_1}}{\cdots\\
&&& \pletin{\var{1}_k}{\var{1}_k'}{\Dsynrevarg[\vGamma]{\trm{1}_k}}{\\
&&&\tPair{\op(\var{1}_1,\ldots,\var{1}_k)}{\lfun\lvar \letin{\lvar}{{D\op}(\var{1}_1,\ldots,\var{1}_k)(\lvar)}{\\
&&&
\phantom{\tPair{\op(\var{1}_1,\ldots,\var{1}_k)}{\lfun\lvar }}
\lapp{\var{1}_1'}{
    \tPrj[1]{\lvar}
    }
+\cdots+
\lapp{\var{1}_k'}{
    \tPrj[k]{\lvar}
    }
}}}}
\end{flalign*}

\begin{flalign*}
    &\Dsynrevarg[\vGamma]{\var{1}} \defeq\nndots\nndots\ndots\ndots\cdot &&  \tPair{\var{1}}{\lfun{\lvar} \tCoProj{\idx{\var{1}}{\vGamma}}(\lvar)}
    \end{flalign*}
\begin{flalign*}
\Dsynrevarg[\vGamma]{\letin{\var{1}}{\trm{1}}{\trm{2}}}
\defeq \nndots\ndots\cdot\cdot&&
\pletin{\var{1}}{\var{1}'}{\Dsynrevarg[\vGamma]{\trm{1}}}{\\
&&
    \pletin{\var{2}}{\var{2}'}{\Dsynrevarg[\vGamma,\var{1}]{\trm{2}}}{\\
    &&
        \tPair{\var{2}}{\lfun\lvar
        \letin{\lvar}{\lapp{\var{2}'}{\lvar}}{
            \tFst\lvar+\lapp{\var{1}'}{(\tSnd \lvar)}
        }}
    }}
\end{flalign*}
\begin{flalign*}&
    \Dsynrevarg[\vGamma]{\tTuple{\trm{1}_1,\ldots,\trm{1}_n}} \defeq \nndots\ndots\cdot\cdot\cdot&&
    \pletin{\var{1}_1}{\var{1}_1'}{\Dsynrevarg[\vGamma]{\trm{1}_1}}{\cdots \\ &&&
    \pletin{\var{1}_n}{\var{1}_n'}{\Dsynrevarg[\vGamma]{\trm{1}_n}}{\\ &&&
    \tPair{\tTuple{\var{1}_1,\ldots,\var{1}_n}}{\lfun\lvar \lapp{\var{1}_1'}{(\tPrj[1]\lvar)}+\cdots\\
    &&& \phantom{\tPair{\tTuple{\var{1}_1,\ldots,\var{1}_n}}{}}+ {\lapp{\var{1}_n'}{(\tPrj[n] \lvar)}}}}}
    \end{flalign*}
\begin{flalign*}
    \Dsynrevarg[\vGamma]{\tMatch{\trm{1}}{\var{1}_1,\ldots,\var{1}_n}{\trm{2}}} \defeq \ndots\ndots\cdot\cdot\cdot\!\!&&
    \tMatch{\Dsynrevarg[\vGamma]{\trm{1}}}{\var{1},\var{1}'}{}\\&&
    \tMatch{\var{1}}{\var{1}_1,\ldots,\var{1}_n}{}\\&&
    \tMatch{\Dsynrevarg[\vGamma,\var{1}_1,\ldots,\var{1}_n]{\trm{2}}}{\var{2},\var{2}'}{}\\&&
    \tPair{\var{2}}{\lfun\lvar\letin{\lvar}{\lapp{\var{2}'}{\lvar}}{\tPrj[1]\lvar+\lapp{\var{1}'}{(\tPrj[2]\lvar)}}}
\end{flalign*}
\begin{flalign*}&
\Dsynrevarg[\vGamma]{\tIni[i]\trm{1}} \defeq \nndots\ndots\ndots\ndots&&
\pletin{\var{1}}{\var{1}'}{\Dsynrevarg[\vGamma]{\trm{1}}}{\tPair{\tIni[i]\var{1}}{\var{1}'}}\end{flalign*}
In each sum branch, value pattern matching specializes the input type of $\var{2}'$ along the selected injection.
\begin{flalign*}&
    \Dsynrevarg[\vGamma]{\vMatch{\trm{1}}{\tIni[1]\var{1}_1\To\trm{2}_1\mid\cdots\mid \tIni[n]\var{1}_n\To\trm{2}_n}} \defeq \cdot\cdot\!\!&&
        \pletin{\var{2}}{\var{2}'}{\Dsynrevarg[\vGamma]{\trm{1}}}{\\ &&&
        \vMatch{\var{2}}{\tIni[1]\var{1}_1\To
        \\ &&&\quad \pletin{\var{3}_1}{\var{3}_1'}{\Dsynrevarg[\vGamma,\var{1}_1]{\trm{2}_1}}{\\ &&&
        \quad \tPair{\var{3}_1}{\lfun\lvar
        \letin{\lvar}{
            \lapp{\var{3}_1'}{\lvar}
        }{\tFst\lvar+\\ &&&
        \qquad\qquad\lapp{\var{2}'}{(\tSnd\lvar)}}}}\\ &&&
        \qquad\qquad\mid\cdots\mid\\ &&&
        \qquad\qquad\tIni[n] \var{1}_n\To
        \\ &&&\quad \pletin{\var{3}_n}{\var{3}_n'}{\Dsynrevarg[\vGamma,\var{1}_n]{\trm{2}_n}}{\\ &&&
        \quad \tPair{\var{3}_n}{\lfun\lvar
        \letin{\lvar}{
            \lapp{\var{3}_n'}{\lvar}
        }{\tFst\lvar+\\ &&&
        \qquad\qquad\lapp{\var{2}'}{(\tSnd\lvar)}}}}}}
    \end{flalign*}
For the iteration clause, write $f_1(\var{1})$ and $f_2(\var{1};\lvar)$ for the primal and linear components $\Dsynrevarg[\var{1}]{\trm{1}}_1$ and $\Dsynrevarg[\var{1}]{\trm{1}}_2$ of the transformed body. These are abbreviations for open terms, rather than bindings of functions in the target language.
\begin{flalign*}
\Dsynrevarg[\var{1}]{\tIt{\var{1}}{\trm{1}}} \defeq \ndots\cdot\cdot\cdot\cdot\cdot&&
\tPair{\tIt{\var{1}}{f_1(\var{1})}}{\lfun\lvar  \tFold{\lvar}{\lvar}{f_2(\var{1};\lvar)}}.
    \end{flalign*}
The folder starts with a cotangent at the final result and applies the backward maps of the successive loop steps in reverse order. Its type is precisely the reindexing along the primal iteration supplied by the iterative indexed structure.

Similarly, we derive the derivative rule for iteration with context by carrying the context in the loop state, exactly as in Section~\ref{ap:source-language}. Write $\Gamma=\var{2}_1:\ty{3}_1,\ldots,\var{2}_n:\ty{3}_n$ and let
\[
\begin{aligned}
h\defeq{}&\tMatch{\var{3}}{\var{2}_1,\ldots,\var{2}_n,\var{1}}{}\\
&\quad\vMatch{\trm{1}}{\tIni[1]u\To\tIni[1]u\mid\tIni[2]v\To\tIni[2]\tTuple{\var{2}_1,\ldots,\var{2}_n,v}}.
\end{aligned}
\]
Then the derived clause is
\begin{align*}
\Dsynrevarg[\vGamma,\var{1}]{\tIt[\Gamma]{\var{1}}{\trm{1}}}
\defeq{}&
\letin{\var{3}}{\tTuple{\var{2}_1,\ldots,\var{2}_n,\var{1}}}{
\Dsynrevarg[\var{3}]{\tIt{\var{3}}{h}}}.
\end{align*}
Here the cotangent of the enlarged state contains both the cotangent of the context and that of the loop state. Keeping both components accounts for the dependence of every loop step on its parameters.

\begin{lemma}\label{lem:paired-CHAD-representation}
For each source judgement $\Gamma\vdash\trm{1}:\ty{1}$, the macro $\Dsynrevarg[\vGamma]{\trm{1}}$ is well typed in context $\Dsynrevarg{\Gamma}_1$ at $\Sigma p:\Dsynrevarg{\ty{1}}_1.\,\Dsynrevarg{\ty{1}}_2\multimap\Dsynrevarg{\Gamma}_2$. Under the correspondence above, its interpretation is that of the morphism $\DSynrevt^c(\trm{1})$ determined in Theorem~\ref{theo:macro-iterative-Freyd-category-morphism}.
\end{lemma}
\begin{proof}
Typing follows by induction. Sequential pairing types the primitive and iteration clauses, using the assumed type of $D\op$ and the folder rule, respectively. Sequencing, tupling and injections use value pairing and linear abstraction after their subcomputations return. Product elimination substitutes the tuple pattern through the type of the backward map. For a sum branch, put $X=\Dsynrevarg{\Gamma}_2$, $P_i=\Dsynrevarg{\ty{2}_i}_2[x_i/p]$ and $Q=\Dsynrevarg{\ty{1}}_2[z_i/p]$. The branch context gives $y':P_i\multimap X$ and $z_i':Q\multimap(X\t*P_i)$. Hence the displayed backward term has type $Q\multimap X$, and the pair has the claimed result type, independent of the intermediate bindings. The contextual iteration clause uses the canonical product reassociation.

For the interpretation, we again induct on the source term. Variables and primitive operations satisfy the claim by definition. For sequencing, suppose that the first transformed computation returns $(b,k)$ and that the second, in the extended context, returns $(c,h)$. The displayed clause returns $c$ together with the linear map
\[
 v\longmapsto \pi_1(h(v))+k(\pi_2(h(v))).
\]
This is precisely composition with context in the op-Grothendieck construction. Both computations must return for the pair to be defined. Tuples sum the component backward maps after the corresponding cotangent projections; injections and case distinctions select the cotangent fibre of the returned branch. Thus these clauses agree with the corresponding categorical structure.

For iteration, suppose the primal starting at $a_0$ terminates after body evaluations at $a_0,\ldots,a_n$, with result $b$. Write $K_i:X(a_{i+1})\to X(a_i)$ for the backward map of a continuing step, $i<n$, and $K_n:Y(b)\to X(a_n)$ for that of the returning step. The interpretation of the folder gives
\[
 K_0\circ\cdots\circ K_n:Y(b)\to X(a_0),
\]
by induction on this finite trace. Hence the paired clause returns $(b,K_0\circ\cdots\circ K_n)$, exactly the paired representation of the lifted iteration. If the primal has no terminating trace, the pair is undefined. Carrying the context in the loop state gives the contextual clause, with its parameter cotangents included.
\end{proof}

\subsection{Correctness of the iterative CHAD}

Finally, we are in a position to state and prove the correctness of CHAD in the iterative language setting described above. Recall that our specification is given by the $\ww$-Freyd category morphism
\begin{equation}
	\DSemCHAD{} \coloneqq \left( \DSemtotal{}, \DSempartial{} \right) : \FFMAN \to \FFFVECT,
\end{equation}
which we refer to as the \emph{CHAD-derivative}, as established in~\eqref{eq:Partial-CHAD}. That is, the CHAD-derivative of the semantics of a program, defined by Theorem~\ref{the:semantics-source-language-iterative-Freyd-category-morphism} in the source language, should coincide with the semantics, as given in Corollary~\ref{theo:concrete-target-language-partially-defined-functions-containers}, of the macro $\Dsynrevarg{-}$ applied to that program.

There are two ways around the diagram: differentiate the meaning of a program, or first transform its code and then interpret the result. The work of Part II ensures that both ways preserve iteration, including its dependent cotangent component. Their agreement on primitives therefore propagates to every program by the universal property of the source language.

\begin{theorem}[Correctness of the Iterative CHAD] \label{theo:fundamental-correctness-total-partial-language}
Let $\DSynrevt$ denote the iterative Freyd category morphism defined in Theorem~\ref{theo:macro-iterative-Freyd-category-morphism} that implements the CHAD code transformation. Let $\sem{-}$ be the iterative Freyd category morphism giving the semantics of the source language, as introduced in Theorem~\ref{the:semantics-source-language-iterative-Freyd-category-morphism}. Finally, let $\semt{-}$ denote the iterative Freyd category morphism interpreting the target language, as described in Corollary~\ref{theo:concrete-target-language-partially-defined-functions-containers}.

	The diagram of \iterativeF morphisms below commutes.
	\begin{equation}\label{deag1:correctness-total-CHAD-partial}
		\diag{correctness-total-language-partial}
	\end{equation}
\end{theorem}
\begin{proof}
	By the universal property of the \iterativeF $\Syn$ established in Theorem \ref{theo:universal-property-source-Freyd-language}, since $\FFFVECT $ is an \iterativeFn{,} we have that there is only one \iterativeF morphism
	\begin{equation}
		\mathtt{C} = \left( \mathfrak{C} , \hat{\mathfrak{C}} \right)  : \Syn \to \FFFVECT,
	\end{equation}
	such that
	\begin{enumerate}[c1)]
		\item for each $n\in \NN $, $\hat{\mathfrak{C}} \left( \reals ^n\right) = \left( \RR ^n, \overline{\RR ^n} \right)  $;\label{condition1:correctness-total-functor-partial}
		\item $ \hat{\mathfrak{C}} \left( \op \right) = \DSempartial{\sem{\op}} = \semt{\DSynrevt ^c \left( \op \right)}   $ for each $\op \in \Op $.\label{condition2:correctness-total-functor-partial}
	\end{enumerate}
	Since both
	$ \semt{\DSynrevt\left( -\right) }$ and $\DSemCHAD\circ\sem{-} $    are \iterativeF morphisms $ \Syn\to \FFFVECT $
	such that \ref{condition1:correctness-total-functor-partial} and \ref{condition2:correctness-total-functor-partial} hold,
	we get that $\semt{\DSynrevt\left( -\right) } = \mathtt{C} = \DSemCHAD\circ\sem{-}$.
	That is to say, Diagram
	\eqref{deag1:correctness-total-CHAD-partial} commutes.
\end{proof}

We identify a morphism of the concrete op-Grothendieck construction with its paired representation, using Lemma~\ref{lem:paired-CHAD-representation}.
\begin{corollary}\label{coro:correctness-theorem-partial-CHAD}
	For any well-typed source program
	$	x: \ty{1} \vdash t: \ty{2}, $
	we have that $\semt{\DSynmacrot{t}} = \DSempartial{\sem{t}} $.
\end{corollary}
\begin{proof}
	By Lemma~\ref{lem:paired-CHAD-representation}, the semantics of the macro is the paired representation of $\semt{\DSynrevt^c(t)}$. The theorem identifies this morphism with $\DSempartial{\sem{t}}$. The canonical correspondence preserves equality and domains of definition, giving the claimed equality for the paired target program.
\end{proof}

The equality includes partiality: the transformed primal has the same domain of definition as the source program, and, on that domain, the backward map is the transpose derivative at the input, acting on cotangents at the returned primal. No derivative is asserted at a point where the source computation is undefined.

\section{Discussion and related work}\label{sec:related-work}
The present work brings together two questions: how iteration interacts with substitution in an indexed category, and how this interaction supplies the structure needed for iterative CHAD. Parameterized initial algebras have a long history in the semantics of data types \cite{lehmannSmyth1981,freyd1991algebraically,fiore2004axiomatic}. The initial-chain construction and its preservation properties are likewise established tools \cite{adamek1974,fioreSaville2017}. Our use of this theory concerns a particular class of functors, namely those given by substitution along a loop body. Iteration-extensivity relates their parameterized initial algebras to substitution along the resulting loop, while fibred iteration expresses the corresponding operation on containers.

This distinction also explains the role of iterative folders. A universal property is useful for constructing and understanding the concrete semantics. A target program, however, needs operations with which to express the transformation. The folder operations isolate this computational structure. Their correspondence with fibred raw iteration allows us to present the target language without requiring its syntax to carry all the universal properties used to construct the concrete model. The enriched development then establishes compatibility between the chosen operations and the least-fixed-point semantics.

CHAD was introduced for simply typed total languages in \cite{vakar2021chad,DBLP:journals/toplas/VakarS22}, following the structure-preserving perspective on AD of Elliott \cite{DBLP:journals/pacmpl/Elliott18}. CHAD for expressive total languages \cite{lucatellivakar2021chad} uses linear dependent types to accommodate sum, inductive and coinductive types. Efficient CHAD \cite{2023arXiv230705738S} studies implementation and complexity. We retain the dependent target and the principle of structure preservation of this development, and extend the first-order source language with partial operations and iteration.

Subsequent work on simply typed reverse-mode AD with variants \cite{lucatelliSimmVakar2026simply} represents cotangents by ambient types together with idempotents determined by primal values. Splitting these idempotents recovers the dependent cotangent structure. This gives a different way of treating the dependence already present in total CHAD. A natural next question is how iteration and iterative folders interact with that representation, and whether the iterative transformation developed here can be expressed in the resulting simply typed setting. The total-language result does not by itself establish this extension.

Further questions concern richer source languages and equations for iteration. The present account isolates the structure required for a first-order language with raw iteration; its concrete interpretation is least-fixed-point iteration. Extending the indexed development to general recursion and to languages featuring algebraic effects, including the finite probabilistic effects studied in \cite{simmLucatelliNunesVakar2026backpropagation}, remains a direction for further work.

\begin{acks}
We are deeply grateful to the anonymous referees for the exceptional
care and effort they devoted to reviewing this work. Their detailed
comments and constructive suggestions helped us clarify the motivation
and organisation of the paper and substantially improve its exposition.

	This research was supported by the ERC project FoRECAST. 	The second author acknowledges support from Google DeepMind.

	The first author gratefully acknowledges support from the Fields Institute for Research in Mathematical Sciences through a Fields Research Fellowship in 2023, as well as support from the Centre for Mathematics of the University of Coimbra (CMUC, \href{https://doi.org/10.54499/UID/00324/2025}{doi:10.54499/UID/00324/2025}), funded by the Funda\c{c}\~ao para a Ci\^encia e a Tecnologia (FCT) through the grants UID/00324/2025 and UID/PRR/00324/2025. He also acknowledges support from the Deutsche Forschungsgemeinschaft (DFG, German Research Foundation) through the project Higher-Order Monad-based Programming and Reasoning (HOMBRe), project number 501369690, led by Sergey Goncharov, during his research fellowship appointment at the School of Computer Science, University of Birmingham. The author also expresses sincere gratitude to Henrique Bursztyn and the Instituto Nacional de Matem\'atica Pura e Aplicada (IMPA) for their generous hospitality.

\end{acks}

\section*{Competing interests}
The authors declare none.

%\section*{Use of generative AI}
%The authors deliberately refrained from using %generative AI tools during the writing and revision %of the present paper.

\end{document}